\documentclass[preprint2,trackchanges]{aastex63}
\hypersetup{linkcolor=red,citecolor=green,filecolor=cyan,urlcolor=magenta}
\begin{document}
\newcommand{\vdag}{(v)^\dagger}
\newcommand\aastex{AAS\TeX}
\newcommand\latex{La\TeX}
\newcommand{\ct}{$^{13}$C}[
\newcommand{\ctb}{$^{13}$C~}
\newcommand{\cd}{$^{12}$C}
\newcommand{\cdb}{$^{12}$C~}
\newcommand{\nf}{$^{14}$N}
\newcommand{\nfb}{$^{14}$N~}
\newcommand{\cta}{$^{13}$C($\alpha$,n)$^{16}$O}
\newcommand{\ctab}{$^{13}$C($\alpha$,n)$^{16}$O~}
\newcommand{\nean}{$^{22}$Ne($\alpha$,n)$^{25}$Mg}
\newcommand{\neanb}{$^{22}$Ne($\alpha$,n)$^{25}$Mg~}
\newcommand{\fe}{$^{56}$Fe}
\newcommand{\feb}{$^{56}$Fe~}
\newcommand{\fes}{$^{60}$Fe}
\newcommand{\fesb}{$^{60}$Fe~}
\newcommand{\al}{$^{26}$Al}
\newcommand{\alb}{$^{26}$Al~}
\newcommand{\oq}{\textquotedblleft}
\newcommand{\cq}{\textquotedblright}
\newcommand{\cqb}{\textquotedblright~}
\newcommand{\ms}{M$_{\odot}$}
\newcommand{\msb}{M$_{\odot}$~}
\received{}
\revised{}
\accepted{}
%% Adds "Submitted to " the argument.
\submitjournal{ApJ}
\shorttitle{Nuclear and Stellar AGB Parameters constrained by SiC grains}
\shortauthors{Palmerini et al.}
\title{Presolar grain isotopic ratios as constraints to nuclear \\
and stellar parameters of AGB nucleosynthesis}
\correspondingauthor{Sara Palmerini}
\email{sara.palmerini@unipg.it, sara.palmerini@pg.infn.it}

\author{Sara Palmerini}
\affiliation{Department of Physics and Geology, University of Perugia, Via A. Pascoli snc, I-06123 Perugia, Italy} 
\affiliation{INFN, section of Perugia, Via A. Pascoli snc, I-06123 Perugia, Italy}

\author{Maurizio Busso}
\affiliation{Department of Physics and Geology, University of Perugia,  Via A. Pascoli snc, I-06123 Perugia, Italy} 
\affiliation{INFN, section of Perugia, Via A. Pascoli snc, I-06123 Perugia, Italy}

\author{Diego Vescovi}
\affiliation{Goethe University Frankfurt, Max-von-Laue-Strasse 1, Frankfurt am Main 60438, Germany;}
\affiliation{INFN, section of Perugia, Via A. Pascoli snc, I-06123 Perugia, Italy}

\author{Eugenia Naselli}
\affiliation{INFN, Laboratori Nazionali del Sud, Via S. Sofia, I-95129 Catania, Italy}

\author{Angelo Pidatella}
\affiliation{INFN, Laboratori Nazionali del Sud, Via S. Sofia, I-95129 Catania, Italy}

\author{Riccardo Mucciola}
\affiliation{Department of Physics and Geology, University of Perugia, Via A. Pascoli snc, I-06123 Perugia, Italy} 
\affiliation{INFN, section of Perugia, Via A. Pascoli snc, I-06123 Perugia, Italy}

\author{Sergio Cristallo}
\affiliation{INAF, Osservatorio Astronomico d'Abruzzo, Via Mentore Maggini snc, I-64100 Collurania, Teramo, Italy} 
\affiliation{INFN, section of Perugia, Via A. Pascoli snc, I-06123 Perugia, Italy}

\author{David Mascali}
\affiliation{INFN, Laboratori Nazionali del Sud, Via S. Sofia, I-95129 Catania, Italy}

\author{Alberto Mengoni}
\affiliation{ENEA, Agenzia Nazionale per la nuove Tecnologie, l’Energia e lo Sviluppo Economico Sostenibile, Via Martiri di Monte Sole 4, I-40129 Bologna , Italy}
\affiliation{INFN, Istituto Nazionale di Fisica Nucleare, Sezione di Bologna, Viale Berti Pichat 6/2, I-40127 Bologna, Italy}

\author{Stefano Simonucci}
\affiliation{Physics Division, School of Science and Technology, University of Camerino, Via Madonna delle Carceri 9B, I-62032 Camerino, Macerata, Italy} 
\affiliation{INFN, section of Perugia, Via A. Pascoli snc, I-06123 Perugia, Italy}

\author{Simone Taioli}
\affiliation{European Centre for Theoretical Studies in Nuclear Physics and Related Areas, Str. delle Tabarelle, 286, 38123 Villazzano, Trento, Italy}
\affiliation{Trento Institute for Fundamental Physics and Applications, Via Sommarive, 14, 38123 Povo ,Trento, Italy}
\affiliation{Peter the Great St. Petersburg Polytechnic University, Russia}

\begin{abstract}
\small

Recent models for evolved Low Mass Stars (with  $M \lesssim 3$ \ms), undergoing the AGB phase assume that magnetic flux-tube buoyancy drives the formation of \ctb reservoirs in He-rich layers. We illustrate their crucial properties, showing how the low abundance of \ctb generated  below the convective envelope hampers the formation of primary $^{14}$N and the ensuing synthesis of intermediate-mass nuclei, like $^{19}$F and $^{22}$Ne. In the mentioned models, their production is therefore of a purely secondary nature. Shortage of primary $^{22}$Ne has also important effects in reducing the neutron density. Another property concerns AGB  winds, which are likely to preserve C-rich subcomponents, isolated by magnetic tension, even when the envelope composition is O-rich. Conditions for the  formation of C-rich compounds are therefore found in stages earlier than previously envisaged. These issues, together with the uncertainties related to several nuclear physics quantities, are discussed in the light of the isotopic admixtures of s-process elements in presolar SiC grains of stellar origin, which provide important and precise constraints to the otherwise uncertain parameters. By comparing nucleosynthesis results with measured SiC data, it is argued that such a detailed series of constraints indicates the need for new measurements of weak interaction rates in ionized plasmas, as well as of neutron-capture cross sections, especially near the  N = 50 and N = 82 neutron magic numbers. Nontheless, the peculiarity of our models allows us to achieve fits to the presolar grain data of a quality so far never obtained in previously published attempts.
\end{abstract}

\vskip 0.5cm
\normalsize
\section{Introduction.} \label{sec:intro}
Several attempts are currently made in the literature for inferring which mixing mechanisms
be responsible for the formation of \ctb reservoirs in He-rich layers below the convective 
envelopes of evolved stars. These reservoirs have been proven to be present in
stars ascending for the second time the Red Giant Branch, called Asymptotic 
Giant Branch stars \citep[hereafter AGB, see e.g.][]{b+9,her1,straniero+06}. The availability of freshly produced \ctb concentrations 
is there crucial to understand the production of neutron-rich elements observed in their 
photospheres \citep{abia3, a20} as well as to account for the enrichment of heavy elements 
beyond Sr in the Galaxy \citep{mai1, mai2,m21}, especially those belonging to the so-called 
{\it main component} of the slow neutron-capture process, i.e. the {\it s-process} \citep{ka11}. 
Among the mentioned attempts, recent publications 
\citep{diego, b+21} underlined the relevance of Magnetohydrodynamics(MHD)-based  mixing schemes, made possible by the peculiar 
physics prevailing below the convective envelopes \citep{nor, nb14}, especially during the repeated 
downward penetrations of their convective envelopes, collectively indicated as the {\it Third Dredge-Up} \citep[hereafter TDU,][]{kl14}, 
which mix freshly produced elements to the surface. In this note, we want to discuss some
properties of the above models that immediately affect the ensuing nuclear yields for 
heavy nuclei. These topics will be analyzed, in conjunction 
with relevant issues concerning the nuclear input parameters (cross sections and 
weak-interaction rates), using as a test-bench the requirement of reproducing isotopic ratios 
of heavy elements measured in presolar SiC grains \citep{Zinner14, lugaro8}. We intend to show that, 
in order to account for the observations and to decide among different model possibilities,
we need a better assessment of neutron-capture cross sections near magic neutron numbers as well as 
new experimental data on $\beta$-decay rates in ionized plasmas. In order to demonstrate 
the above issues, in section \ref{sec:models} we illustrate some peculiar properties of recent models for evolved low-mass stars, while in section \ref{sec:nuclear} we discuss nuclear input parameters relevant for the synthesis of neutron-rich nuclei in the mentioned atomic mass region. Section \ref{sec:rates} then discusses the improvements possible today in the treatment of weak interactions. In particular, using as a guiding example the case of 
$^{134}$Cs, we show briefly
the requirements for nucleosynthesis models, as well as what can be obtained by modern theoretical approaches and what will be soon available in the field of experimental
verifications, through the project PANDORA \citep{Mascali2017, Mascali2020}. Subsequently,  in section \ref{sec:sic}, we briefly outline the experimental database of presolar SiC grain measurements and then in section \ref{sec:mod} we discuss how the model results compare with
these detailed constraints, examining the most important evidence emerging from the comparison. Preliminary conclusions from this work are then drawn in section \ref{sec:the-end}.  

\section{Relevant properties of the adopted stellar models.}\label{sec:models}
\subsection{Stellar layers hosting neutron captures}\label{sec:shell}
It is known since many years \citep{h+85, gb91} that the isotopic abundances shown by red giant photospheres
can be accounted for only if non convective mixing mechanisms below the envelope are at play.
Their nature has been an important object of research in stellar physics 
over the past decades \citep[see e.g.][and references therein]{w+95, lan99, her03, dv03, dt03, e+06, e+08, cri09, cl10, b+10, nb14, cri+18}. Some of 
these mechanisms must operate also below the
TDU in AGB stars,
in order to let protons from the envelope be mixed into the He-rich layers. They will
subsequently produce \ctb at H-burning reactivation and then neutrons
thanks to the \ctab reaction \citep{ar9, b+01, cris0, cri09, bi12, t+14}. Later on (several 10$^4$   yr later, in stars 
below about 3 \ms), a thermal instability occurs ({\it Thermal Pulse, or TP}) in the He shell that lays 
at the bottom of the He-rich zone  below the convective envelope. %into which TDU penetrates. 
This instability is due to the combined effects of the gradual compression of these layers, operated 
by H-shell burning, by the degeneracy of 
the C-O core and  by the natural difficulty of maintaining stability in a  thin burning shell 
\citep{y04}. An intermediate convective buffer 
then fills the He-rich region, while the temperature increases rapidly. For low-mass stars 
(1 $\lesssim$ M/\ms $\lesssim$ 3) this growth does not achieve values significantly larger than
$T \simeq$ 3$\cdot10^8$K, but in more massive red giants it can lead to quite 
higher temperatures (above 3.5 $\cdot 10^8$K). Then the complementary neutron 
source \neanb can be activated: it is restricted at rather high temperature regimes 
because its $Q$ value is negative \citep[$Q_{22}^{(\alpha,n)}$ = -478 keV, see e.g.][]{janis,adsley}. Hence, in low-mass stars (where, as said, $T \lesssim 3\cdot 10^8$ K) its effects are not large: they mainly control the freeze-out of reaction branchings for which 
nuclear parameters depend on $T$. The cycle of \ctab and \neanb activation is 
repeated several times in AGB stars: actually, repeated exposures are known since  
decades to be required to feed adequately the whole distribution from Fe to Pb \citep{s+65,k+90}.
     
\begin{figure}[t!!]
\includegraphics[width=\columnwidth]{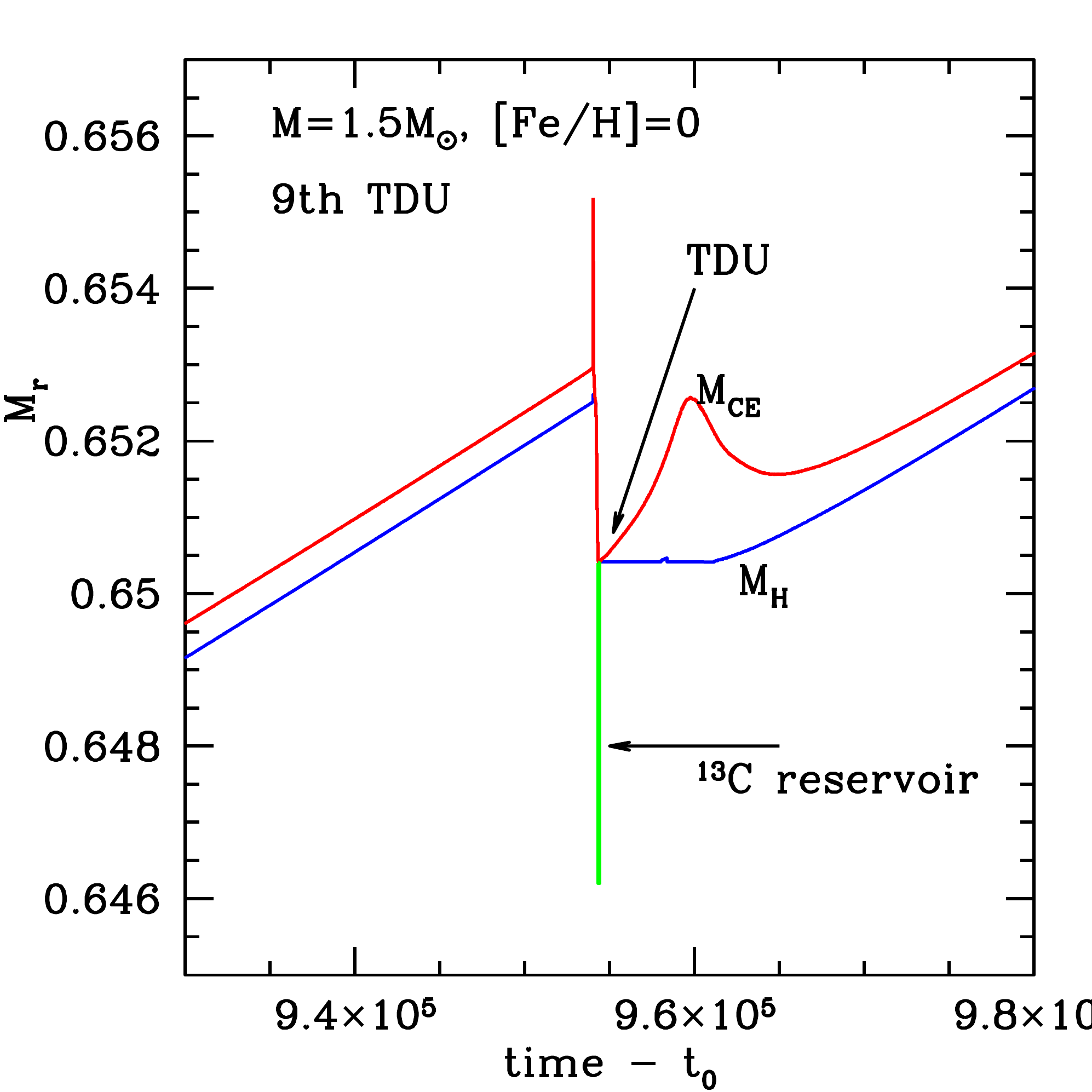}
\caption{The occurrence of the envelope penetration in a Third Dredge Up episode,
for a mass M=1.5 \msb of solar metallicity. The figure shows in red the innermost 
border of the convective envelope ($M_{\rm CE}$) and in blue the position of the 
H/He interface ($M_{\rm H}$). Its minimum before H-burning restarts is called 
{\it post-flash dip}. The parameter $t_0$ is the stellar age at the moment of 
the first TDU episode. Note how, of the rather long duration of 
the post-flash dip ($\sim 10^4$   yr), only a short fraction (about a century) is really occupied by TDU. The \ctb pocket (here extending for 4.2x10$^{-3}$ \ms) is represented
in green. Its structure, as well as that of the previous H-rich pocket, are expanded in Figure \ref{fig:pocket}.  
\label{fig:tdu}}
\end{figure}

\begin{figure}[t!!]
\includegraphics[width=\columnwidth]{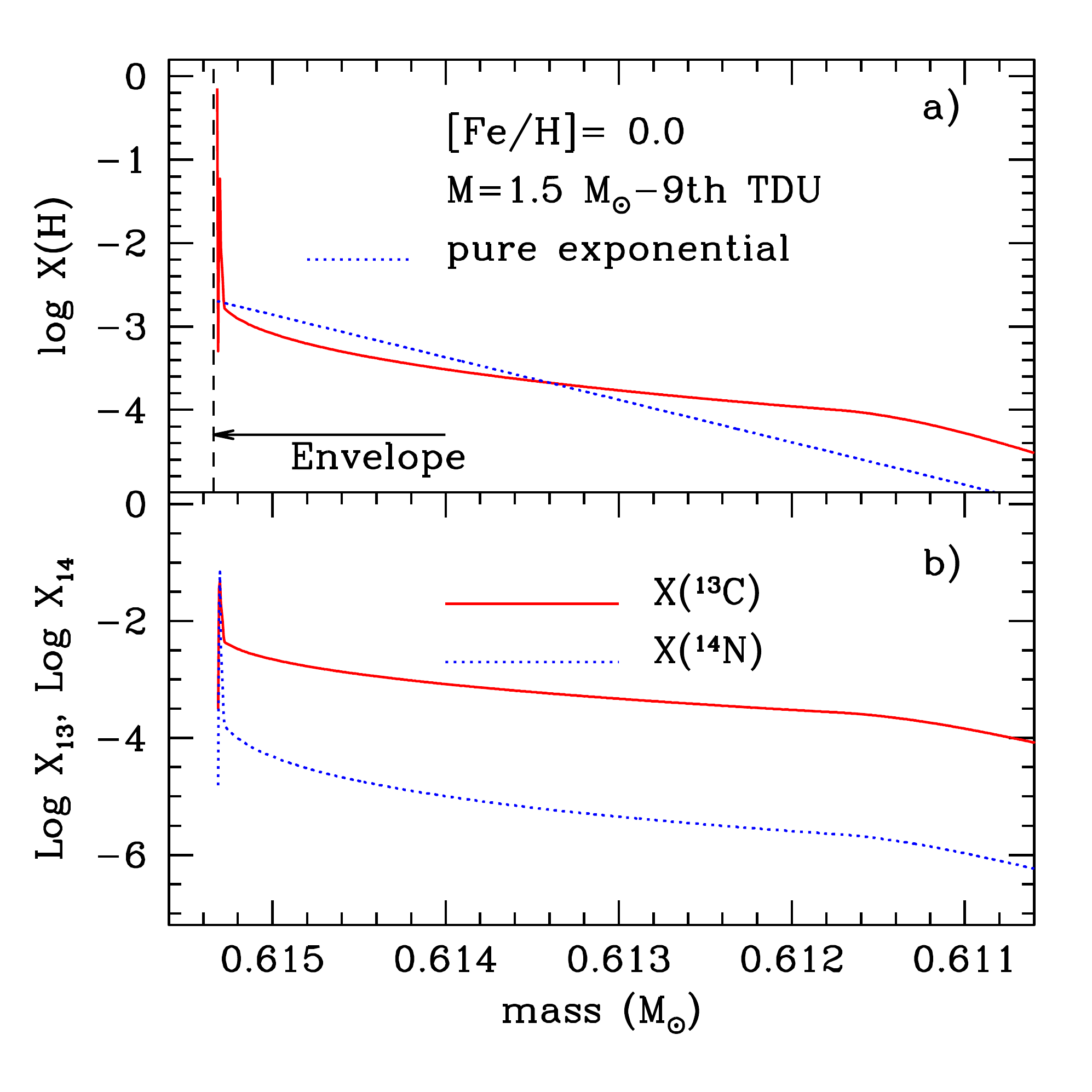}
\caption{The abundances in the \ctb pocket at the TDU represented in Figure \ref{fig:tdu}. Panel
{\it a} shows the profile of the proton abundance, penetrated according to equations 
from (14) to (17) by \citet{t+16}, as computed in \citet{b+21}. 
Panel {\it b} shows the ensuing abundances of \ctb and \nfb formed after hydrogen burning
in the shell restarts. Later, in these layers the \ctab reaction will release 
neutrons for $s-$processing. \label{fig:pocket}}
\end{figure}

In the framework outlined above, previous MHD models for mass circulation in 
evolved stars \citep{b+07, nb14, t+16,p+18} have been applied recently \citep{diego, b+21} 
to the production of \ctb
and then of neutron-rich elements beyond Fe through slow neutron captures. 
The basic properties of those recent computations can be understood with 
reference to Figures \ref{fig:tdu} and \ref{fig:pocket}. The first one 
illustrates the occurrence of a TDU episode, with the underlying layers that
must be affected by extra-mixing, while the second shows the peculiar 
distribution of protons (and later of \ctb and \nf) ensuing from  this specific 
mixing scheme. Extensive discussions of the computations that lay behind the formation
of reservoirs similar to the one of Figure \ref{fig:pocket} were presented in  
the above mentioned works.

As Figure \ref{fig:pocket} shows, the new distribution has some peculiar characteristics.
Namely, the abundance of protons remains always very 
low; so low, actually, that H burning cannot efficiently proceed to \nf, because 
it is consumed almost completely by the \ctb production. This is so everywhere, except
in a thin top layer where \ctb is abundant. By comparison, a purely exponential
distribution is shown in Figure \ref{fig:pocket} (see panel {\it a}). It 
presents a rather wide layer in mass where protons are sufficiently abundant 
to synthesize considerable \nfb concentrations. Since \nfb is an efficient 
{\it poison}, or neutron-absorber, the number of neutrons available for being 
captured by $^{56}$Fe and its progeny is enhanced in our new models
and so is the neutron exposure per cycle, in the various pulse-interpulse 
cycles mentioned for AGB stars.

By contrast, in our computations the shortage of \nfb cuts drastically the nuclear channels
starting from nitrogen, thus reducing the production of several isotopes in the mass 
range 15-30, including $^{18}$O, $^{19}$F and $^{22}$Ne. The reduced production of this 
latter, in particular, hampers the efficiency of the \neanb source in the $TP$s, leading to a type 
of $s$-processing never achieving high neutron densities, hence feeding only 
marginally nuclei at the neutron-rich side of crucial branching points of the 
$s$-process, like $^{86}$Kr, $^{87}$Rb, $^{96}$Zr, etc., even during the neutron flows 
available in the convective buffers accompanying $TP$s. Another consequence 
of the reduced \nfb abundance is that the production of $^{19}$F is inhibited sharply, 
which fact promises to account well for its low abundance in AGB photospheres \citep{carlos1, carlos}. 
This last point deserves a closer scrutiny, which goes beyond the scopes of this note 
{ and is presented by \citet{ves21}}.
  
\subsection{Magnetic effects in stellar winds}\label{sec:magwinds}

Our mixing scheme was originally built from an analogy
based on the physics of magnetic flux tubes in the Sun
\citep{b+07}. Solar magnetic structures cross the 
convective layer emerging from the regions in the tachocline 
where the dynamo mechanism has its roots and a similar
behavior can be at least partially envisaged in evolved stars
\citep{Ay+81}. A substantial difference { lies} in the 
fact that stars of classes IV, III and II, with spectral types
later than K3 stay, in the Hertzsprung-Russell diagram, at the cool side of the so-called 
{\it Coronal Dividing Line}, or $CDL$ \citep{lh79}, 
i.e. of the border separating bluer, $X$-ray emitting active stars (where
high-energy radiation is due to charged particles trapped in coronal
loops) from redder giants not displaying evidence of a magnetized corona
in space-borne $X$-ray observations \citep{h+91}. 
Following \citet{hs01}, the dividing 
line can actually be seen as a border beyond which the large 
convective envelopes tend to trap the buoyant flux tubes, 
generally hampering their outward emergence in a corona.  

The above scenario is largely accepted; for our 
purposes, trapping and breaking of magnetic flux tubes in 
the envelope guarantees the mixing of nucleosynthesis products 
\citep{t+16, diego, b+21}. 

Despite the lack of a real corona as a large-scale 
structure, observations of magnetic fields at the surface of AGB
stars are numerous 
\citep[see e.g. reviews by][and references therein]{vl11,vl12}. 
In particular, \citet{j+05} reported field 
values of the order of a kilogauss in central stars of Planetary 
Nebulae (PNe), while fields of lower intensity (up to 10 G) were observed
through VLBI techniques since the nineties \citep{kd97}. 
More recently, \citet{her06, her08}
reported measurements of the magnetic Zeeman effect on SiO masers 
at 2-10 AU from the central stars, suggesting that the corresponding
fields act as catalysts for dust formation and as collimators for the 
winds. Fields of similar intensity (from 0.6 to 10 G)
were observed  around C-rich evolved stars \citep{du+17} and post-AGB 
objects \citep{s15b, s15a}.

MHD models of a dynamo effect in AGB  stars and of its consequences
on mass outflows were performed by 
\citet{pl8, pl10, p20}. They suggest that, 
from the point of view 
of a distant observer, the nebula generated around an AGB star would
not show signs of deformation from a spherical shape up to and 
through the slow superwind phase. By contrast, anisotropic structures
would be already developing, induced by magnetic sources; these 
ones would remain hidden in the innermost regions, creating a  
bipolar cavity during the early superwind phases. Then, the pre-PNe
stage would begin when the fast wind emitted by the core engulfs 
this cavity and increases the anisotropy of the gas distribution.

The development of anisotropic structures was early suggested by 
\citet{ros91, ros95}, who
assumed a change in the topology of the atmospheric magnetic
field from large, coherent coronal loops (that would appear at 
the hot side of the $CDL$) to an open field-line geometry
where only small loops would remain (so that the star would move
to the cool side of the $CDL$). 
The larger surface fraction covered by open field lines would then
allow for the escape of a cool, massive stellar wind driven by 
Alfvén waves. Among subsequent suggestions of persisting links
between magnetic activity and AGB winds, one has to mention the work
by \citet{sk03}. According to them, magnetic 
flares can occur in AGB stars, above photospheric cool regions 
similar to solar dark spots. Their presence would enhance locally
both the wind efficiency and the dust formation. 
Another important confirmation on the presence of magnetic fields in low mass stars comes from recent white dwarf (WDs) studies. Magnetism in WDs is determined via polarization measurements and the observation of characteristic distortions and shifts of spectral lines due to the Zeeman effects. %or characteristic Zeeman patterns of spectral lines observed in intensity. 
In the past, the estimated frequency of the occurrence of magnetic
fields in WDs was rather low, because it was mainly based on discoveries made with low resolution classification spectroscopy, which is mainly sensitive to field 
strength between 1 and 100 MG. In recent years, new high-resolution spectroscopic and spectropolarimetric measurements have shown that at least 20\% of WDs host magnetic 
 fields, with typical strengths from a few kG up to about 1000 MG. Recently, \citet{bl20} discovered a significant number of new WDs with strong magnetic 
 fields (from 5 to 200 MG strength) in the 20 pc volume around the solar system, highlighting again the importance of spectropolarimetry in detecting magnetic fields 
 in these compact objects. This result is in agreement with previous suggestions \citep{ac04,kw14}, confirming the suspicion that 
 magnetism is more a common than a rare characteristic of WDs.

For the purposes of this work, we can then recognize that there is
ample support in the literature to the existence of open 
magnetic structures at the AGB surface (relics of the original
flux tubes survived to turbulence in the convective layers) 
where dust formation can occur and even become very efficient. 
This possibility is
broadly in line with the present picture of AGB mass loss, where the 
traditional idea of a rate steadily increasing with time, up to the final 
superwind stages where most of the mass would be lost, has been substituted 
by a very complex scenario, characterized by a non-monotonic growth of
the wind efficiency, affected by pulsation and other dynamical phenomena
and in which the total mass ejected in final hot superwind stages is possibly
reduced by their rather short duration \citep[see e.g.][for an updated review]{ho18}

\begin{figure}[t!!]
\includegraphics[width=\columnwidth]{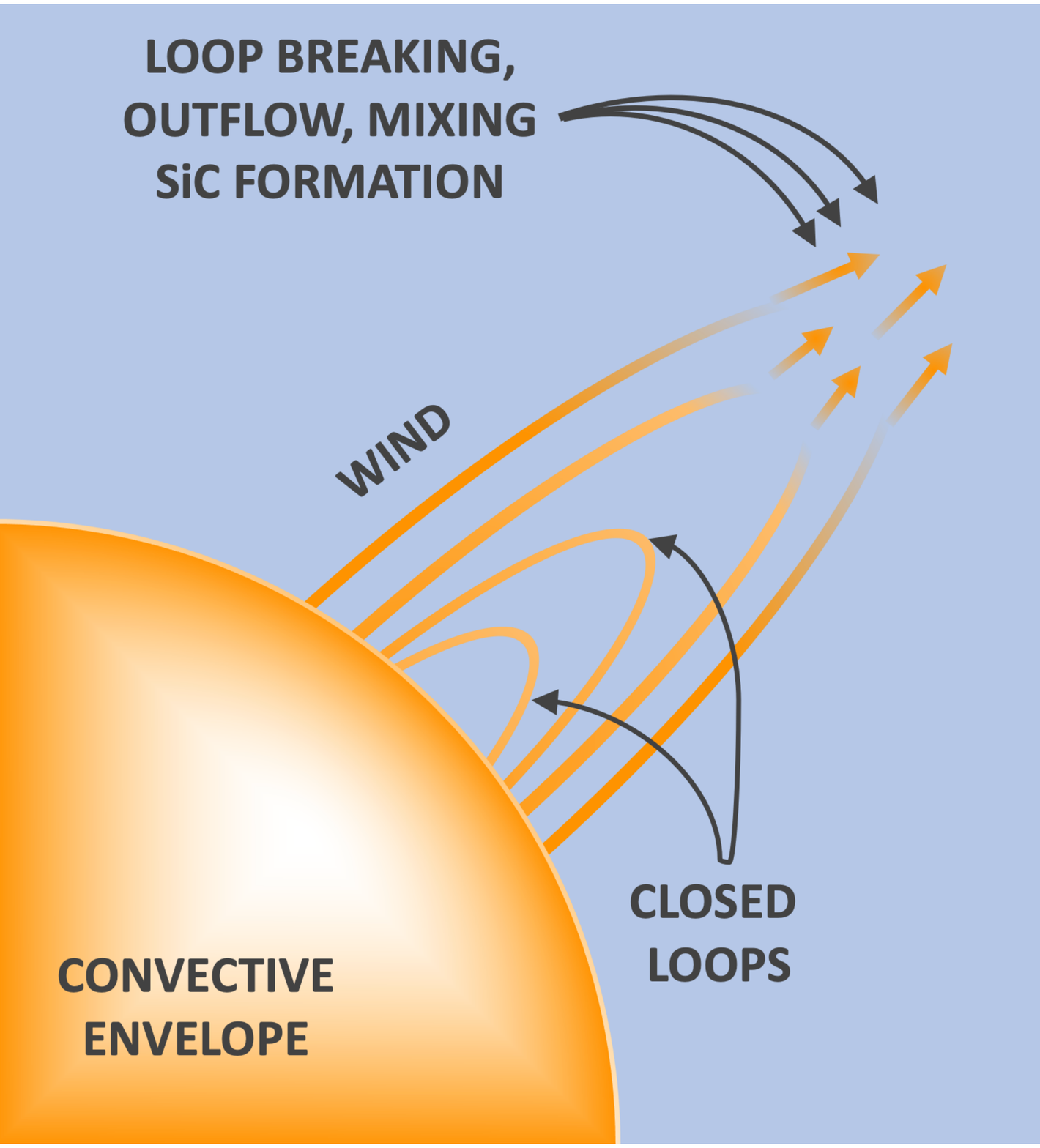}
\caption{A cartoon representing schematically the breaking of magnetic structures in the stellar winds (e.g. coronal flux 
tubes, carrying C-rich matter and $s$-process nuclei). 
\label{fig:cartoon}}
\end{figure}

In our approach, among the dynamical phenomena affecting the winds in the TDU phase there would be the formation of regions in which a few magnetic structures, 
carrying C-rich material outward, would ultimately open, accompanying 
and possibly guiding gaseous and dusty winds. A schematic representation of this phenomenon is shown as a cartoon in Figure
\ref{fig:cartoon}.

The above scenario offers an important tool to the comparison of 
model results with observations. In particular, C-rich structures, 
breaking within the wind and mixing there with the ambient composition, 
would be present at any moment in the evolution of an AGB star. 
As the material they transport from the He-shell regions is very 
largely enriched in carbon (about 25\% of abundance in mass) 
tiny traces of it remaining in the 
wind would be sufficient to permit formation of blobs with a C/O ratio 
exceeding unity, so that carbon-based dust like SiC can form, 
even when the general composition of the envelope is O-rich.

In order to account for the above possibility, in section \ref{sec:mod} 
and in the accompanying figures we shall 
compare the isotopic composition measured in presolar SiC 
grains both with model envelope abundances and with 
those of a phase (called {\it magnetic wind}, or $MW$)
obtained by mixing the envelope with a limited percentage (from 0.5\% to 5\%)
of material having the He-shell composition most recently
transported outward by flux tubes. This would mimic the effects
of magnetic flares or of other instabilities breaking the
originally closed loops into open wind structures. A preliminary 
version of this idea was presented earlier, in a simplified form, by
\citet{b+21}, where the composition of this C-rich wind phase was roughly
approximated through the so-called {\it G-component} of the $s$-processed material \citep{zin98}.

\section{Status of the set of nuclear parameters adopted}\label{sec:nuclear}

As mentioned, detailed information on the isotopic admixture of $s$-processed
materials entering the winds of AGB stars and forming dust there are contained in the
record of presolar SiC grains, especially of the {\it Main-Stream} group. From there
we can infer relevant constraints on the nuclear parameters controlling stellar neutron
captures, especially for nuclei near the magic neutron numbers $N = 50$ and $N = 82$.
Recent measurements for them can be found in the updates of the presolar grain database at the 
Washington University of Saint Louis \citep{wustl,wustl2020}. We shall therefore use the most
recent sets of data from this repository in order to try an iterative process for 
improving our knowledge of the nuclear physics that lays behind $s$-processing. We 
start describing here below the status of the nuclear inputs adopted from the 
literature, in the aim of finding required and possible improvements. 

Except for special cases, mentioned when necessary, Maxwellian-averaged neutron-capture 
cross sections (MACS) are taken from the KADONIS on-line repository,
in its version 1.0 \citep[hereafter K1, see][and the KADONIS webpage\footnote{https://exp-astro.de/kadonis1.0/}]{dil}, 
whose recommended values are in any case compared with those proposed 
by the {\it National Nuclear Data Center} of the {\it Brookhaven National Laboratory}\footnote{https://www.nndc.bnl.gov/astro/} \citep{bnl}. Choices from this
repository will be indicated briefly as coming from {\it BNL}. When pertinent, the K1 repository 
presents recommended values that derive from theoretical calculations of the energy dependence, 
normalized to an actually measured value. They are computed as an average over recent 
compilations, among which one must cite {\it ENDF/B-7.1} \citep{endfb}, {\it TENDL15}\footnote{https://tendl.web.psi.ch/tendl$\_$2015/tendl2015.html},
{\it JEFF3.2}\footnote{http://www.oecd-nea.org/dbforms/data\-/eva/\-evatapes/jeff$\_${32}/} and {\it JENDL4.0}
\citep{jendl}. When renormalizations to gold are necessary, the suggestions by \citet{au2} and \cite{au1} 
are adopted. All the recent measurements from the n${\_}$TOF collaboration are also included. 
For unstable nuclei not present in the K1 repository we used theoretical Hauser-Feshbach 
computations from the TALYS package\footnote{https://tendl.web.psi.ch/tendl$\_$2019/talys.html}, in its 2008 
version, as discussed by \citet{gor08}. Corrections for stellar 
conditions of the cross sections were introduced using the traditional
{\it Stellar Enhancement Factors} ({\it SEF}), instead of the recently suggested X-factors from \citet{rauscher}. Reasons for this have been discussed in some detail 
in \citet{b+21}. Rates for weak interactions as a function of temperature 
and density are assumed from the work by \citet{ty87}. The ensemble of the
above choices will be indicated as our {\it standard} (ST) case.

For various nuclei discussed in this work considerable uncertainties 
are still present and are outlined in what follows. In the aim of clarifying
some crucial points of the $s$-process path for which new measurements would
be important, we shall consider a series of possible local modifications 
to the input data, as suggested by the comparison with either the presolar 
grain data or the average solar-system inventory of purely-$s$-process isotopes. 
The results of computations made using sets of parameters modified ad-hoc  
to improve the fit to experimental SiC data will be referred to as models of {\it version 2} (V2).

\subsection{The branching points at $^{84}$Kr and $^{85}$Kr}\label{sec:krbranch}
\begin{figure}[t!!]
\includegraphics[width=0.7\columnwidth,angle=-90]{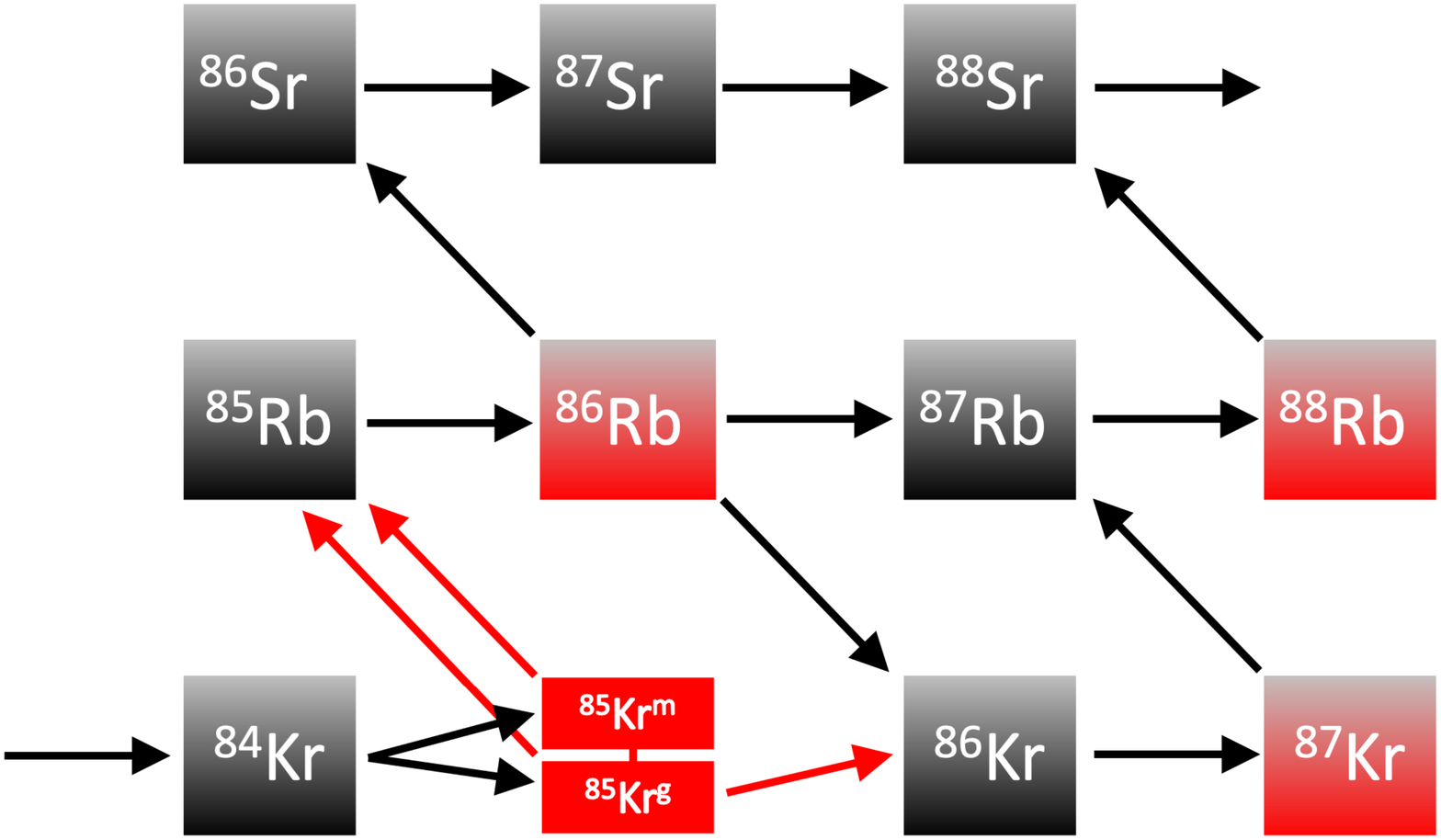}
\caption{The s-process nucleosynthesis path in the region around $N = 50$; the $^{85}$Kr branching is highlighted by the red arrows. 
Stable isotopes are shown in black boxes while unstable ones have a red background.  \label{fig:85kr}}
\end{figure}

Several nuclei, produced by neutron captures near the magic neutron number $N = 50$
depend more or less remarkably on the operation of the crucial branching points of the 
$s$-process chain at $^{85}$Kr and at its parent $^{84}$Kr \citep[see Figure \ref{fig:85kr} and e.g. the paper by][]{w86}. 
A recent discussion of this and  other branching points on the {\it main 
component} in the light of parameterized AGB models was presented by \citet{b15}.
In the K1 repository the 30 keV recommended cross section for $^{84}$Kr is 33.1 mb. This 
compares very well with the data listed in the {\it BNL} site, being close to a sort 
of average of them. The channel pointing to the isomeric state of
$^{85}$Kr has a branching ratio of 0.586, which means that almost 60\% of the 
flux goes to the isomer ($^{85}$Kr$^m$, 
at 305 keV and with $t_{1/2} \simeq $ 4.5h). This estimate is 
in line with more recent evaluations by Tessler et al. (2021, submitted); it is
however higher than in previous standard choices; e.g., in the 
previous release of KADONIS \citep[v0.3, or K03,][]{dil1}, where
this branching ratio was suggested to be about 40\%. Concerning the unstable $^{85}$Kr, 
its ground state has a half-life of about 10.5  yr, sufficient to effectively capture 
neutrons before the decay, in stellar conditions. Its cross section has only a 
theoretical estimate and is affected by a high uncertainty (of the order of 50\%, 
typically). The K1 recommended value (73 $ \pm $ 34 mb) represents again a sort of average
of those reported in the {\it BNL} repository.  A further clarification of all these
points is vital, because the flow through the isomeric state subsequently 
feeds primarily (at 80\%) the channel passing through $^{85, 86}$Rb and then 
ending up at $^{86,87}$Sr, at the expense of $^{86}$Kr and $^{87}$Rb, which 
feed only $^{88}$Sr. This implies that the ratio of Sr isotopes shown by 
presolar grains depends not only on their own cross sections, but also 
on those of $^{84,85}$Kr, on the branching ratio to $^{85}$Kr$^m$ and on the
decay rates of unstable isotopes through which the flow passes (like $^{85}$Kr$^m$ itself
and $^{86}$Rb). New cross section measurements in this mass region would be
important also for the understanding of the Rb/Sr ratio in AGB stars and their relatives
\citep[see e.g.][and references therein]{rbmaria}. With new facilities for 
measuring weak interactions in ionized plasmas only a couple of years from finalization, 
it will certainly be worth fixing all these issues on experimental grounds.

\subsection{Sr}\label{sec:Sr}

The precise values for the cross sections of the Sr isotopes that are
fed by slow neutron captures ($^{86}$Sr, $^{87}$Sr, $^{88}$Sr) are crucial 
for interpreting the isotopic admixtures of Sr itself, as well as their ratios
to Ba isotopes, as measured in presolar grains \citep{liu0, liu1, liu3, stephan}. 
For $^{86}$Sr, the 30 keV recommended value is 60 mb, again in good agreement with the {\it BNL}
data and only slightly lower than the previous choice (64 mb) made by K03. 

For $^{87}$Sr, the recommended value is from experimental measurements \citep{bauer} and
the energy dependence is derived from the repositories mentioned above. The choice by 
K1 (93.8$ \pm $3.8 mb) is again only slightly smaller than the one of the previous K03 compilation and compatible
(somewhat in the higher part of the distribution) with {\it BNL} data.

For $^{88}$Sr, K1 recommends a re-normalization of previous weighted averages, 
from measurements such as those by \citet{Koe00, kzb90}. However, the presence of discrepancies 
in the published data from the {\it Time of Flight} (TOF) and {\it Activation} methods 
is noticed and this is a special case in which new experimental efforts are needed. 
In this respect, we notice that in the most recent measurements by \citet{katabuchi},
the 30 keV reference value (9.4 mb) is much larger than recommended by 
K1 (6.3 mb). This last datum is then slightly larger than most of those from {\it BNL} 
(that group around 5.2 mb). This is important especially if considered together with the uncertainties
already discussed, affecting the previous branching points of the $s$-chain at 
$^{84,85}$Kr (see above). One can in particular notice that very similar relative 
production factors for the isotopes of Sr can be obtained in two different ways; 
namely: (i) by adopting, in neutron captures on $^{84}$Kr, the lower value of the 
branching ratio to $^{85}$Kr$^m$ (40\%) and taking the neutron capture cross section of 
$^{88}$Sr from the K1 recommendations; or (ii) adopting the higher branching ratio to 
$^{85}$Kr$^m$ (60\%), but then using, for the cross section of $^{88}$Sr, the measurements by 
\citet{katabuchi}. Our standard choice here will be that of choosing the K1 recommendations, 
however one has to { remember} that a lower production of $^{88}$Sr is possible using either
of the previously mentioned choices (whose effects are mimicked in our case V2). It is clear 
that on these issues an experimental clarification
is urgent, { as is a proper treatment of the decay rates} of $^{85}$Kr and $^{86}$Rb in ionized plasmas. 

\begin{figure*}[t!!]
\includegraphics[width=0.75\textwidth,angle=-90]{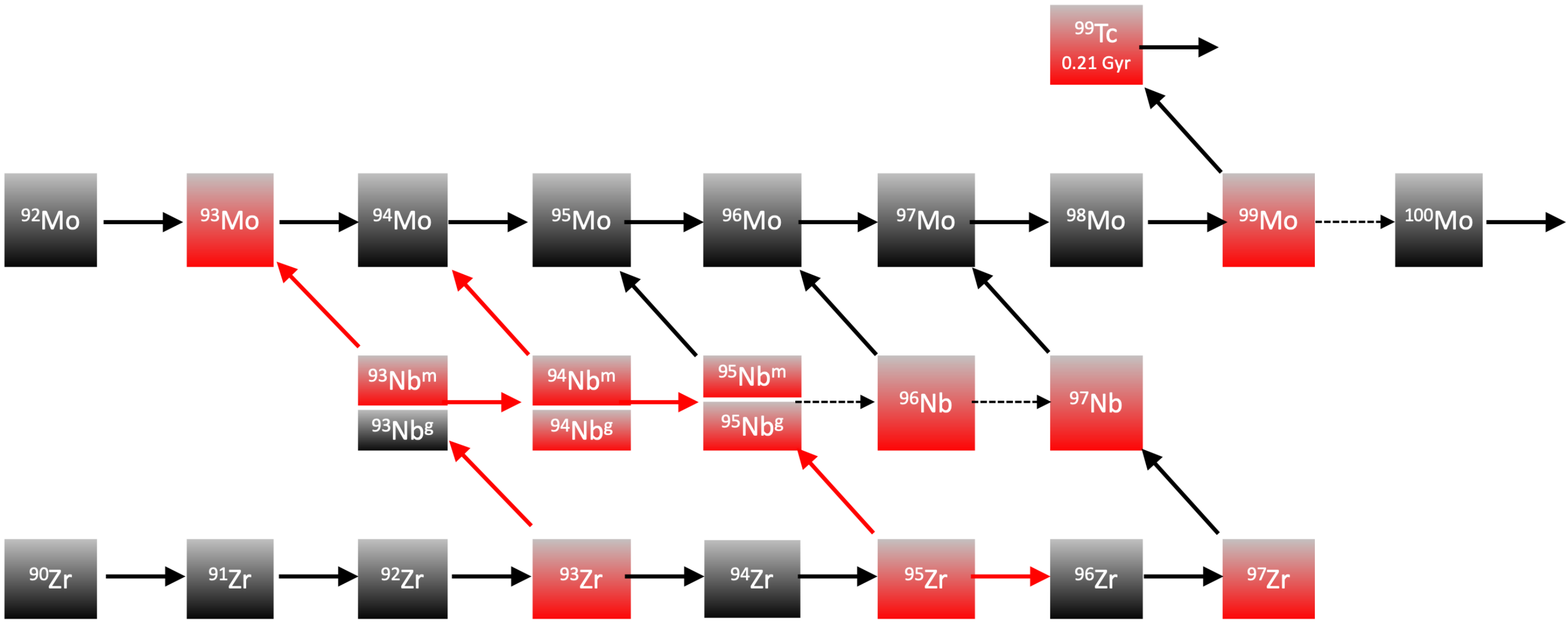}
\caption{The s-process nucleosynthesis path trough Zr, Nb and Mo isotope. Same color code as in Figure \ref{fig:85kr}.  \label{fig:93Nb}}
\end{figure*}

\subsection{Zr and Nb}
An analysis of the possible uncertainties affecting the reproduction of Zr isotopes in SiC grains 
was early presented by \citet{lu03}, to which we refer the reader for a general assessment of the problem.
In our work, for stable Zr isotopes and for the rather long-lived $^{93}$Zr, the K1 recommended 
cross sections include the measurements of the n$\_$TOF collaboration \citep{ta08a, ta08b, ta10,ta11,t+11b,ta13}.
In particular, the values provided for MACS from 5 to 25 keV were recalculated and normalized to 
the experimental data at 25 keV. For the unstable $^{95}$Zr MACS from 5 keV to 100 keV are from 
theoretical calculations. Uncertainties are indicated to be between 25 and 50\%. The data reported by
{\it BNL} are on average larger by 30\% (these choices would produce a larger amount of $^{96}$Zr and a lower contribution to $^{95}$Mo. We shall mimic this behavior in our cases V2 by using there the K1 cross section at the upper limit of its reported error bar).

From $^{93}$Zr an interesting reaction branching departs, based on its $\beta^-$ decay 
\citep[early recognized by][who indicated it as a source of information on the stellar physical conditions]{bahcall}. 
This decay leads 
to $^{93}$Nb via two channels, feeding directly the ground state (with 27\% probability) or passing through
its isomer $^{93}$Nb$^m$ (73\% probably). The resulting half-life is reported in the {\it ENSDF} tables\footnote{https://www.nndc.bnl.gov/ensdf/} to be 1.6$ \cdot $10$^6$ yr, hence of the same order as the whole duration of the TP-AGB phase. 
The $^{93}$Nb half-life does not decrease with T below about $4\cdot 10^8$, according to \citet{ty87}. During the development of $TP$s, the temperature quickly
increases up to 280-300 MK, where $^{93}$Zr is strongly produced, so that at the end of the AGB phase its abundance is between one fourth and one
half of those for the nearby nuclei ($^{92}$Zr and $^{94}$Zr).  Due to its long half-life, this isotope behaves almost as a stable species during s-processing and $^{93}$Nb is largely bypassed. Indeed, normal AGB stars, while showing in their spectra the unstable $^{99}$Tc, which is shorter-lived with respect to $^{93}$Zr, are instead Nb-poor. On this basis one would find that AGB stars contribute minimally to the nuclei  that immediately descend from $^{93}$Nb itself, like the unstable $^{94}$Nb and its decay daughter $^{94}$Mo, whose s-process contribution was recently found to be of 1 to few percents \citep{ste19, b+21}.
However, due to the high abundance contrast ($^{92}$Zr and $^{94}$Zr are more abundant than $^{94}$Mo in the Sun by about a factor of 10), even a small leakage from the $^{93}$Zr decay to $^{93}$Nb (say, 1$\%$) would contribute { significantly} to the chain $^{93}$Nb$-^{94}$Nb$-^{94}$Mo. In summary, although $^{94}$Mo remains a mainly-p nucleus, it cannot be excluded that it may receive a contribution from the s-process larger than estimated by \citet{b+21}. On the other hand, the whole remaining abundance of $^{93}$Zr decays to Nb at the end of the TP-AGB stage. 
Therefore, when the star is in a binary system exchanging mass between the components (i.e. is a so-called Ba-star), $^{93}$Nb accumulates on the surface of the companion. When this last evolves to the AGB stage and undergoes $s$-processing in its turn, neutron captures will occur on a material that is Nb-enhanced, albeit by limited amounts (less than about 2-3 times the initial abundance) due to dilution in the envelope at the red giant stage. One of these very peculiar evolved stars, rich in both $^{99}$Tc and $^{93}$Nb, was recently observed by \citet{shetye} and defined to be a {\it bi-intrinsic} AGB star.  In such an object, a peculiar $s$-processing will occur, feeding more effectively $^{94}$Mo. Although the effect is probably not large, due to the mentioned dilution, the $s$-process contribution to $^{94}$Mo remains for the above reasons quite uncertain. 
This fact may affect the attempts at accounting for the Mo isotopic admixture of presolar SiC grains, where the ratio $^{94}$Mo/$^{96}$Mo plays an important role. Experimental clarifications of the nuclear data 
(decay and cross sections) for this rather peculiar channel ($^{93}$Zr-$^{93}$Nb-$^{94}$Nb-$^{94}$Mo) would 
therefore be welcome.

A further source of uncertainty for the Zr isotopes concerns the $\beta^-$unstable $^{95}$Zr. As mentioned, its 
(n,$\gamma$) cross section is still purely theoretical, although an experimental method to determine it (and 
the ones of other short-living $s$-process nuclei acting as branching points of the flow) was discussed by \citet{sonna} using data of the inverse ($\gamma$,n) reactions. According to this
discussion and to the one presented by \citet{lu03}, improvements can be expected essentially only from
new estimates of the cross section itself, as the decay rate seems to be well determined. In particular, the 
ground level (a 5/2$^+$ state with a half-life of 64.02 days) decays to $^{95}$Nb and then to $^{95}$Mo 
{ with probably little to no} dependence on the temperature conditions, because of the very high energy of the excited states.

\subsection{Mo}\label{sec:mo}
The element Mo plays an important role for reconstructing the pollution of the solar system in $s$-process elements and its heterogeneity, as discussed by  \citet{sa21}. The same authors clarify how the trend of isotope ratios in presolar grains 
is separate from the meteoritic ones, very well defined and characteristic of $s$-processing environments in AGB stars. Here, the lightest Mo isotope that can receive some (marginal) contribution from $s$-processing in AGB stars is $^{94}$Mo. Although we already mentioned that it is basically a $p$-process 
nucleus, in evolved low-mass stars it can be fed by neutron
captures destroying $^{92}$Mo, through the daughter of this last, $^{93}$Mo.  As it is shown by Figure \ref{fig:93Nb}, two channels 
then lead to $^{94}$Mo from this source: a direct link via n-captures on $^{93}$Mo itself
(radioactive via e$^-$-captures, with a half-life of 4000y), or an indirect one, through 
the decay product $^{93}$Nb, undergoing n-captures to $^{94}$Nb. This last then can either 
decay to $^{94}$Mo ($\beta^-$-decay, with half-life of 2$ \cdot $10$^4$  yr) or participate to 
further neutron captures, feeding $^{95}$Nb (half-life 35 days) and $^{96}$Nb. In this respect, 
the channel from $^{94}$Nb merges (on one side) with the already mentioned contribution coming 
from the initial $^{93}$Nb abundance (and/or its refurbishing in the rare case of a bi-intrinsic 
AGB star). On the other side, it also starts new branching reactions on further unstable Nb
isotopes. Concerning heavier Mo isotopes, we must notice that the solar system abundance of 
$^{96}$Mo, which is an $s$-only nucleus, is reproduced very well by present $s$-process models
\citep{liu+19,b+21}. Improvements in nuclear data would therefore mainly help in clarifying details 
of the $r$ and $p$ processes, through their contributions to Mo isotopes, which are precisely
predicted by presolar grain measurements \citep{ste19}. The recommended cross sections of Mo isotopes 
from the K1 compilation almost coincide with the values listed in the {\it BNL} site 
(with exclusion of the unstable $^{99}$Mo, where the {\it BNL} value, from the ENDF/B and JEFF collaborations, is higher: 481 against 366$ \pm $92 mb: the effects of this alternative choice is again tested in our V2 cases). The uncertainty is in any case high, as only theoretical estimates are available. For other Mo isotopes, where experimental data exist, they are however quite old, so that new measurements are required and are actually already planned by the 
n${\_}$TOF collaboration \citep{gu+13}. For that purpose, \citet{liu+19} stressed the importance to have precise neutron 
capture cross sections as a function of the temperature even in energy regions
not yet explored by activation techniques. On the other hand, a search for improvements in decay rates may concentrate on 
{ the subtle contribution to $^{94}$Mo from $^{94}$Nb mentioned above}.

\subsection{Cs and Ba}

\begin{figure}[t!!]
\includegraphics[width=0.75\columnwidth,angle=-90]{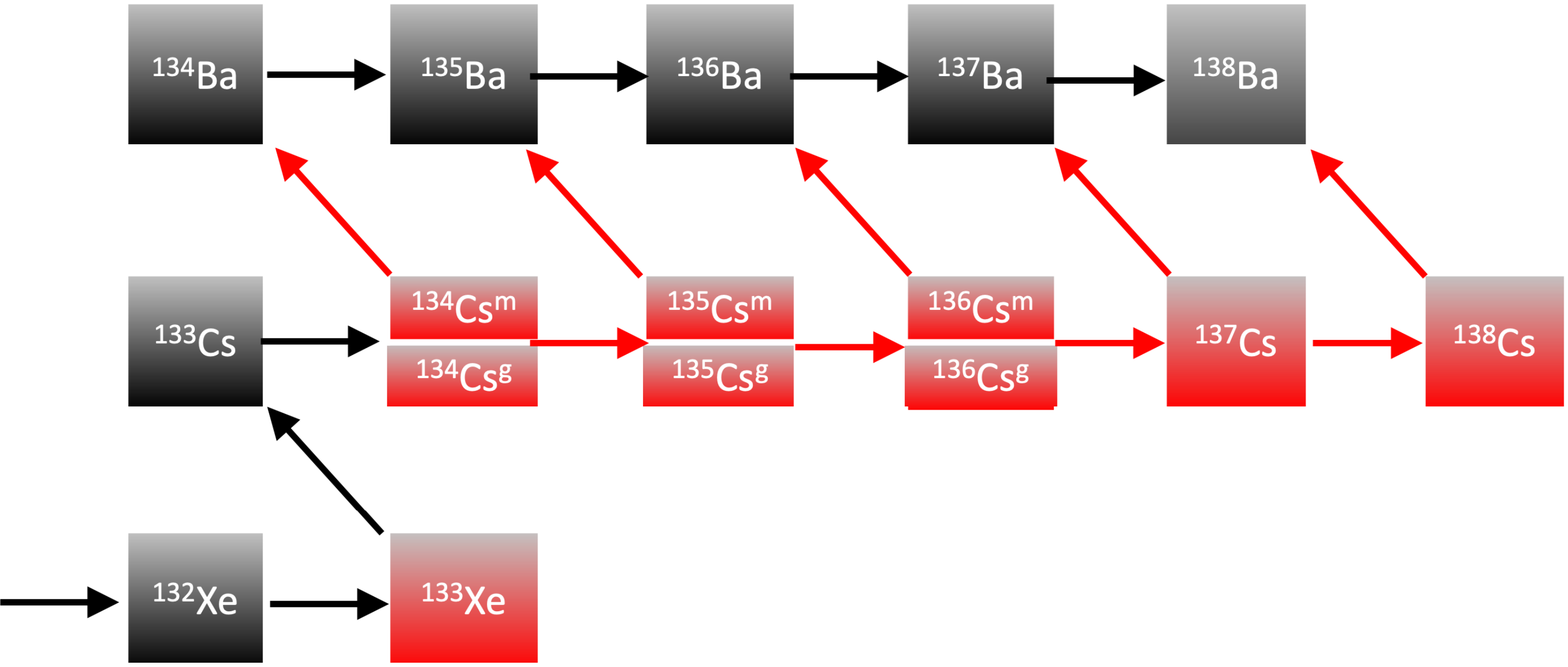}
\caption{The s-process nucleosynthesis path trough Cs and Ba isotopes (thus in the region around $N = 82$). The $^{134}$Cs and  $^{135}$Cs branching points are highlighted by red arrows. Same color code as in Figure \ref{fig:85kr}.  \label{fig:134Csb}}
\end{figure}

The $s$-process contribution to the element Ba starts from neutron captures 
on the stable isotope $^{133}$Cs, whose recommended cross section is indicated to 
be 502$ \pm $28 mb in K1, in agreement within the error bar with most of the 
values quoted at the {\it BNL} site. In nucleosynthesis models, the production of this nucleus
is in its turn affected by the choice one makes for the initial abundance of Xe, whose 
average concentration in the solar system is unknown \citep{lod20}. When knowledge of tiny details is 
required, this implies some remaining ambiguity on the absolute levels of production 
for Ba isotopes. 

As Figure \ref{fig:134Csb} shows, after $^{133}$Cs, the flux proceeds through a branching point at the radioactive $^{134}$Cs, 
where n-captures compete mainly with $\beta^-$ decay (laboratory half-life of 2  yr)
to excited states of $^{134}$Ba and, much less effectively, with electron captures 
to $^{134}$Xe (half-life of 6.8$ \cdot 10^5$ 
 yr). From $^{134}$Cs, neutron captures feed the longer-lived $^{135}$Cs, whose half-life would require a specific reanalysis (see below), and then  $^{136}$Cs (half-life of 13.16 d) and $^{137}$Cs (half-life of 30.07 y), which are sites of branching points for the 
$s$-process path, but whose decay rates remain essentially unchanged 
for varying temperatures. 

This is not so for $^{134}$Cs and $^{135}$Cs. According to the computations (made under conditions of thermodynamic equilibrium)
by \citet{ty87}, at 3$\cdot 10^8$K (a temperature rather typical of $TP$s) the decay rate
of $^{134}$Cs is enhanced with respect to the laboratory by a factor  of about 200. 
{ However, these phenomenological computations are affected by large uncertainties that could be reduced with better nuclear parameters
 obtained through more modern approaches.}
%However,  these phenomenological computations are affected by large uncertainties and this is a  crucial point on which better nuclear parameters would be welcome. 
As discussed in \citet{b+21}, when one assumes the rate of $^{134}$Cs decay from \citet{ty87}, the galactic 
production of $^{134}$Ba and $^{136}$Ba (two $s$-only nuclei) does not fit the solar constraints very well. This point will be analyzed separately in next section, as an example of the general need for better weak interactions along the $s$-path. Also the half-life of $^{135}$Cs might deserve a specific reanalysis and this would be so in particular if the decay rate of
$^{134}$Cs were to be reconsidered, with a larger production of $^{135}$Cs itself 
(see the next two sections). Even following the \citet{ty87} recommendations, in the temperature and density conditions prevailing in {\it TPs}, the half-life of 
$^{135}$Cs would be of the order a few hundred years, thus inducing some branching effect 
in the $s$-process chain. In our computations, we consider the possibility of a reduced
efficiency in the $^{134}$Cs decay, hence of an increased importance of the $^{135}$Cs
branching, in our V2 cases.

Concerning the cross sections of Ba isotopes themselves, the K1 repository quotes for them rather small uncertainties (always less than 10\%) and the Maxwellian-averaged data are either taken directly from recent measurements or renormalized according to \citet{au2, au1}. 

\section{Weak interaction improvements: the case of $^{134}$C{\small s}}\label{sec:rates}
\subsection{Requirements from nucleosynthesis}\label{sec:weak1}
In the analysis of the nuclear parameters presented so far, we mainly concentrated on cross sections, but
occasionally we met crucial decay processes whose accuracy one would like to see improved
for clarifying  important branching points. We consider here the example of the $^{134}$Cs decay, affecting  the isotopic admixture of Ba and the effectiveness of the branching point 
at $^{135}$Cs. As mentioned, the two stable nuclei more heavily affected by the decay at $^{134}$Cs are $^{134}$Ba and $^{136}$Ba, whose status as purely $s$-process nuclei is not reproduced very well by galactic neutron-capture nucleosynthesis, when the decay rate is taken from \citet{ty87}. See also, for this, \citet{pra+18,pra+20}. 

\begin{figure}[t!!]
\includegraphics[width=\columnwidth]{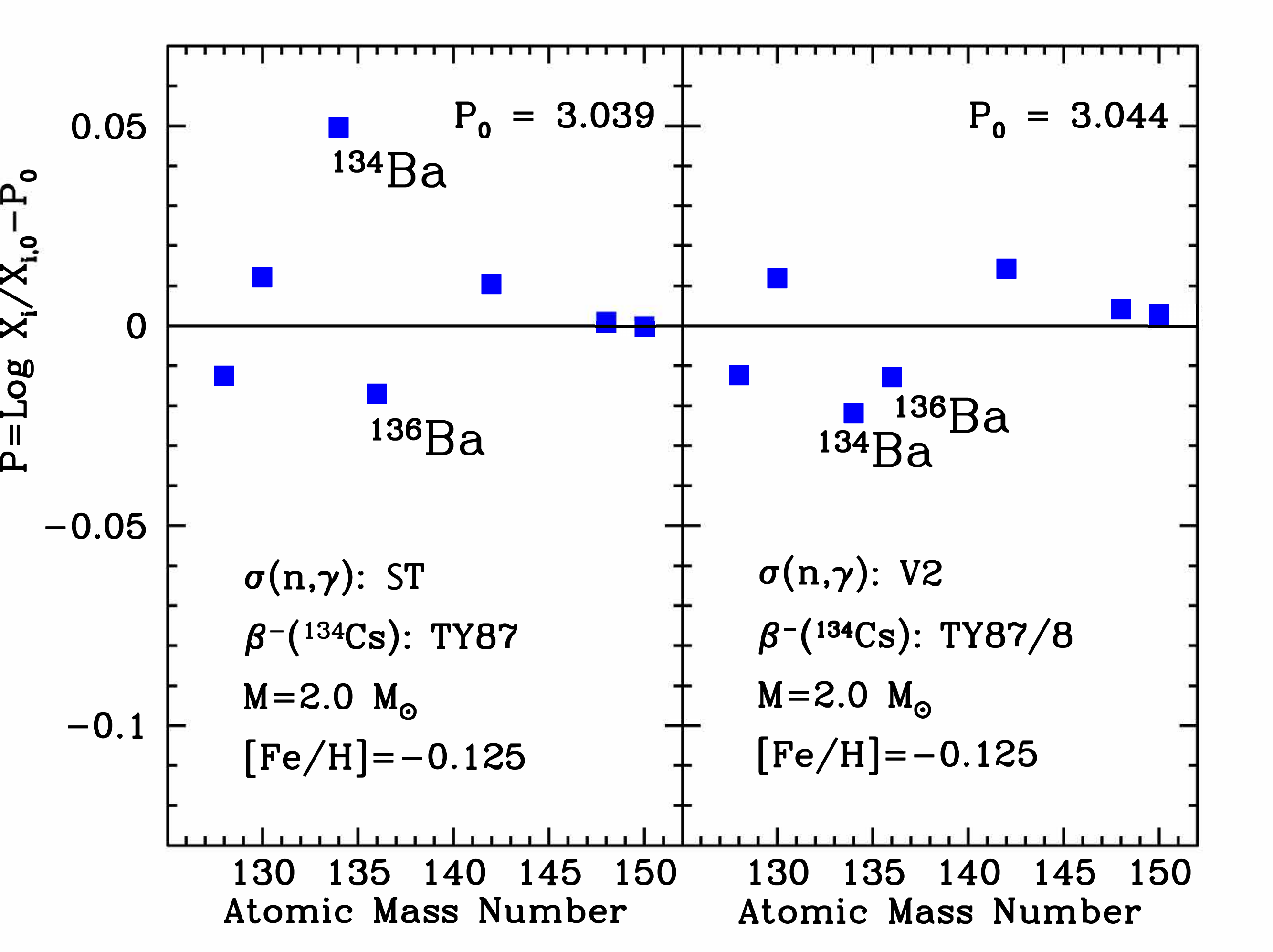}
\caption{The distribution of normalized production factors of $s$-only nuclei in the atomic mass range 
125 to 150. In the left panel, with our standard choice (ST)
of nuclear parameters, $^{134}$Ba and $^{136}$Ba show some noticeable discrepancy. 
In the right panel, with the choice V2 for nuclear parameters, they are reconciled 
(the residual difference of their production factor with respect to unity
is not significant, especially in the light of the absence of a secure datum for the solar
concentration of the precursor Xe and of the uncertainties on the $^{133}$Cs cross section). 
In order to obtain the result of the right panel the decay rate
of $^{134}$Cs to $^{134}$Ba was assumed to have a temperature dependence less steep than 
suggested by \citet{ty87} by a factor of 8. \label{fig:barium}}
\end{figure}

An illustration of this uncertain behavior is attempted in Figure \ref{fig:barium}. It shows 
the production factors from a model AGB star of 2 \ms, for a metallicity 
slightly lower than solar, which represents a sort of average model for galactic $s$-processing,
in the sense that it naturally generates $s$-elements in solar proportions \citep[see discussion of 
this point in][]{t+16,b+21}. In particular, the figure shows purely $s$-nuclei in the atomic mass range
from 125 to 150. They are displayed in logarithmic scale, normalized to the average production factor
of $s$-nuclei, which is indicated on the plot. For the computations of the first panel (left) the nuclear
parameters of the ST choice discussed above were adopted and it is clear that in such a case there is a
discrepancy between $^{134}$Ba and $^{136}$Ba that is on average larger than for nearby $s$-nuclei. To
level their production (right panel) one needs to assume that the temperature dependence of the decay rate 
for $^{134}$Cs is less steep than suggested by \citet{ty87} by a rather large factor (in the illustrative
example shown, we changed it by a factor of 8). This confirms the finding by \citet{cris+15}, who obtained a 5$\%$ decrease of $^{134}$Ba abundance by decreasing the  $^{134}$Cs decay by a factor of three.
We notice that  a similar leveling would have also been 
possible  by changing upward the cross section of $^{134}$Cs by more than a factor of two. Although the 
accepted value is from theoretical calculations, the possibility of such a large variation seems today
unlikely.
\subsection{A revised $\beta^-$ half life for $^{134}$Cs}\label{sec:weak2}

\begin{figure}[t!!]
\includegraphics[width=\columnwidth]{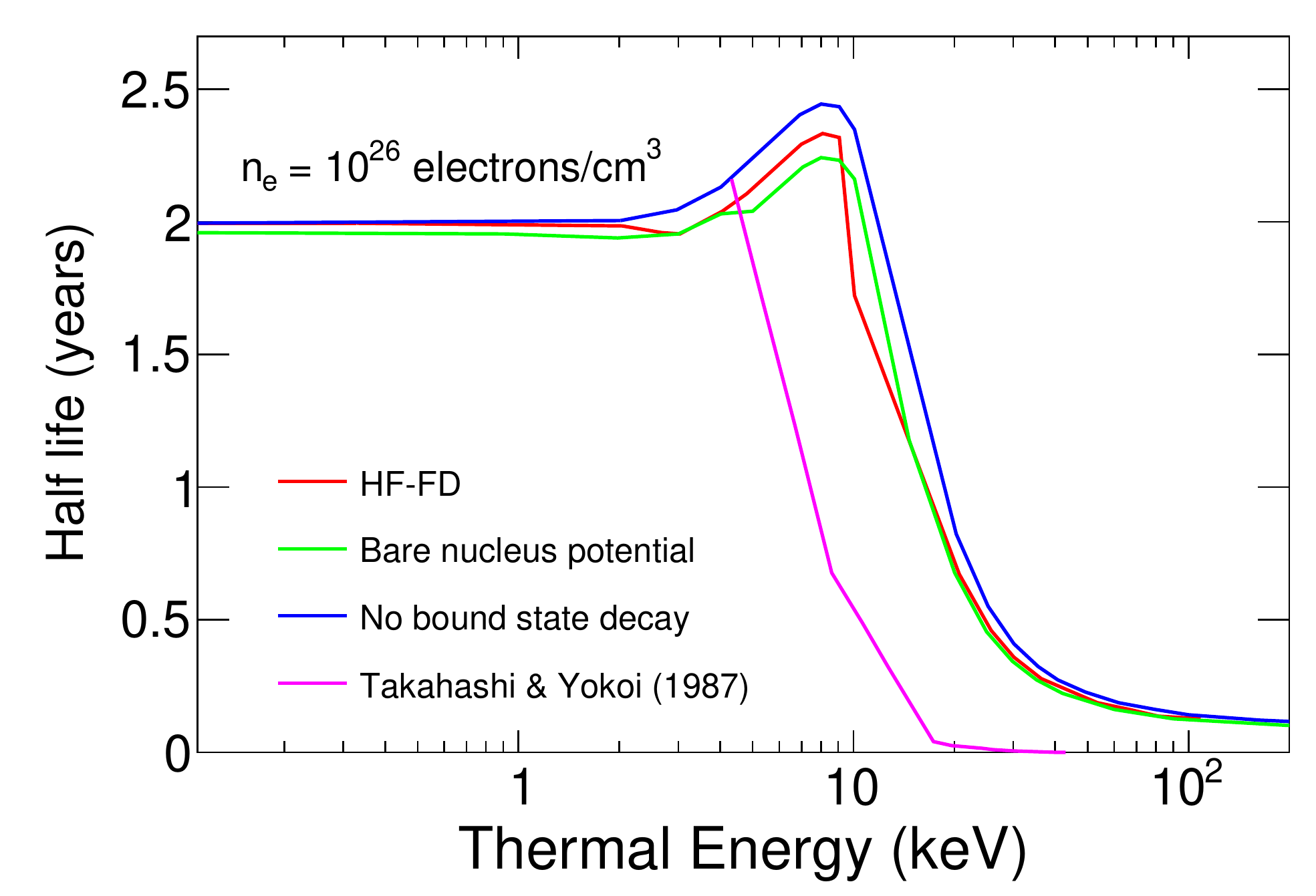}
\caption{The half-life of $^{134}$Cs as a function of  the thermal energy ($k_BT$) estimated by using: (i) a Hartree-Fock (HF) mean-field potential with a Fermi-Dirac (FD) probability distribution (red line); (ii) same, but excluding the decay to bound states (blue line, no Fermi-Dirac temperature, orbitals are occupied in their HF ground state); (iii) \citet [][magenta line]{ty87}; (iv) the bare nucleus potential (green line).} \label{fig:rate}
\end{figure}

Very recently a study by Simonucci et al. (2021, to be submitted) confirmed that the temperature 
dependence of the $^{134}$Cs $\beta-$decay rate should indeed be less steep than so far assumed. In their approach, the 
Dirac-Hartree-Fock equations were solved for both the electron phase and the nucleus, modeling 
the nucleon-nucleon interaction through a relativistic one-body Wood-Saxon potential and factorizing 
the hadronic and leptonic currents as two non-interacting parts, also including the contributions 
from bound-state decays. Figure \ref{fig:rate} shows that in the temperature range relevant for 
He-burning in the AGB stages, the new half-life, as obtained by the authors, is 
longer than in \citet{ty87} by a factor ranging from 2.5 (near 8-10 keV) to more than 30 (at 30 keV).
In particular, the red line includes the contribution to the half-life of $^{134}$Cs of the atomic electrons, which populate the atomic orbitals according to { a Fermi-Dirac} distribution function \citep{m+18}, while in the blue curve the electrons are clamped down to the electronic ground state (zero-temperature, close to laboratory conditions). In the latter case, the half-life is indeed higher as escaping $\beta^-$ electrons are prevented by Pauli's exclusion principle from occupying bound orbitals (that are fully occupied) and can only be emitted in the continuum.

Here one is forced to wait for future measurements of the decay rate in ionized plasmas, hoping that these calculations and the previously-mentioned astrophysical suggestions can find in that way an experimental 
confirmation. 

\begin{figure}
\centering
%{\includegraphics[width=1\textwidth]{Fig9.pdf}}
{\includegraphics[width=1.08\columnwidth]{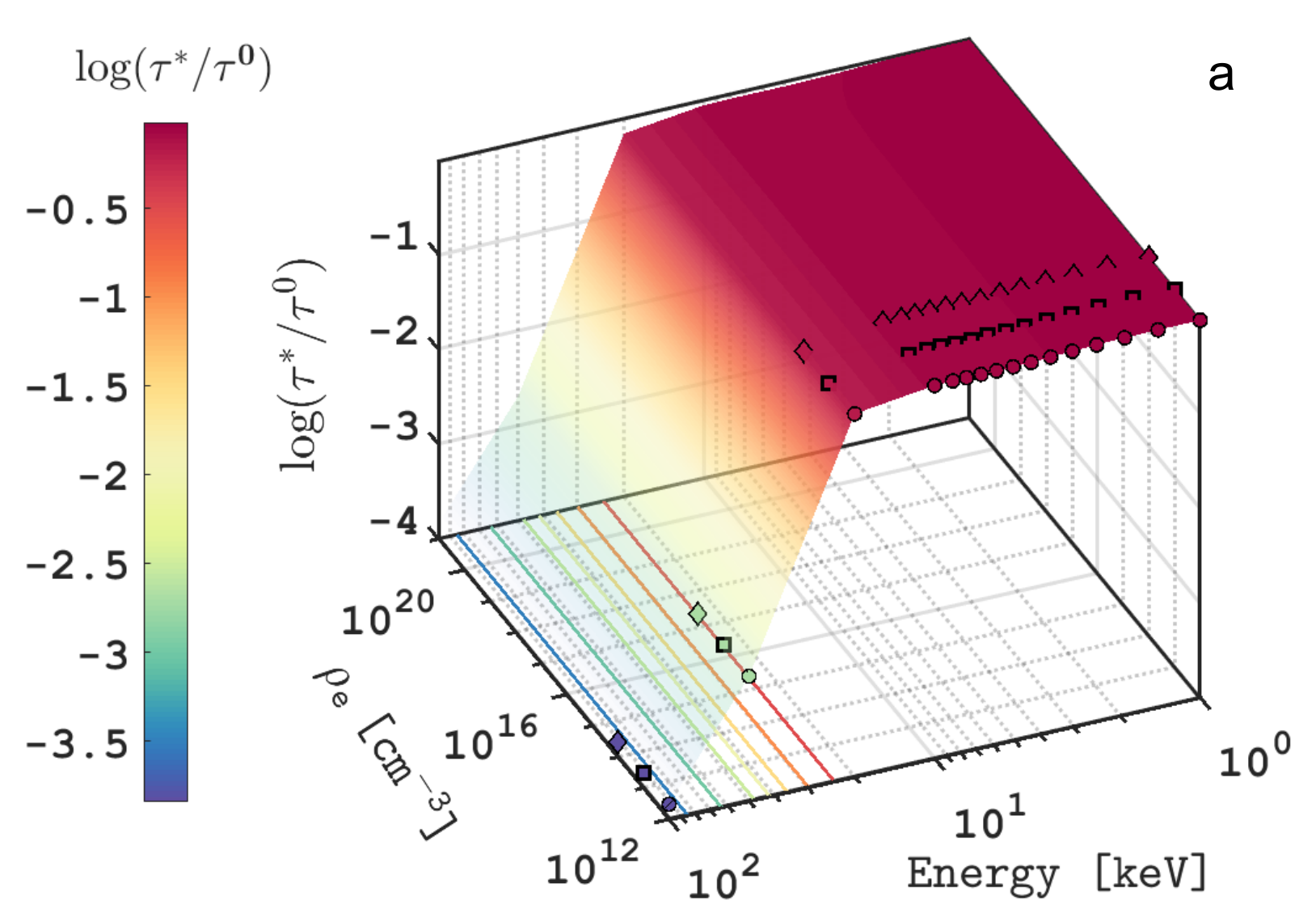}}
{\includegraphics[width=1.08\columnwidth]{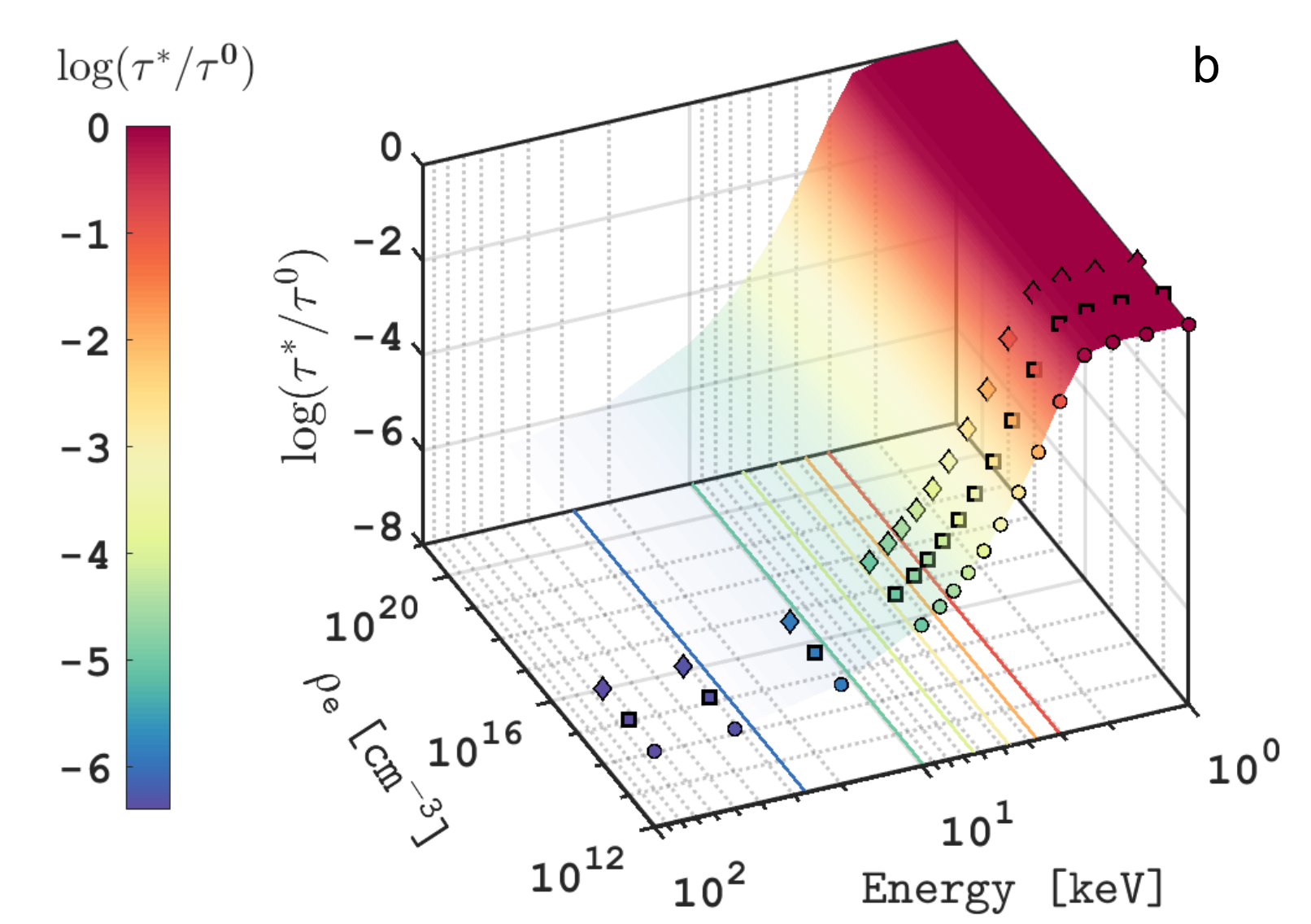}}
{\includegraphics[width=1.08\columnwidth]{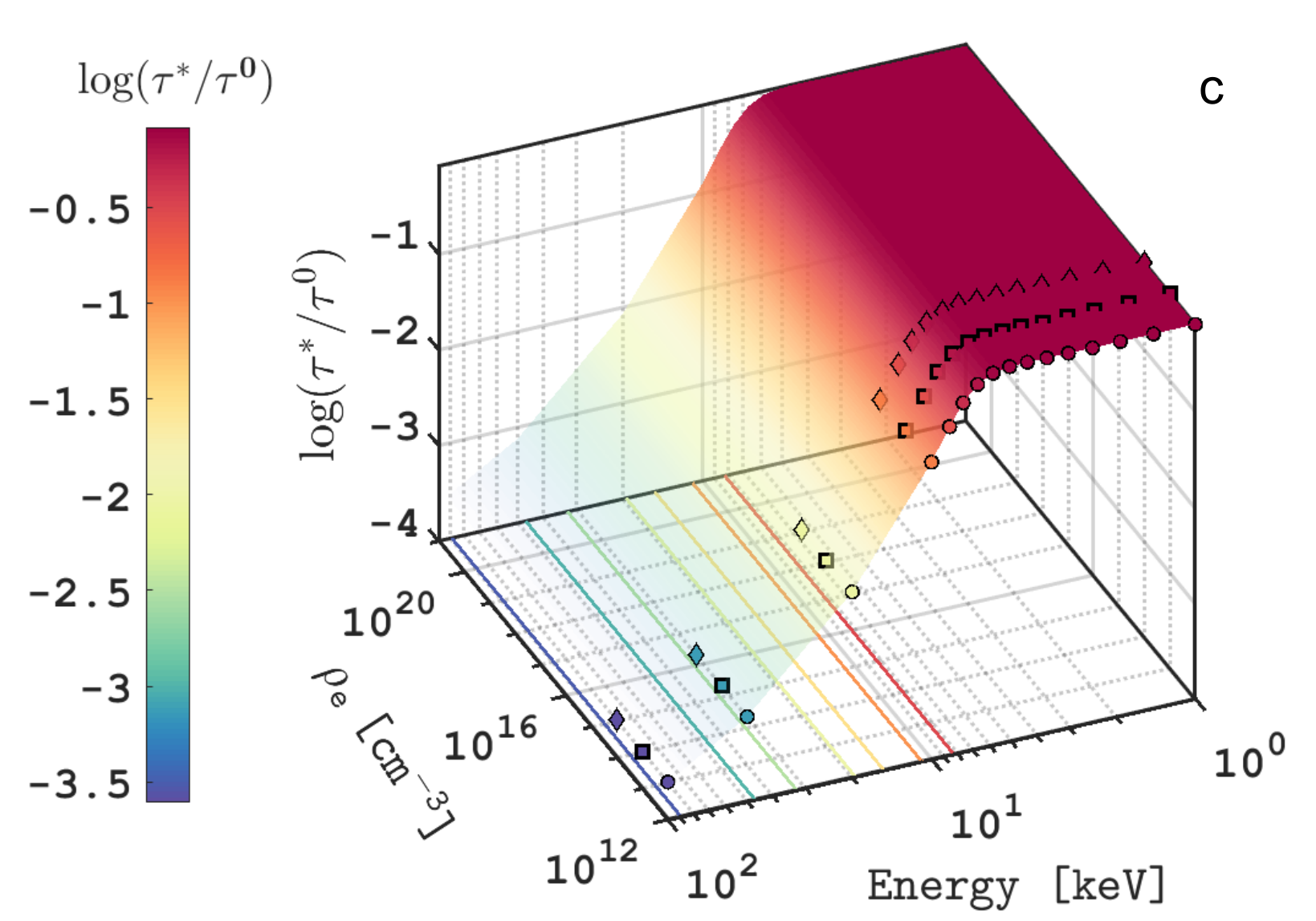}}
\caption{Mean lifetime in ionized plasma as a function of the plasma density (in the range  $\rho_{e}\sim 10^{12} - 10^{21}$ cm$^{-3}$) and of the energy (in the range $k_{B}T_{e}\sim { 1} - 10^{2}$ keV) for $\beta^{-}$ decays of (\textbf{a}) $^{93}_{40}$Zr$\rightarrow^{93}_{41}$Nb, (\textbf{b}) $^{94}_{41}$Nb$\rightarrow^{94}_{42}$Mo, and (\textbf{c}) $^{134}_{55}$Cs$\rightarrow^{134}_{56}$Ba. The color bar (log-scale) indicates the ratio $\tau^{*}/\tau^{0}$. Terrestrial average lifetimes $\tau^{0}$ are (\textbf{a})  $2.32 \cdot 10^{6}$  yr, (\textbf{b}) $2.93 \cdot 10^{4}$  yr, and (\textbf{c}) 2.98  yr.  Marker-points are shown only for densities accessible by the PANDORA plasmas.
} \label{fig:beta}
\end{figure}

\subsection{Foreseen experimental revisions}\label{sec:weak3}

For the scope mentioned in the previous subsection we analyzed in some detail the real possibility that a 
variation in the rate of the $\beta^-$ decay of $^{134}$Cs and of other radionuclei by factors similar 
to the one inferred by the new calculations there shown can be measured in ionized plasma environments.

In this framework, the PANDORA project \citep[Plasmas for Astrophysics, Nuclear Decay Observation and Radiation for Archaeometry,][]{Mascali2017, Mascali2020} aims to measure for the first time $\beta$-decay rates in controlled ionized plasmas made of radionuclides, at different plasma densities and temperatures. The innovative PANDORA plasma trap (presently under construction at LNS-INFN) will be able to produce and confine plasmas with electron-ion densities up to $10^{13}$ cm$^{-3}$ and electron temperatures $T_e$ in the range $0.1-30\ keV$. These conditions would mimic some important aspects of stellar environments,  in particular the charge states. More in detail, the experimental approach consists in determining a direct correlation of the plasma properties with the nuclear decay itself. This will be achieved by simultaneously identifying and discriminating, via a multi-diagnostic system, the $\gamma$-ray products following $\beta$-decays from unstable isotopes and the photons self-emitted by the plasma \citep{Naselli2019,Naselli2020}. PANDORA has been designed in order to maintain the radionuclides in a dynamical equilibrium for several days or even weeks. Indeed, simulations of the $\gamma$-decay detection efficiency  (as a function of the radionuclide lifetime and of the effective activity in the plasma volume) show that the minimum measurement time needed to achieve a sufficiently accurate measure ($\geq 3\sigma$) ranges from hours to several days,  depending on the isotope under investigation.

Virtual experiments have been performed  to study the feasibility of measuring the decay rates of several nuclei in the PANDORA trap and to predict possible enhancements of weak-interaction rates in stellar environments. Among the cases of study there are several  nuclei involved in s-process nucleosynthesis,  $^{93}$Zr, $^{94}$Nb and $^{134}$Cs in particular. The charge state distribution (CSD) in the plasma of these isotopes was investigated, starting from a "stellar-like" scenario and down to the PANDORA density range, by using the FLYCHK population kinetics code \citep{Chung2005}. This was done in both local thermodynamic equilibrium (LTE) and non-LTE conditions. FLYCHK can estimate the probability for each ion stage to be populated, together with the mean charge state of the plasma species, according to the atomic-level population distribution and the plasma thermodynamics. 
Once the radionuclide CDSs have been modeled in non-LTE conditions, the in-plasma $\beta-$decay rates  of $^{93}$Zr, $^{94}$Nb and $^{134}$Cs can be estimated.
This is illustrated in Figure \ref{fig:beta}, which shows { the} behavior of the $\log(\tau^{*}/\tau^{0})$ (namely the log-ratio between the lifetime in stellar conditions and the one in terrestrial conditions $\tau^{0}$)  for the nuclei studied, as a function of the plasma density $\rho_e$ and of the thermal energy $k_B T_e$. Calculations show that the plasma temperature affects largely the isotope decay rates and  that temperature effects dominate over those related to density. This finding strengthens the credibility of the half-lives in stellar conditions that will be estimated from experimental data taken by PANDORA in its operating density range. The results  of this analysis will be discussed in detail for a larger sample of radionuclides in a forthcoming paper. 

%%%%%%
%NUOVO
%Figure  \ref{fig:beta} shows the $\log(\tau^{*}/\tau^{0})$ (namely the log-ratio between the lifetime in stellar conditions and the one in terrestrial conditions $\tau^{0}$,)  for the selected sample of nuclei of $^{93}$Zr, $^{94}$Nb and $^{134}$Cs, as a function of the plasma density $\rho_e$ and of the thermal energy $k_B T_e$. Calculations are based on the phenomenological inputs, as in \citet{ty87}, and have been made for a wide range of thermal energies and plasma densities,  the results in Figure \ref{fig:beta} show that the plasma temperature might largely affect the isotope decay rates and  that temperature effects dominate over those related to density. This finding strengthens the credibility of the half-lives in stellar conditions that will be estimated from experimental data taken by PANDORA in its operating density range. The results  of this analysis will be discussed in detail for a larger sample of radionuclides in a forthcoming paper. 
%%%%

\section{Presolar SiC-grain measurements.}\label{sec:sic}

\subsection{General remarks}\label{sec:gen}
In general, the observed properties of the stars becoming C-rich during their AGB evolution match well those expected in the range 1 to 3 \ms  \citep{a20}, where the lowest masses pertain to older objects, dynamically distributed similarly to the old disk population \citep{cstar2}. There is also a tail for more massive members (up to maybe 4 or 5 \ms), which can be identified 
with very red objects, belonging to the thin disk, as early noticed by \citet{cstar1}. In this rather wide range, the AGB stars that are the main parents of presolar grains have been recently suggested to be { about} 2 \ms, with metallicities close { to the solar one} \citep{sicsergio}. This suggestion, especially for what concerns the metallicity, is in accordance with the ages of presolar SiC grains recovered from the Murchison chondrite, which have been estimated by \citet{heck}, based on the cosmogenic $^{21}$Ne produced inside
those solids, from $^{21}$Na decay. Most (60\%) of the grains turned out to be older than the solar system by less than 300 Myr, with a minority showing ages up to 3 Gyr.
According to recent estimates of the relations between age and abundances for
various representative elements in the galactic disk \citep{agemet1, agemet2, agemet3} we expect a rather
small spread of metallicities across the above mentioned age span at the galactic radius
of the solar system. For these reasons, and in agreement with \citet{sicsergio}, our comparisons between model predictions 
and observed isotopic ratios in SiC grains will be restricted to a small spread, $-0.15 \lesssim [Fe/H] \lesssim 0.1$. In those comparisons, the mentioned  suggestions lead us to 
expect that stellar models of about 2 \msb explain most of the SiC grain constraints.

\subsection{The Main-Stream SiC Grains and their Age}\label{sec:SiC}

SiC crystals form the most abundant sample of presolar grains and are among the most precise 
tools available to constrain $s$-process nucleosynthesis and the ensuing envelope enrichment 
in AGB stars. %These last represent actually the largest factory of dust particles in the Galaxy and some of them were 
Some dust particles of AGB origin were captured in pristine meteorites, showing us the isotopic composition 
of the stellar winds where they were formed. The laboratory measurements of the isotopic admixtures
for trace heavy elements  in such SiC grains offer then diagnostic tools for the
composition of evolved red giants that cannot be obtained from their 
stellar spectroscopy, { which is hampered on one hand by variability of the stellar sources and on the other by restriction to elemental abundances in most cases}.
% which is on one { hand} hampered by variability of the stellar sources and on the other { by restriction to elemental abundances in most cases}.
% normally provides only elemental abundances.  

Among the families of presolar SiC grains so far recovered, the largest one \citep[more than 16500 grains, according to][]{wustl2020} is that of  {\it Main-Stream} (hereafter MS) grains. Their isotopic ratios of 
C, N and Si, along with traces of s-elements measured in  a subset of them, certify their AGB origins \citep{Zinner14}. So far,  Sr, Zr, Mo or Ba isotopic ratios have been measured only  in  214 MS-SiC grains out of the several thousands available in the database maintained at the Washington University in St. Louis \citep{wustl}. For none of the grains the isotopic ratios were measured for all the 4 elements. However, in most cases data about the isotopic mix of Ba and at least of one or two of the other elements are available. 
Many authors determined the isotopic composition for smaller subsets of the 214 grains using different techniques, e.g. Resonance Ionization Mass Spectrometry (RIMS) or  Nano-Secondary Ion Mass Spectrometry (NanoSIMS).

%and CHILI Ma questo non è il primo esempio di RIMS?

\begin{figure*}[t!!]
\centering
{\includegraphics[width=0.25\textwidth]{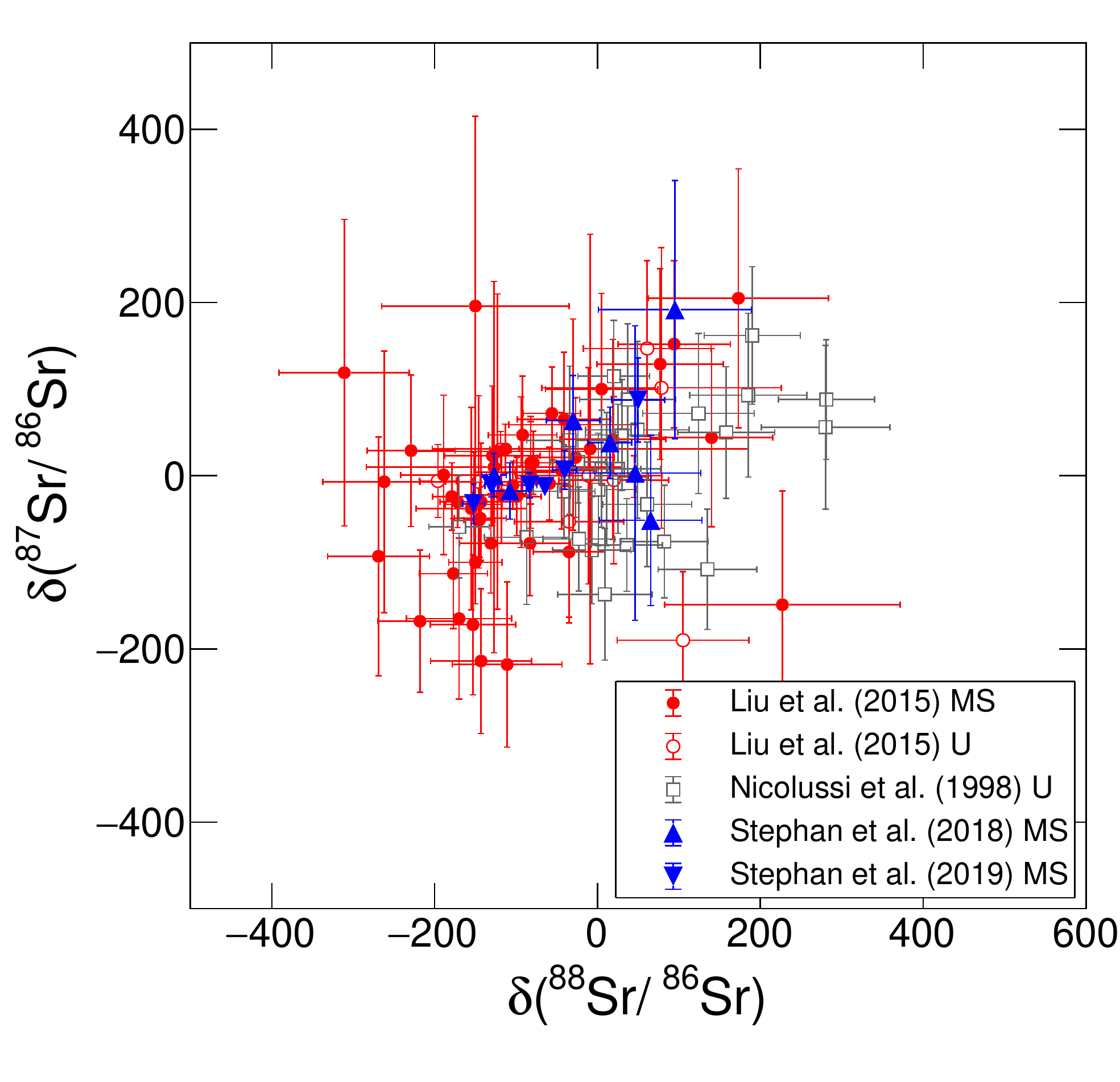}}
{\includegraphics[width=0.25\textwidth]{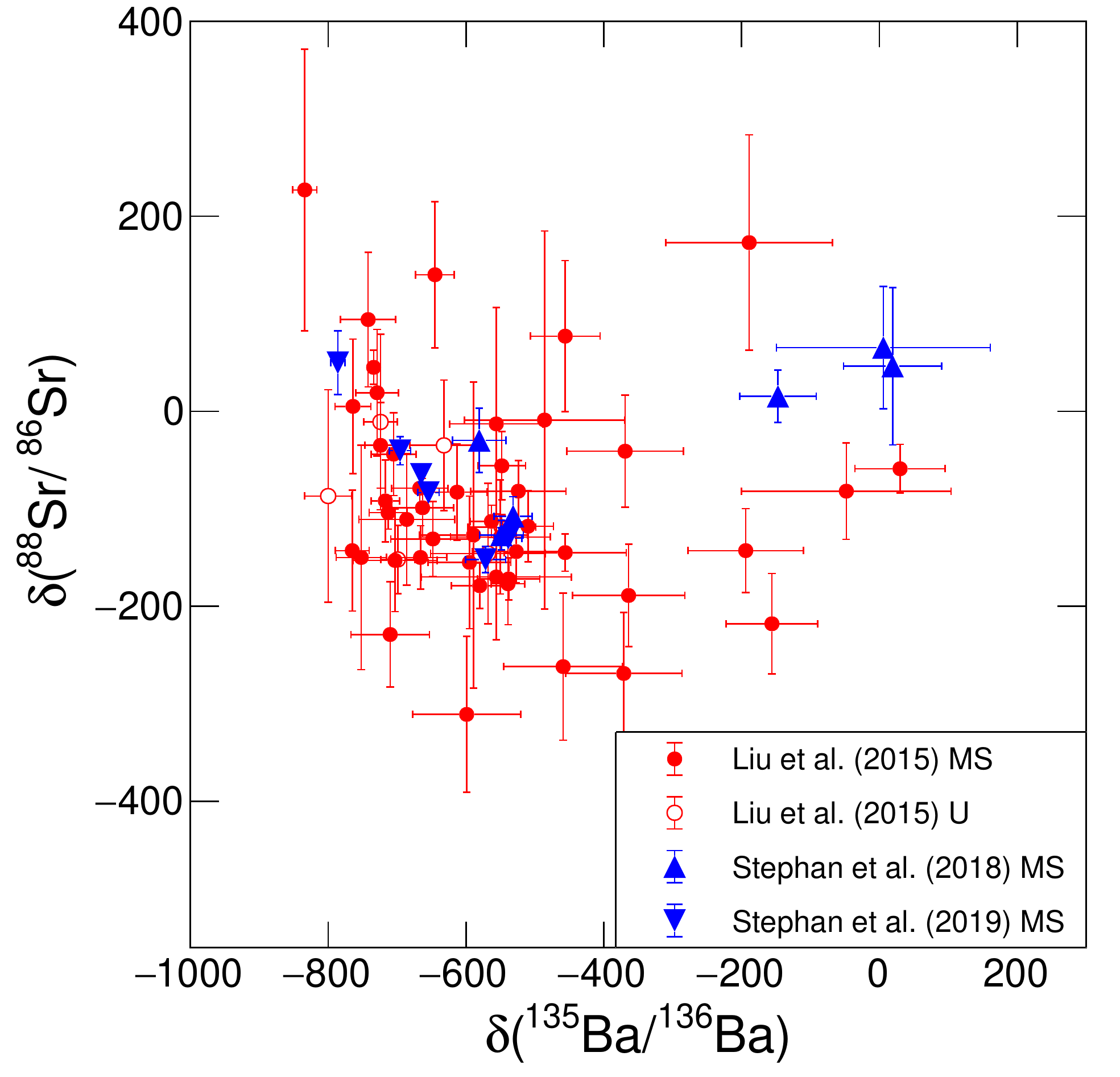}}
{\includegraphics[width=0.25\textwidth]{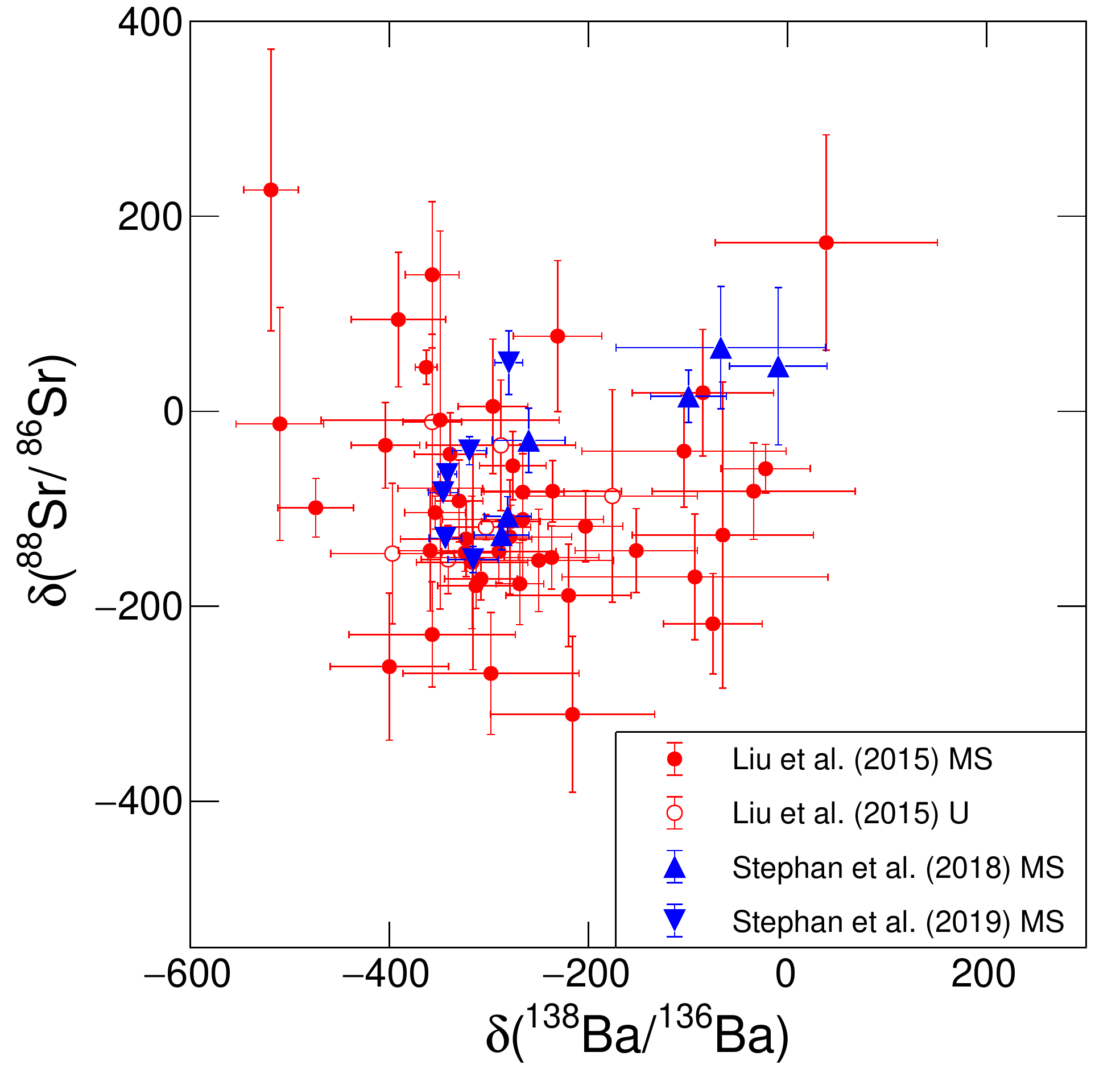}}
{\includegraphics[width=0.25\textwidth]{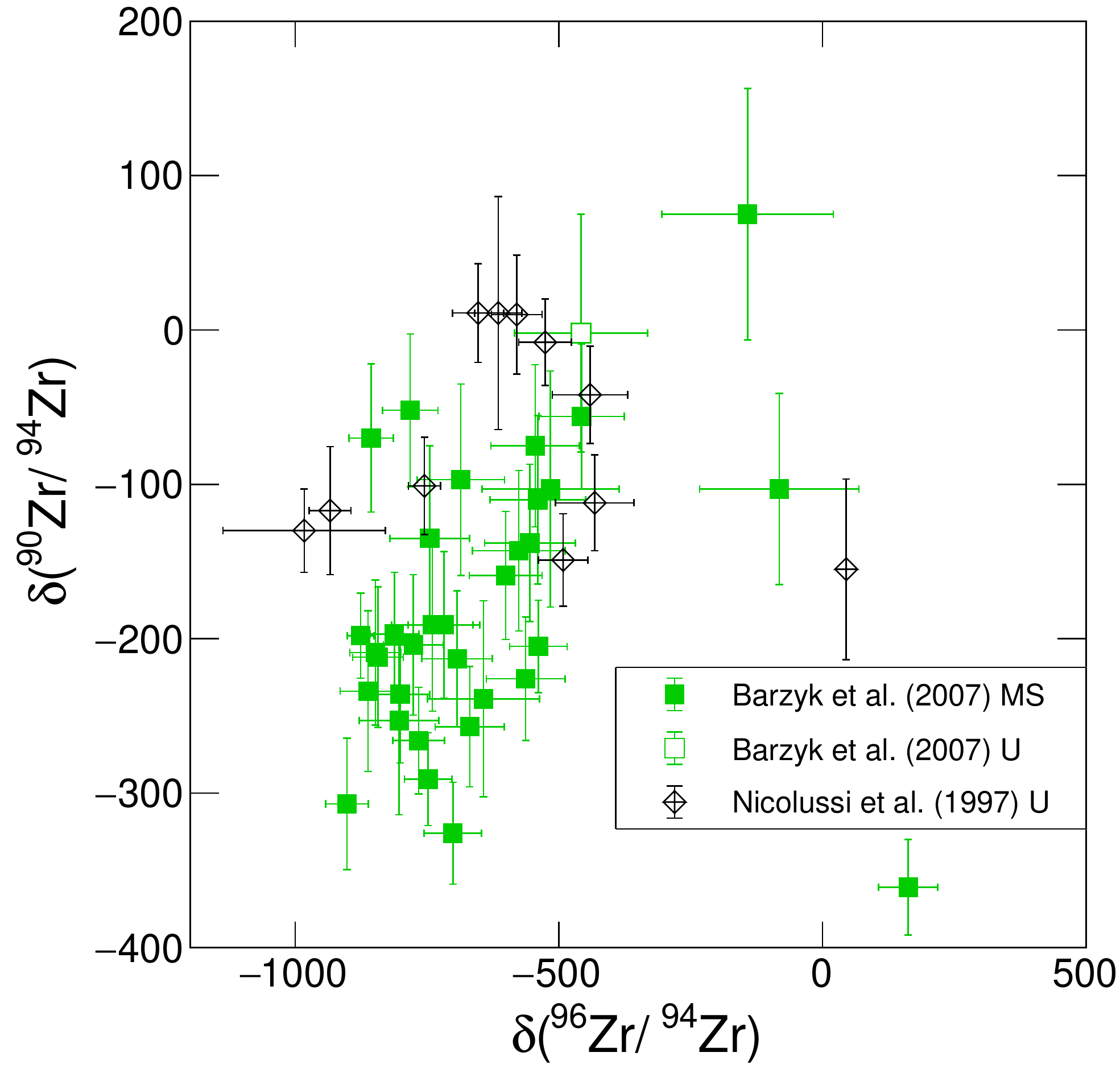}}
{\includegraphics[width=0.25\textwidth]{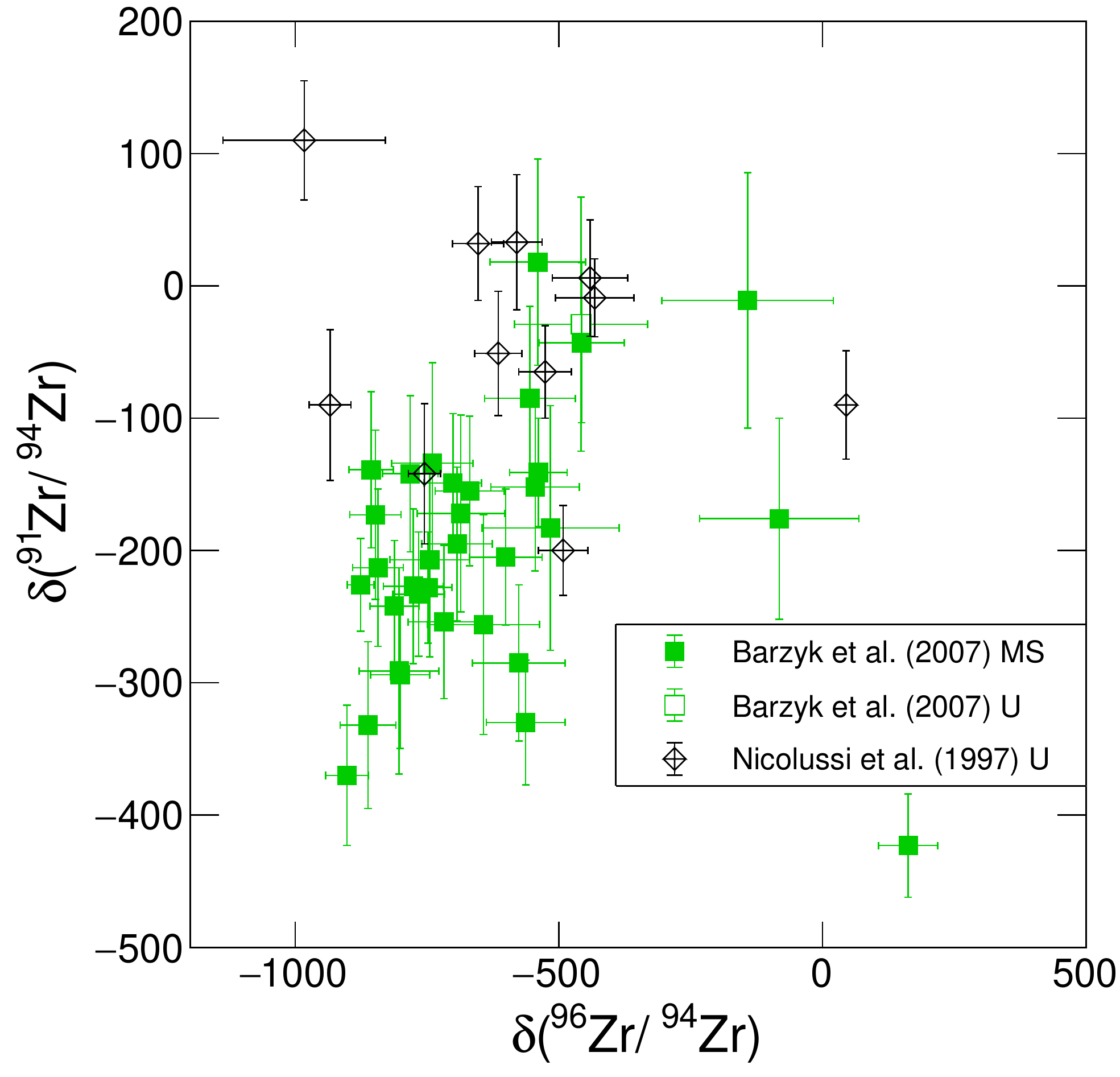}}
{\includegraphics[width=0.25\textwidth]{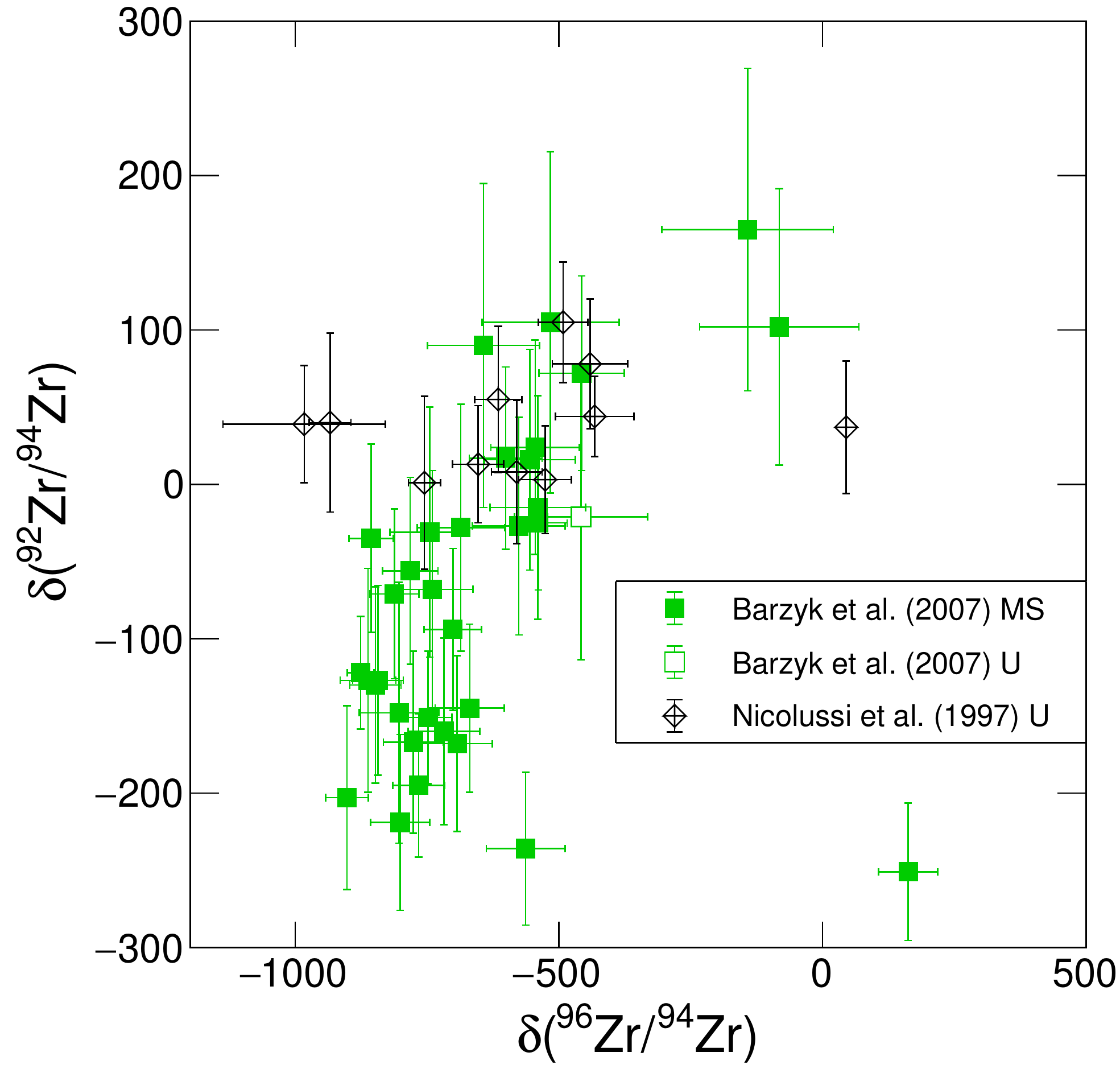}}
{\includegraphics[width=0.25\textwidth]{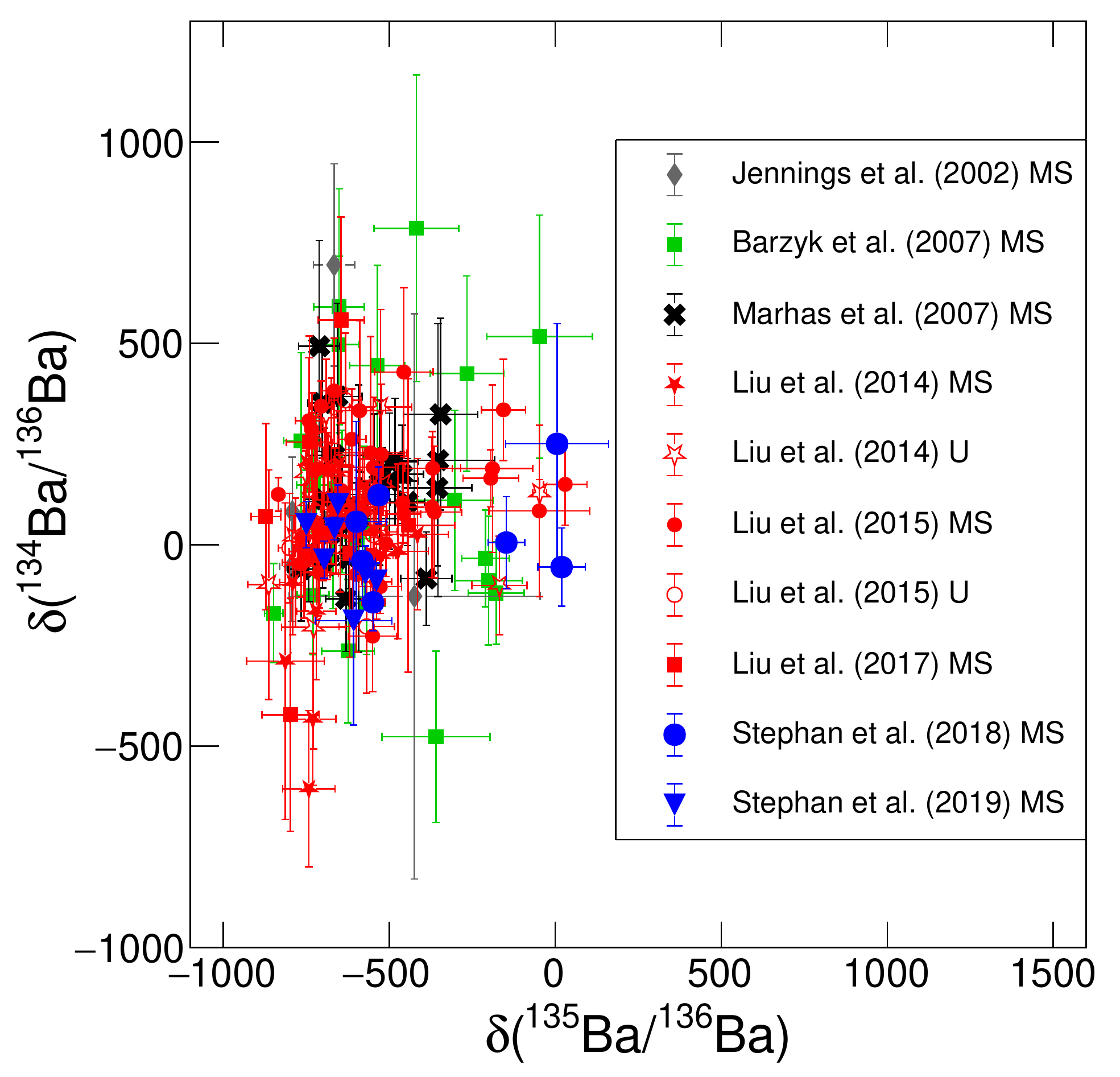}}
{\includegraphics[width=0.25\textwidth]{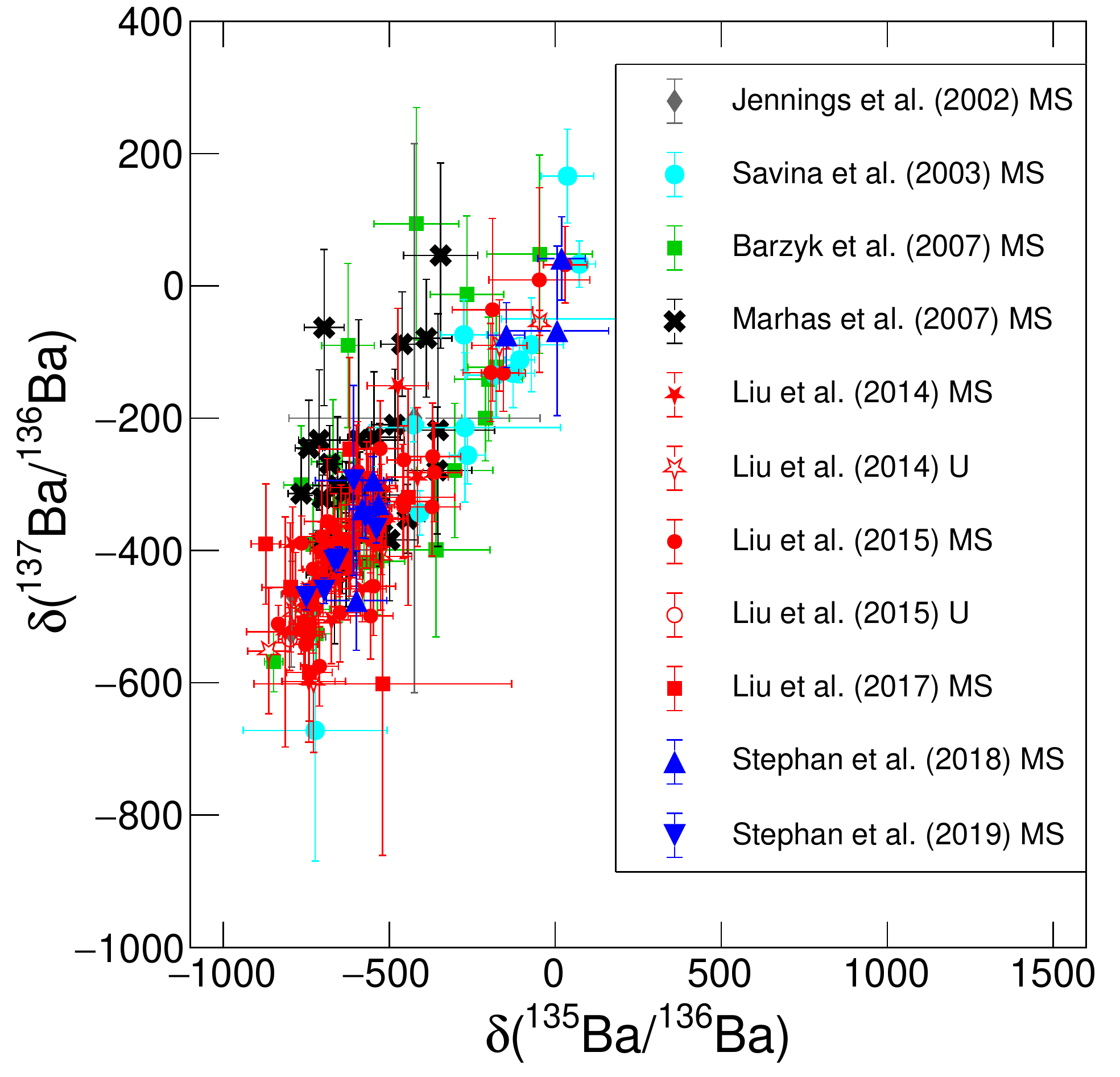}}
{\includegraphics[width=0.25\textwidth]{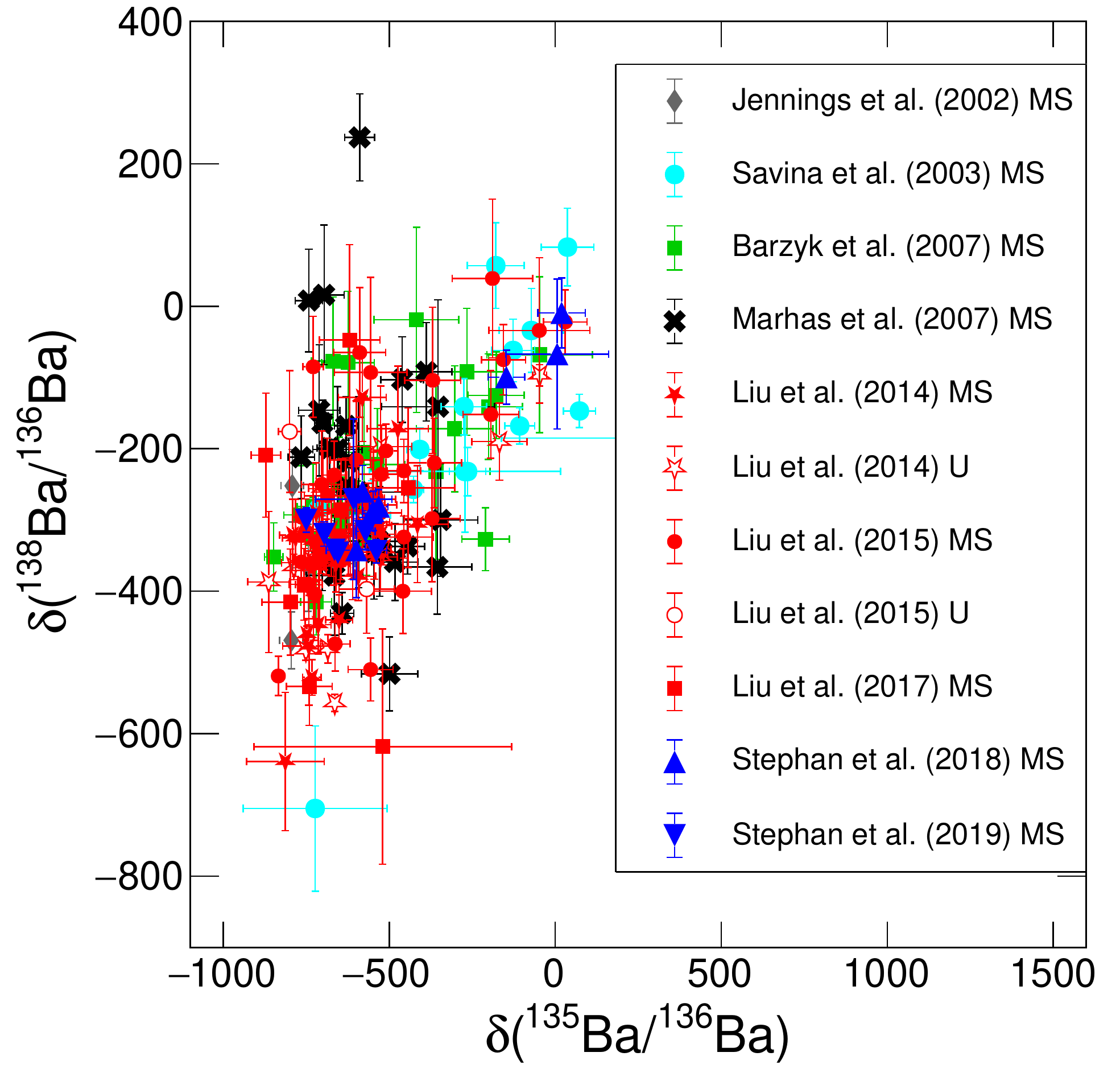}}
{\includegraphics[width=0.25\textwidth]{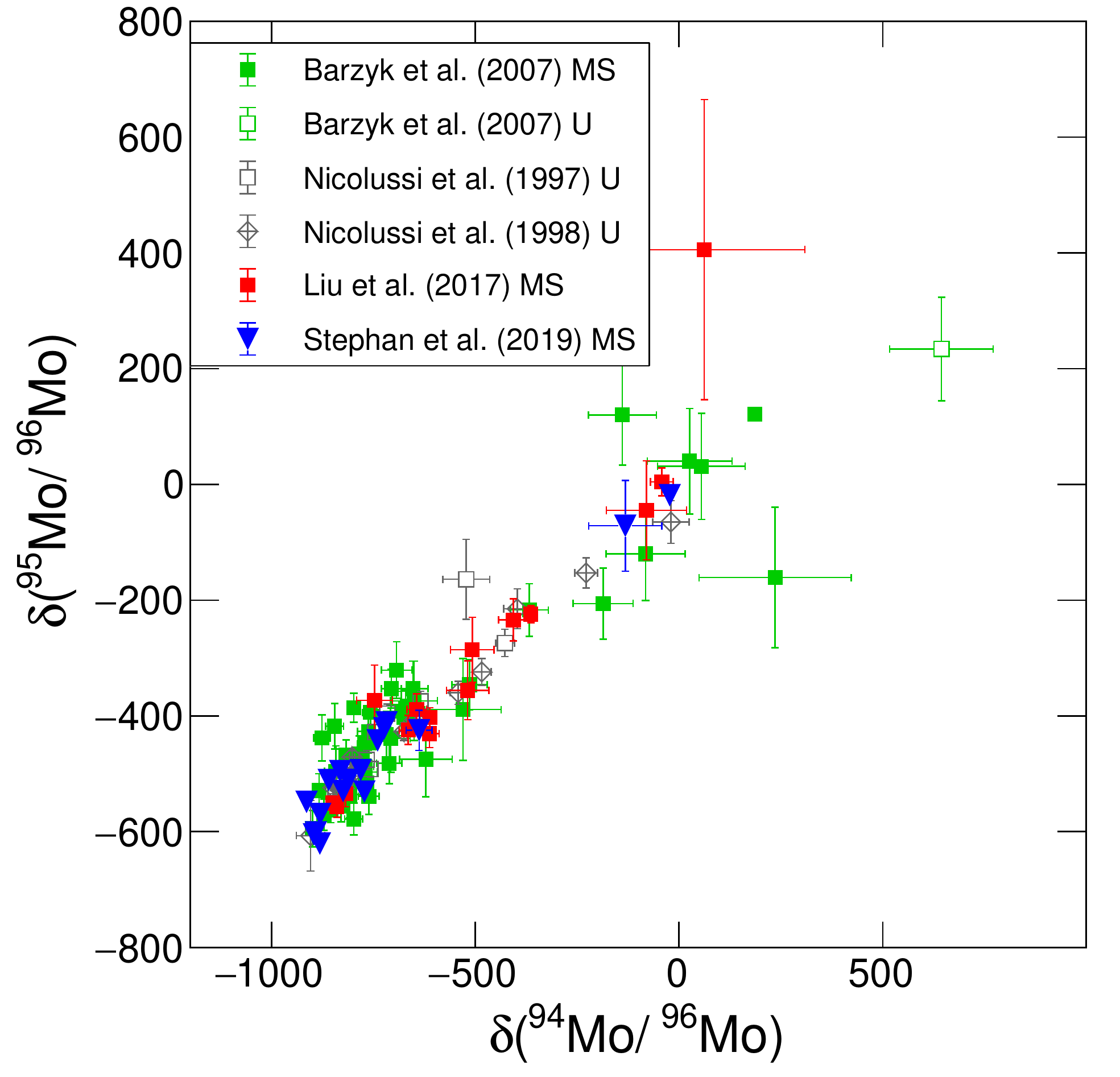}}
{\includegraphics[width=0.25\textwidth]{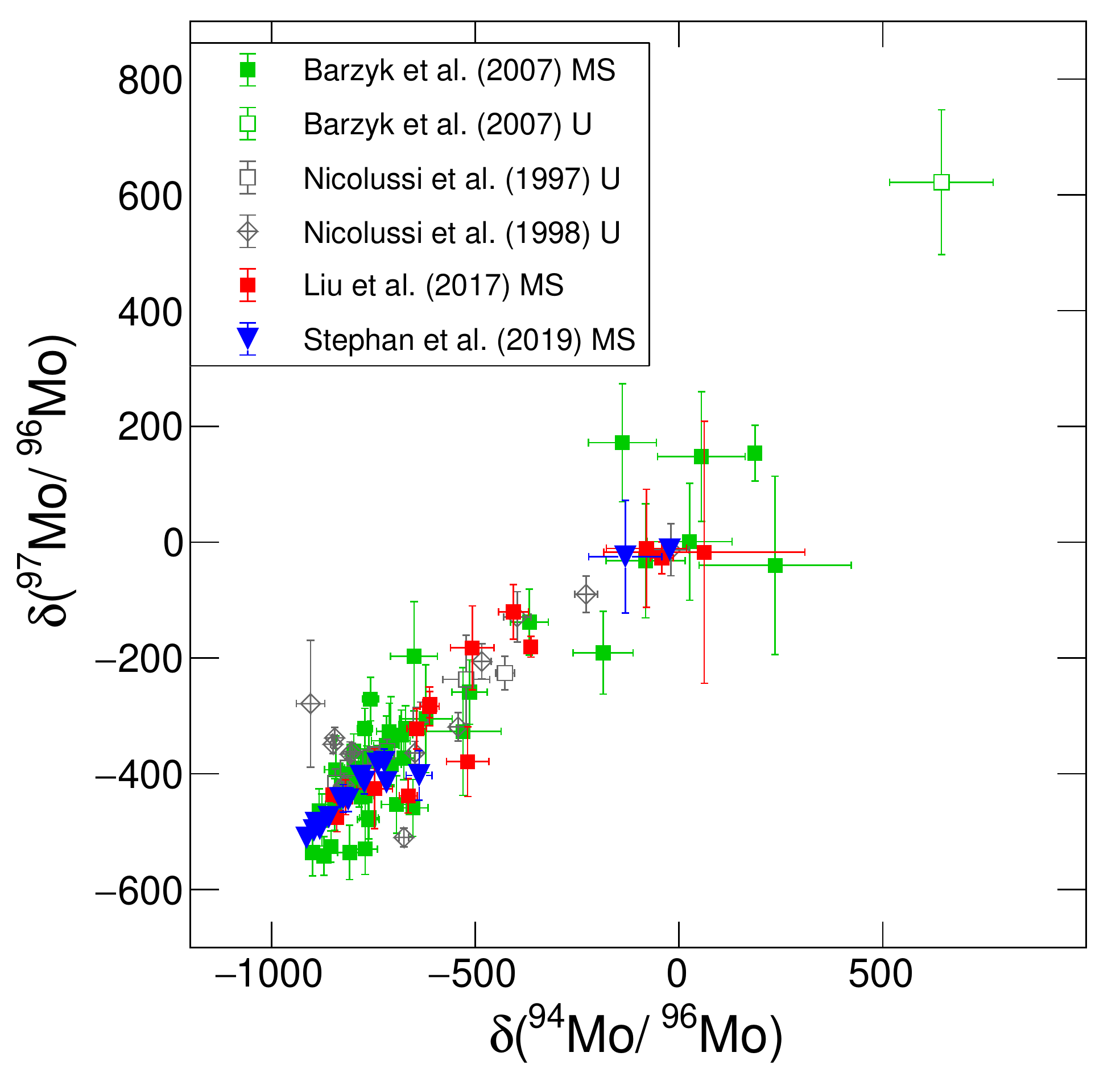}}
{\includegraphics[width=0.25\textwidth]{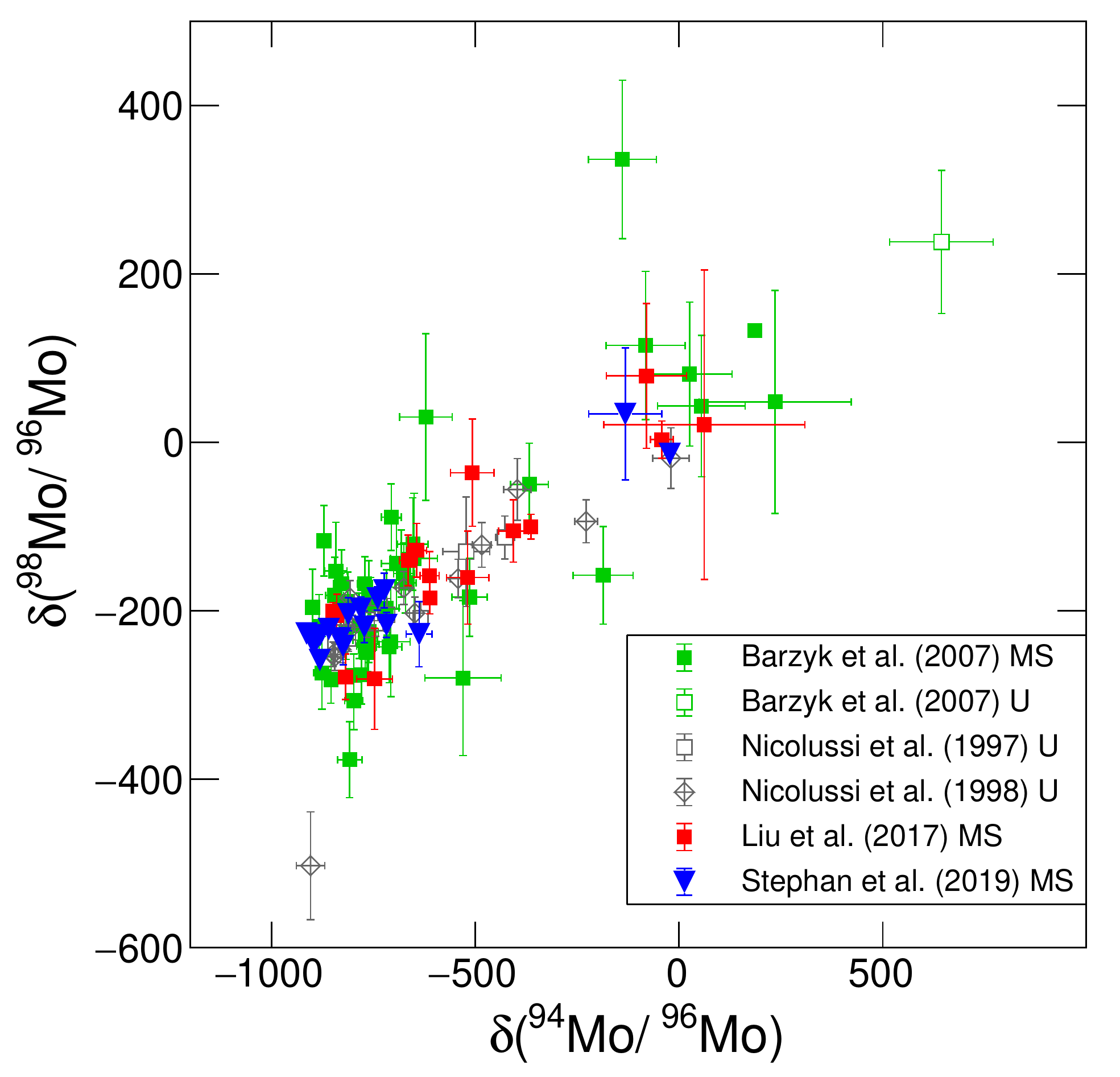}}
{\includegraphics[width=0.25\textwidth]{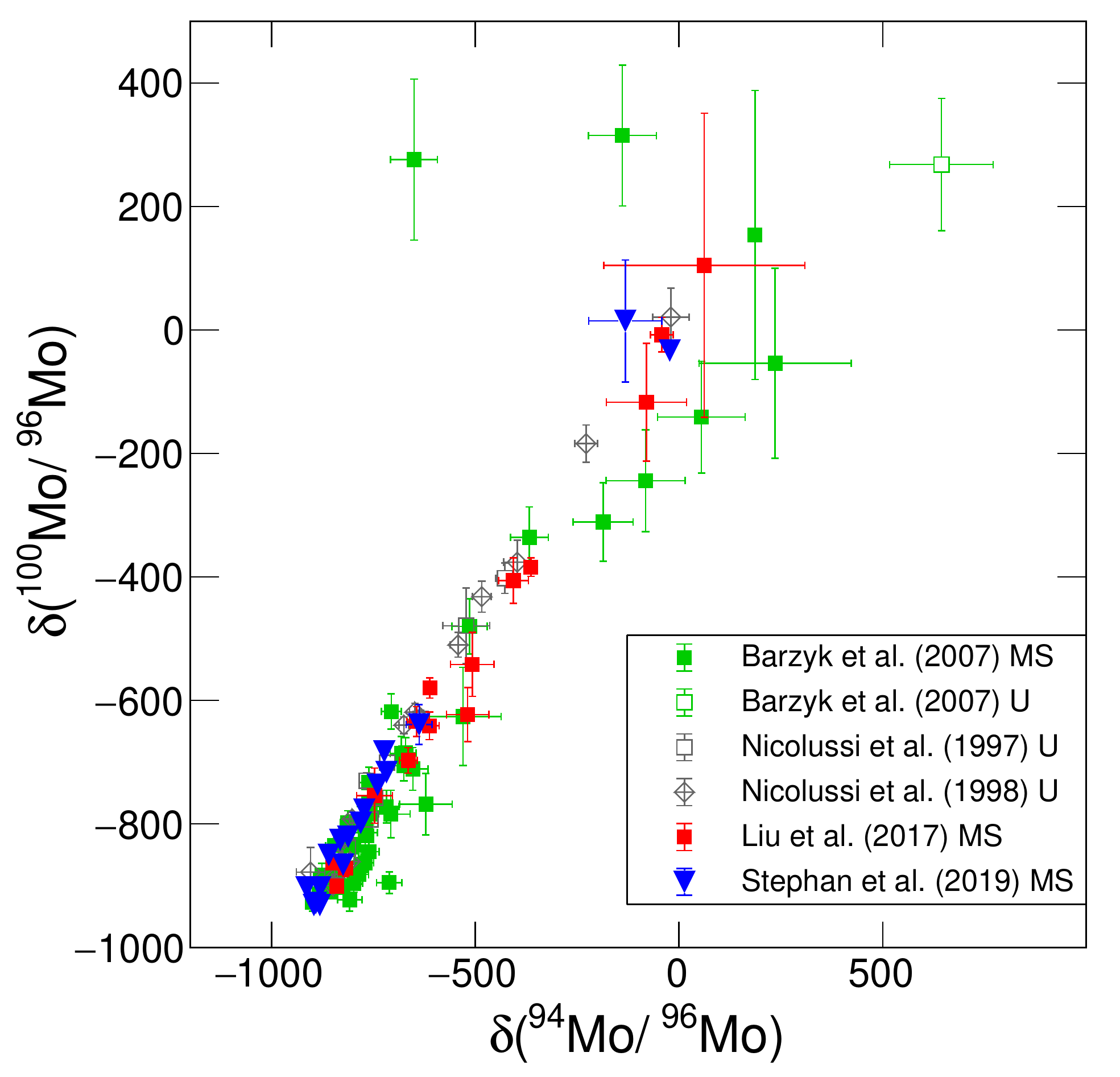}}
\caption{The whole sample of MS and U presolar SiC grains, for which the database of the Washington University in St. Louis reports values for the Sr, Zr, Mo and/or Ba isotopic ratios  \citep[][and references therein]{wustl,wustl2020}. Grain data were measured by several authors, as indicated by the legends. See text for details. \label{fig:SiC}}
\end{figure*}

The results are reported in several papers \citep{Nicolussi97,Nicolussi98,Jennings2002,savina03,Barzyk07,Marhas07,liu0,liu1,liu3,stephan,ste19}. Despite this fragmented scenario, data from the different samples show common trends for the measured 
values of the isotopic ratios, as shown in Figure \ref{fig:SiC}. There (as well as in any
reference to isotopic ratios presented in this work) the data are shown in the so-called 
${\it \delta}$ notation, where, for example:
$$
\delta\left(\frac{^{135}{\rm Ba}}{^{136}{\rm Ba}}\right) = 10^3\cdot\left[\left(\frac{^{135}{\rm Ba}}{^{136}{\rm Ba}}\right)/\left(\frac{^{135}{\rm Ba}}{^{136}{\rm Ba}}\right)_{\odot}-1\right]
$$

The mentioned concordant collective behaviors 
are maintained despite the different techniques adopted and the different research groups involved, thus revealing a remarkable strength in the constraints posed by MS-SiC data to $s$-process nucleosynthesis. 
In addition to the MS samples, the cited authors determined the isotope ratios for Sr, Zr, Mo and Ba also 
in a few tens of further presolar SiC grains, which have been classified as U (unknown), due to lack of 
clear data about C or N isotopic ratios. Since the heavy element isotopic ratios shown by grains of the U group are in agreement with the ones from the MS-SiC family, it is reasonable to believe that they are of AGB origins, too. We shall therefore use both the grain samples for the comparison with nucleosynthesis predictions of our models. For an easier reading, in the figures showing the comparisons we represent the SiC isotopic ratio data with grey dots, independently of the authors who actually did the measurements. The only distinction we maintain is the one between the MS and U families, which are plotted by filled and empty dots, respectively.

\section{Comparing models and measurements: a few results}\label{sec:mod}
\subsection{Sr versus Ba isotopic ratios}\label{sec:hsls}

We start the comparisons of model predictions with measurements
in presolar SiC grains by considering the relative behavior of the isotopes of two crucial elements (Sr and Ba), characterizing the neutron flow at the neutron magic numbers $N  = 50 $ and $N = 82$.

\begin{figure*}[t!!]
\centering
{\includegraphics[width=0.3\textwidth]{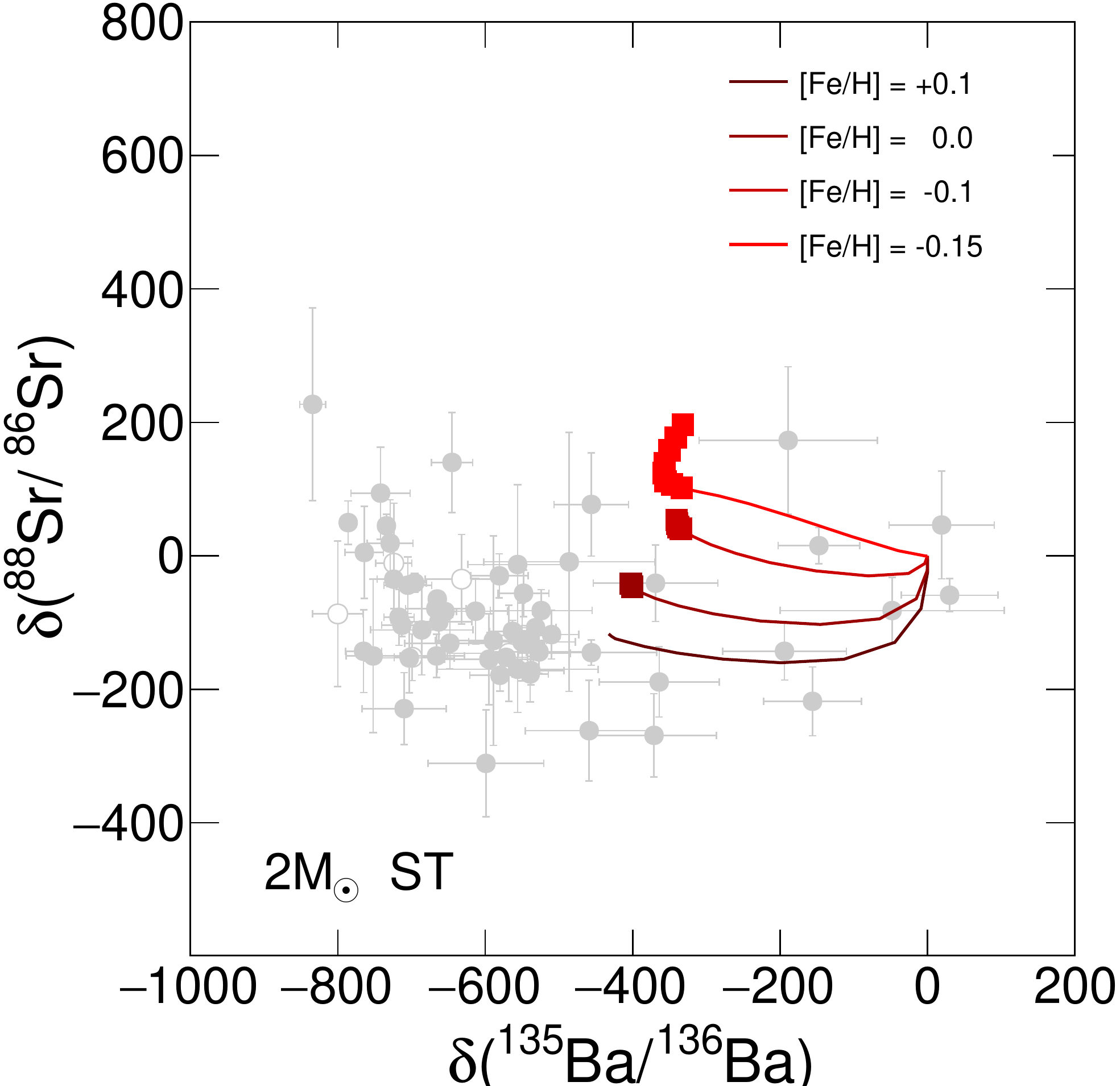}}
{\includegraphics[width=0.3\textwidth]{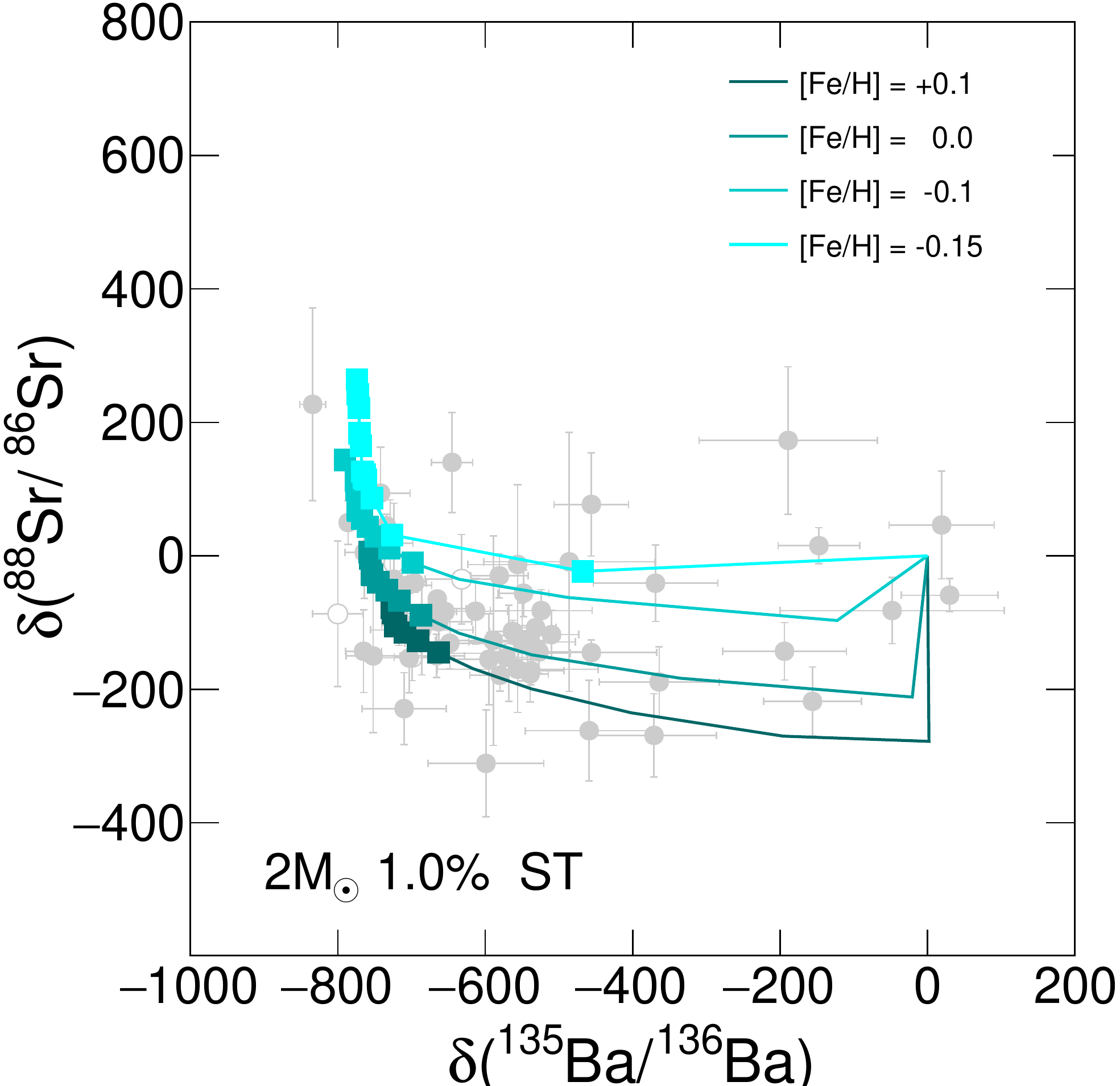}}
{\includegraphics[width=0.3\textwidth]{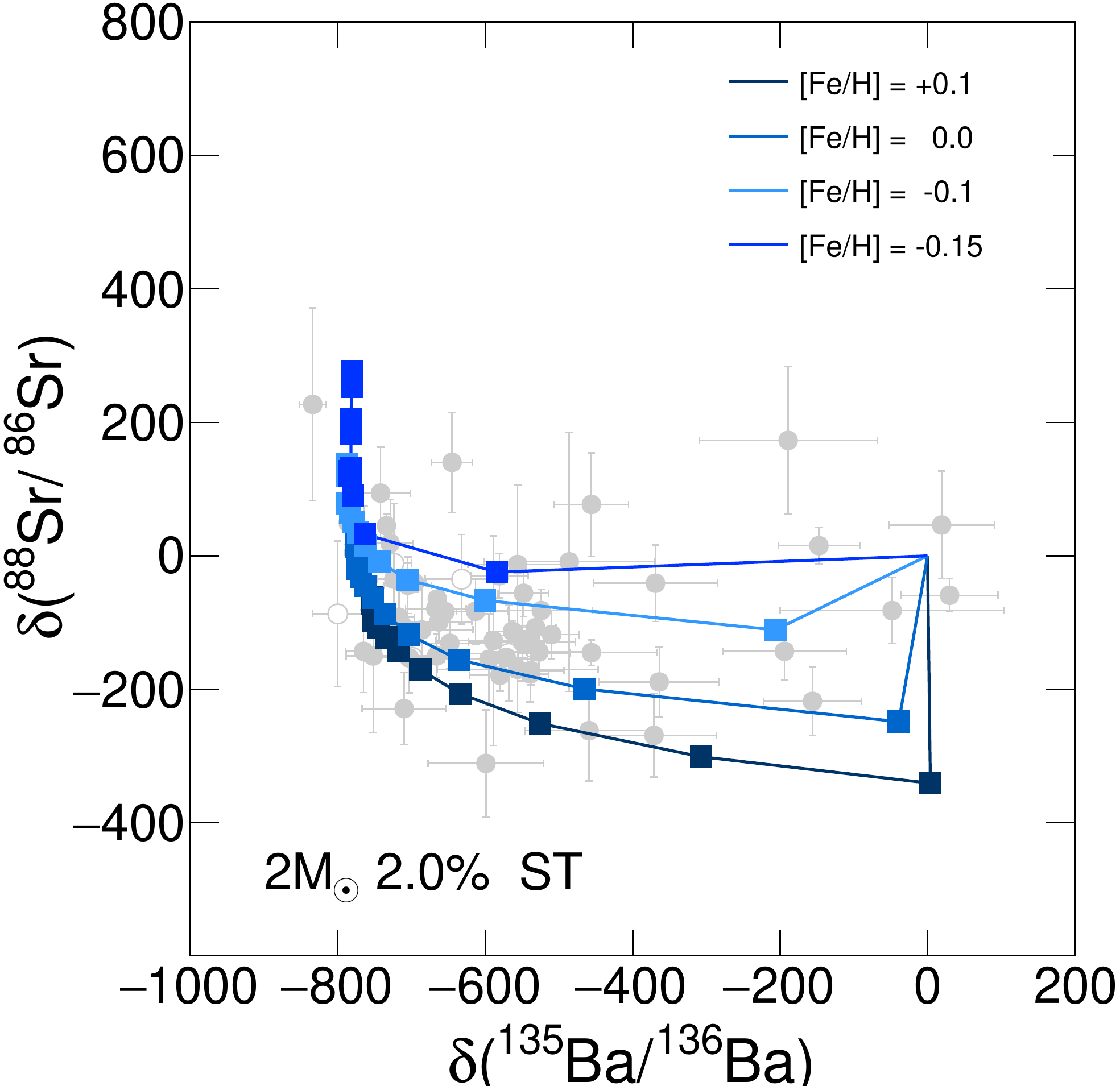}}
\caption{A comparison of nucleosynthesis predictions from  2\msb models (full lines) with SiC measurements (represented by gray dots with error bars, in the delta notation  defined in the text). The case of the isotopic ratios $^{88}$Sr/$^{86}$Sr and $^{135}$Ba/$^{136}$Ba is shown. The first panel (left) shows the envelope composition. There, full squared dots represent model abundance ratios
achieved after TDU episodes that follow  $TP$s, provided the envelope has reached a composition with C/O $>$ 0.8 \citep[see][]{b+21}. The central panel shows the isotopic ratios in the 
magnetized winds, as defined in section \ref{sec:magwinds}, if the percentage of He-shell material introduced in them by breaking magnetic flux tubes it at the level of 1\%. The last panel (right) represents the same situation as the central one, but for a percentage of He-shell material of 2\%. Also in these other two panels full squares indicate the {\it TPs} were the wind has a C/O ratio larger than 0.8.}
\label{fig:stroba1}
\end{figure*}

\begin{figure*}[t!!]
\centering
{\includegraphics[width=0.3\textwidth]{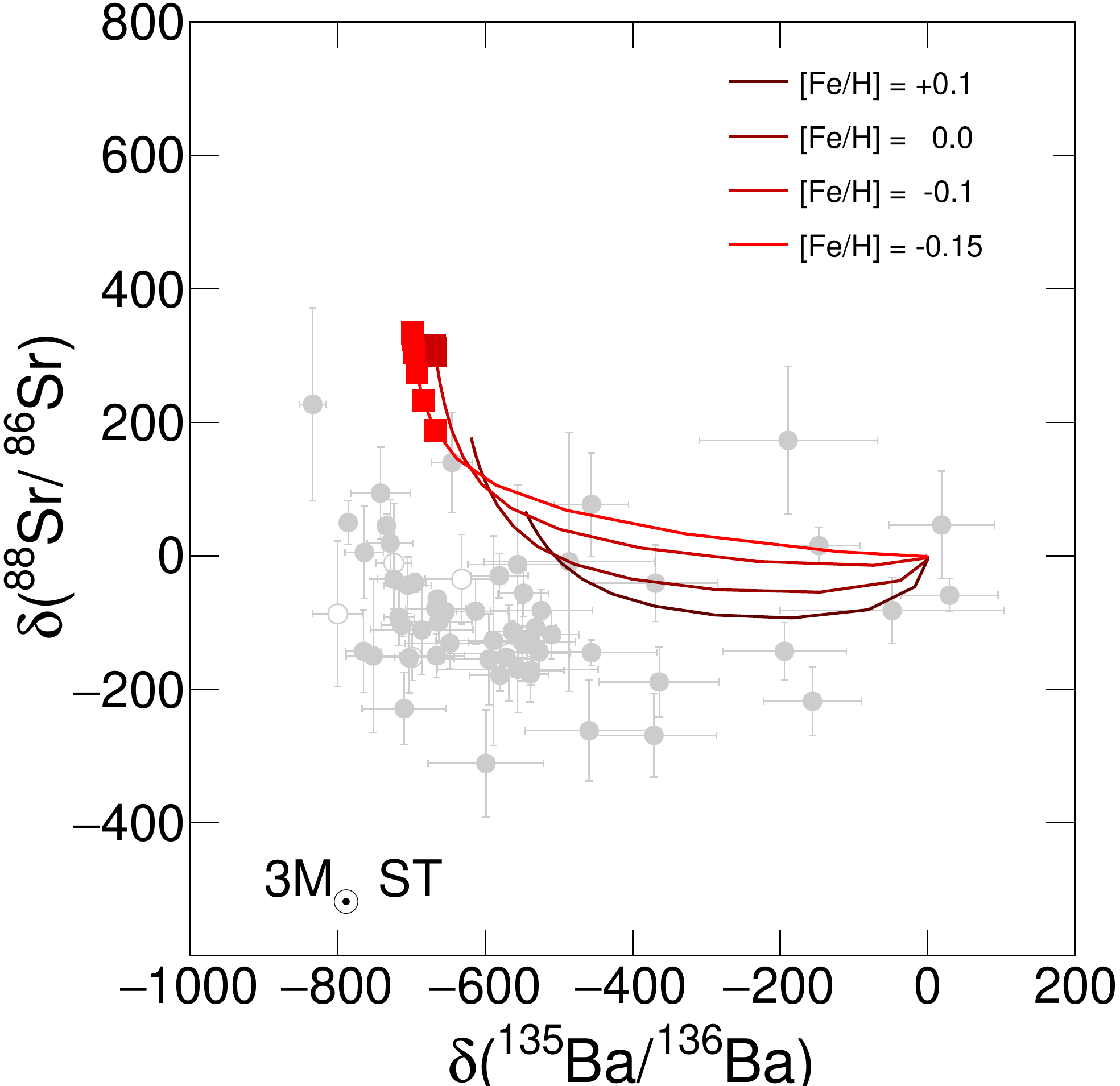}}
{\includegraphics[width=0.3\textwidth]{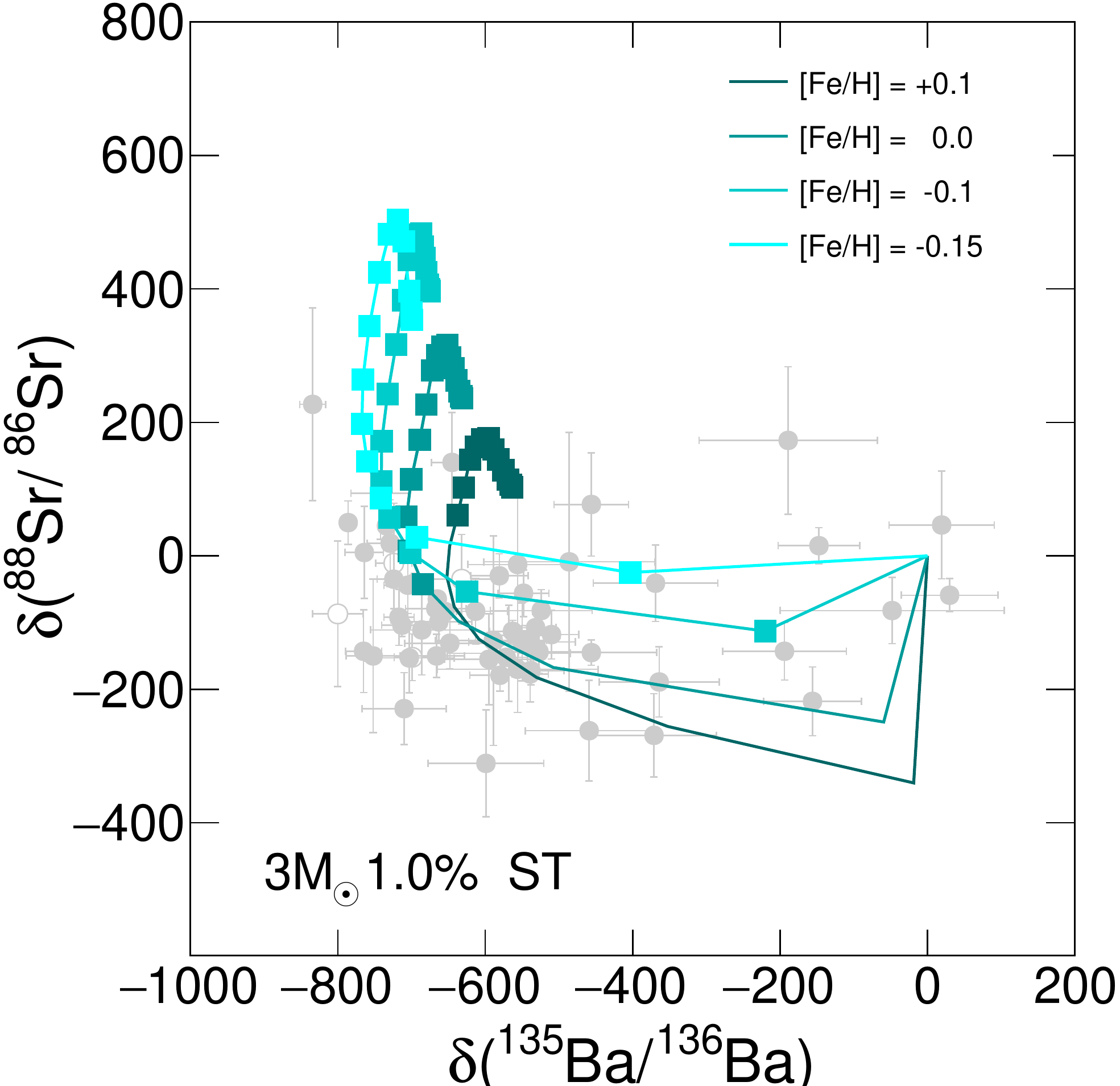}}
{\includegraphics[width=0.3\textwidth]{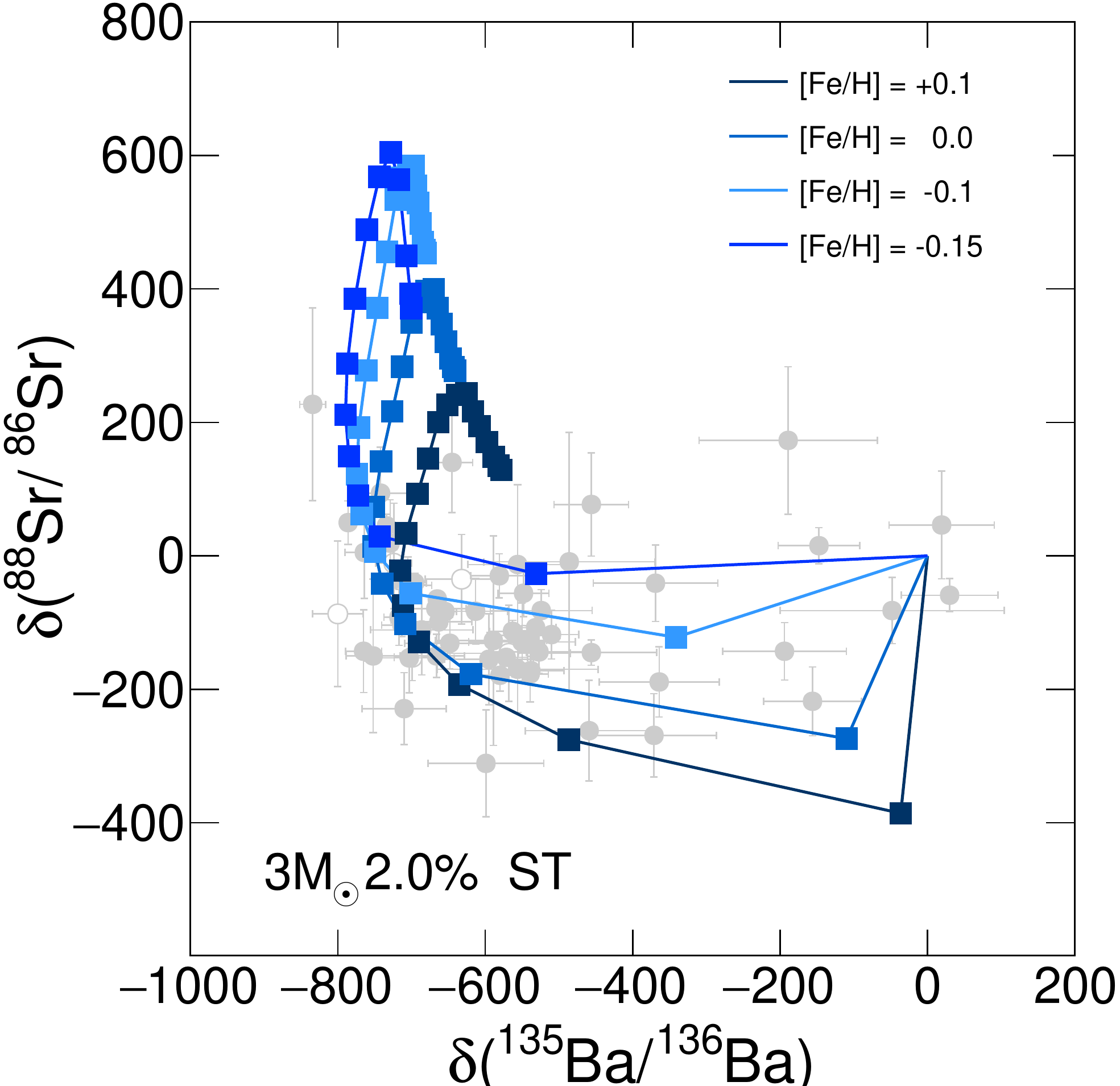}}
\caption{Same as Figure \ref{fig:stroba1}, but for models of 3 \msb stars.
\label{fig:stroba2}}
\end{figure*}

\begin{figure*}[t!!]
\centering{
{\includegraphics[width=0.3\textwidth]{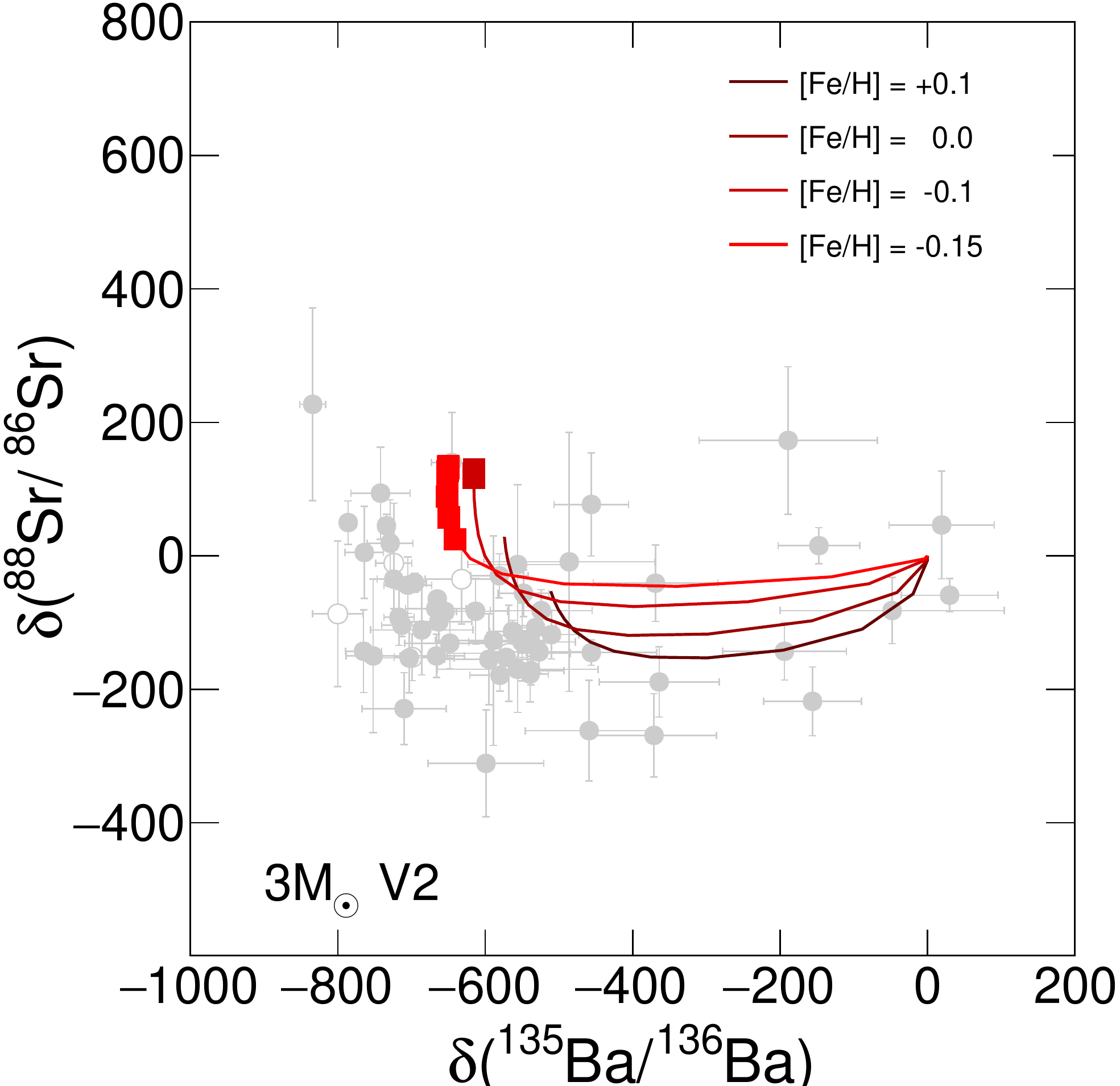}}
{\includegraphics[width=0.3\textwidth]{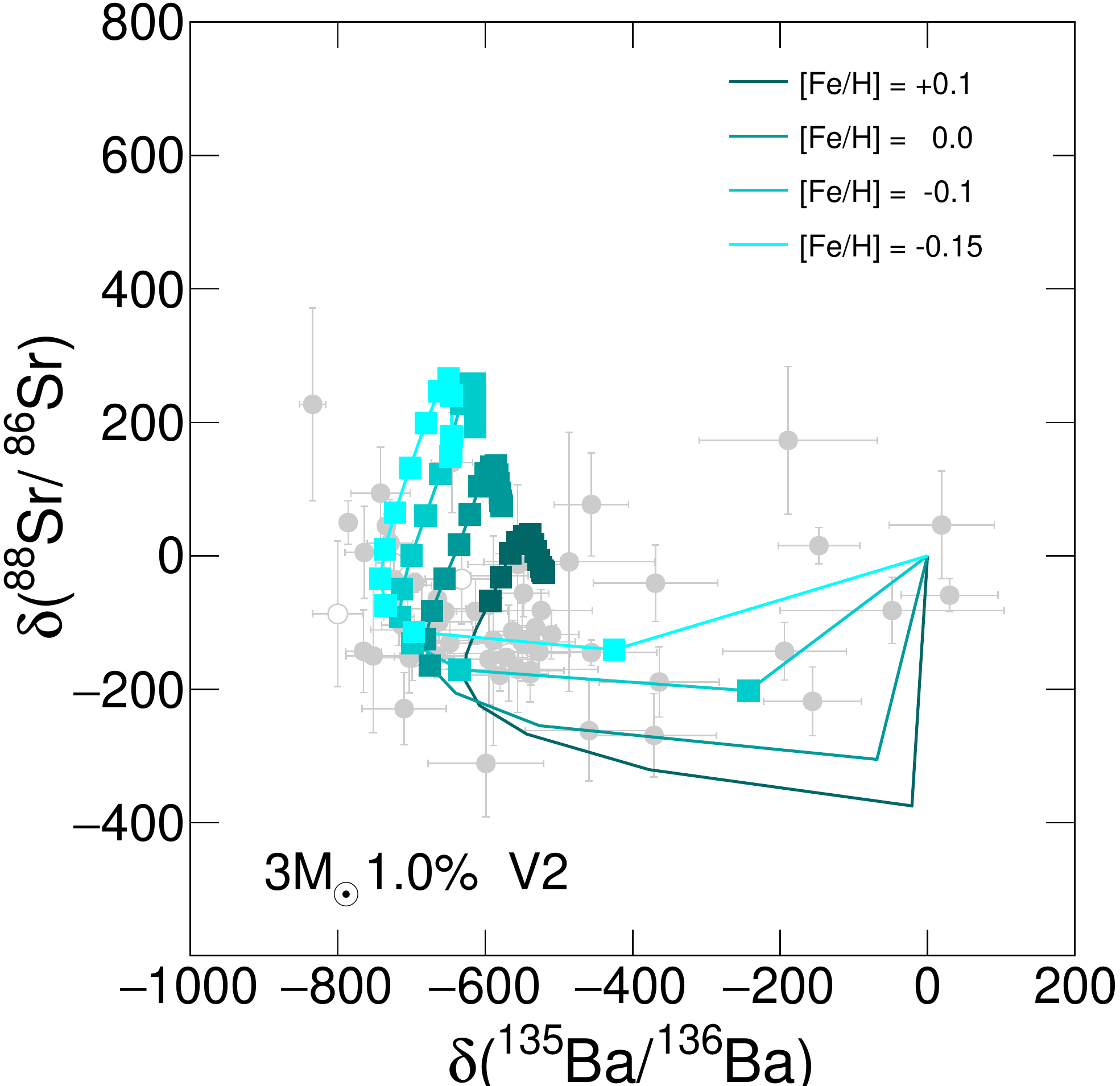}}
{\includegraphics[width=0.3\textwidth]{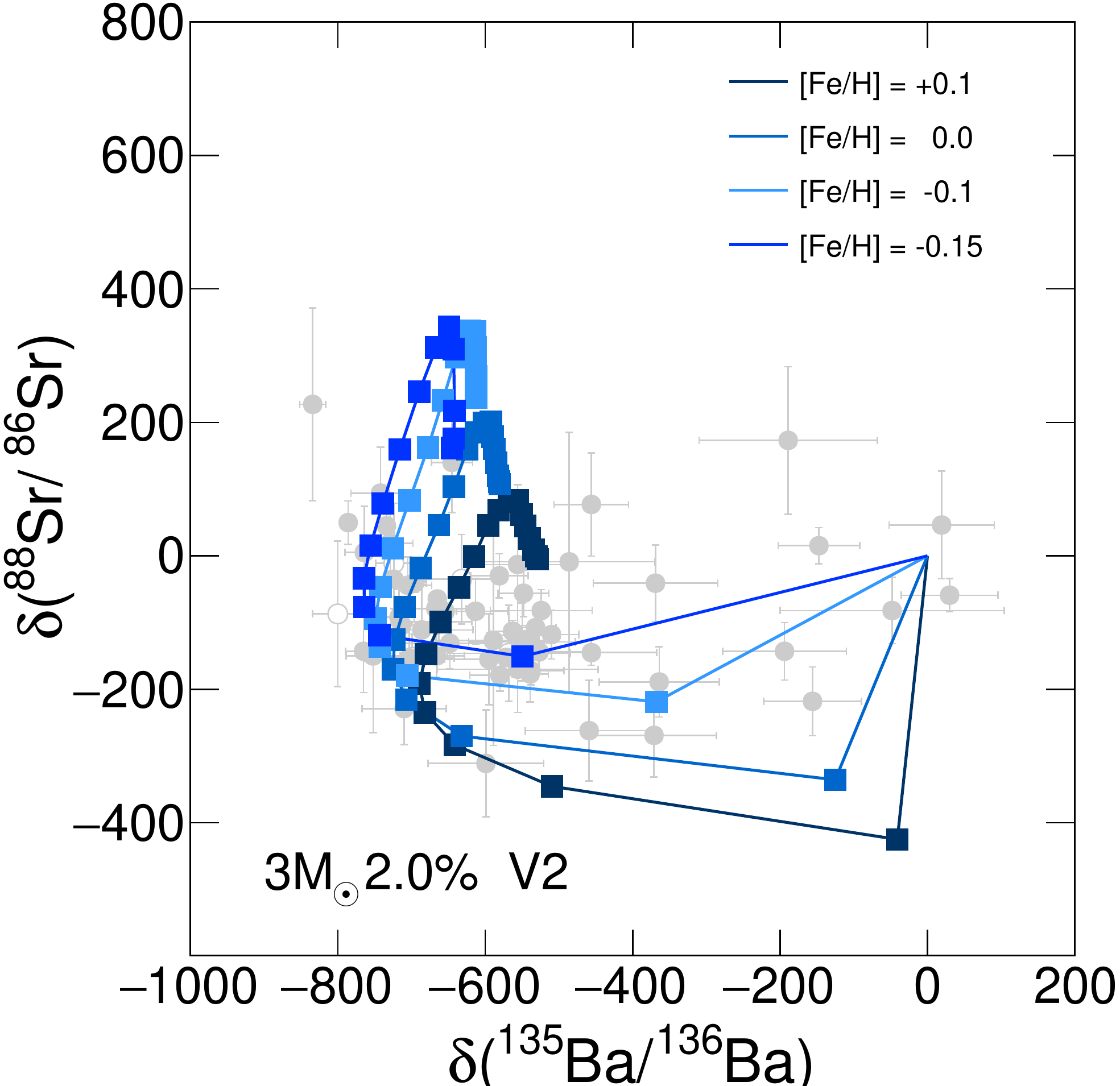}}
\caption{Same as Figure \ref{fig:stroba2}, but for the test models V2, with tentatively modified nuclear inputs (see section \ref{sec:nuclear}).}
\label{fig:bastro}}
\end{figure*}

\begin{figure*}[t!!]
\centering{
{\includegraphics[width=0.3\textwidth]{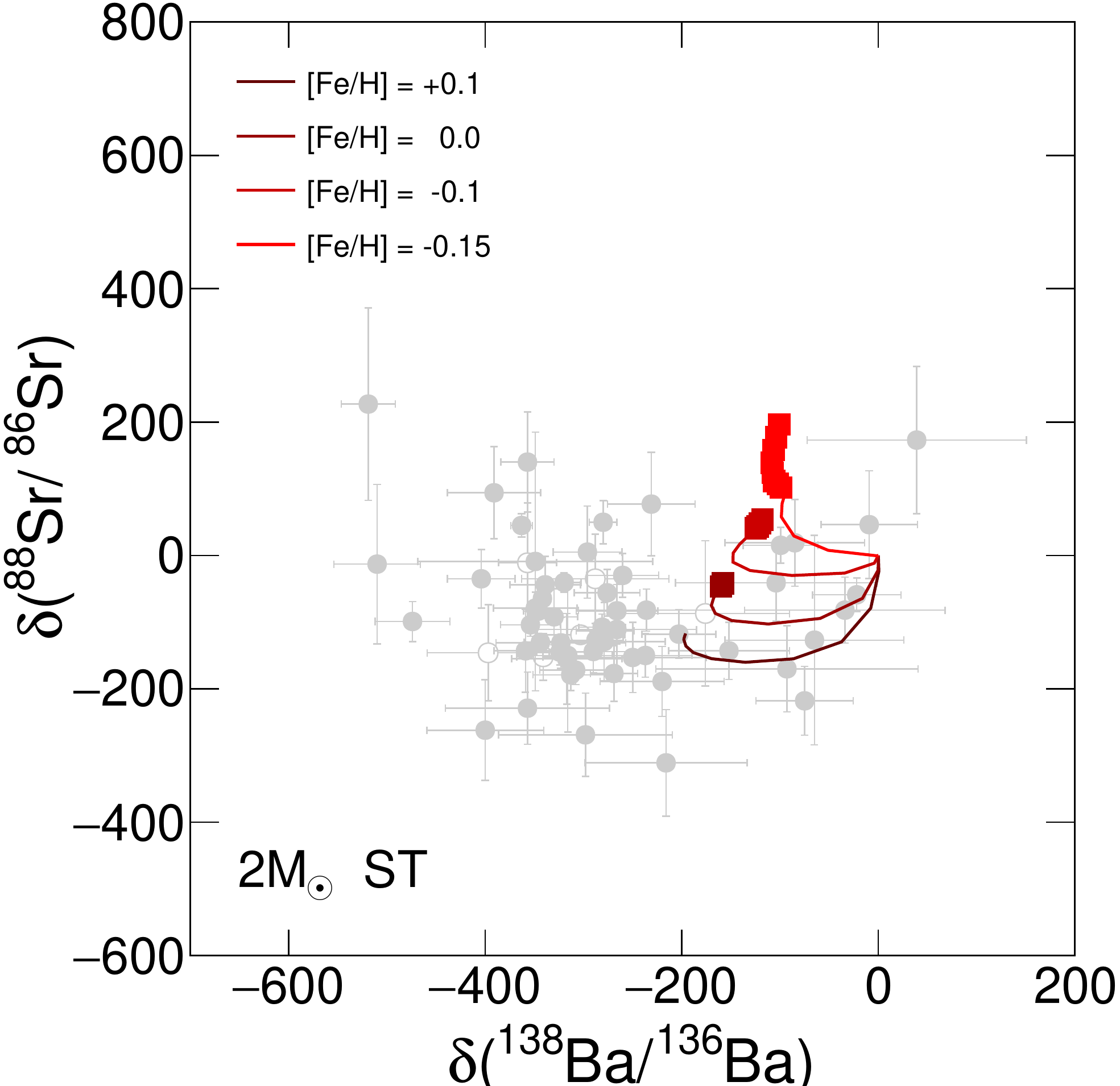}}
{\includegraphics[width=0.3\textwidth]{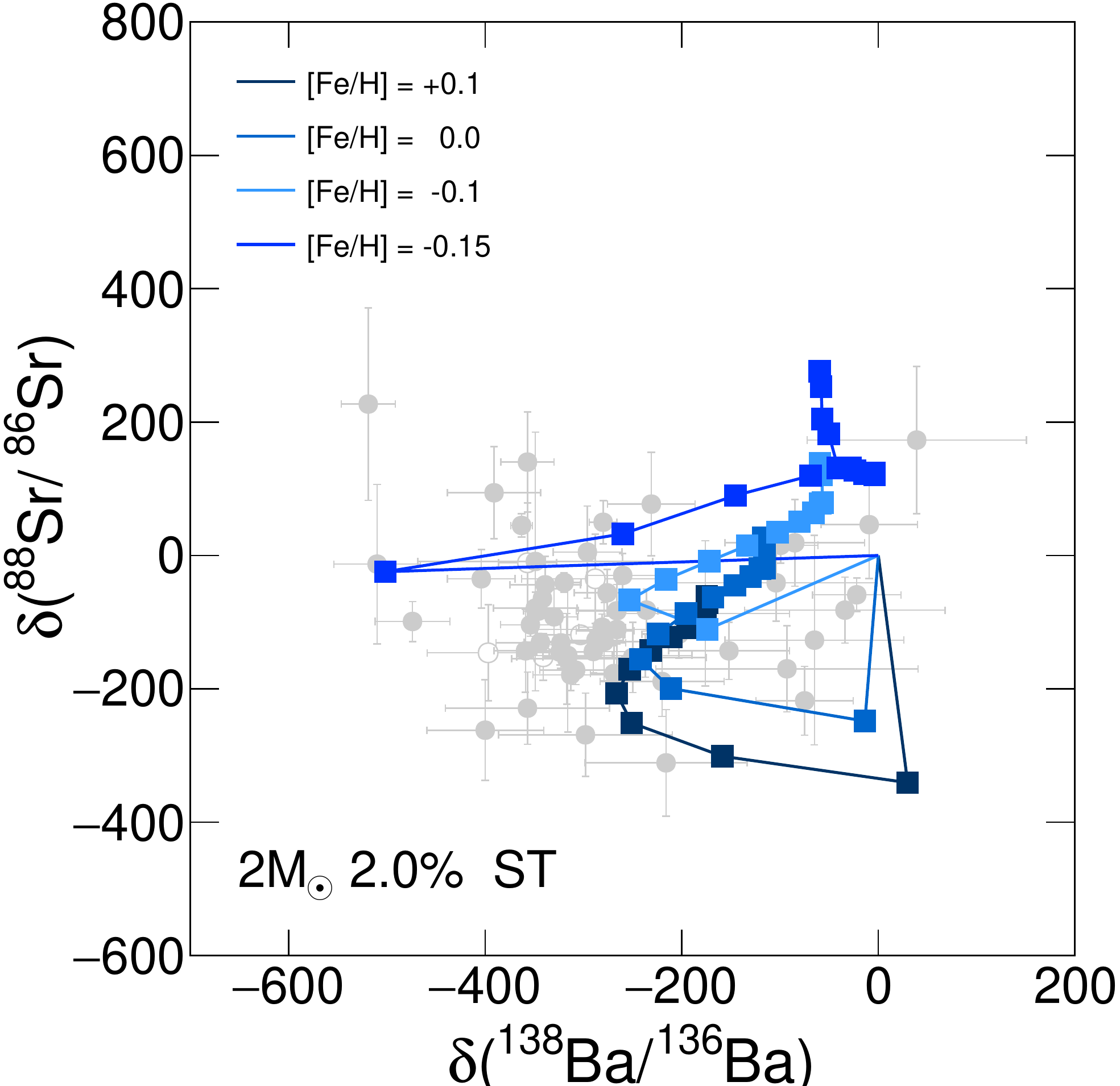}}
{\includegraphics[width=0.3\textwidth]{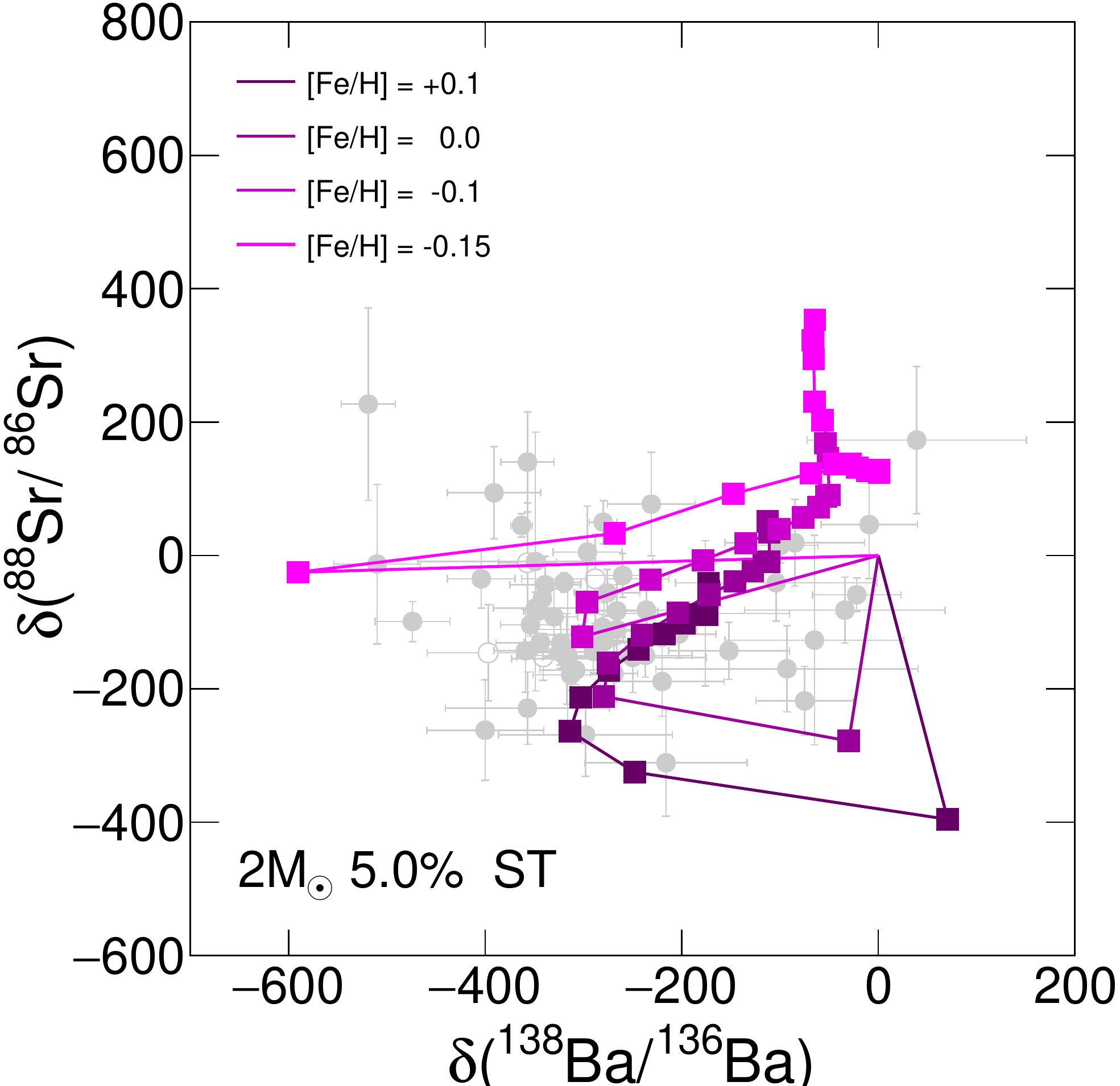}}
\caption{A comparison of model predictions with SiC data (with notations as defined in Figure \ref{fig:stroba1}) for the isotopic ratios $^{88}$Sr/$^{86}$Sr and $^{138}$Ba/$^{136}$Ba.
The first panel (left) shows the envelope composition, the central panel shows the isotopic ratios in the magnetized winds when the percentage of He-shell material in them is 2\%. The panel at the right side refers again to the winds, but for a percentage of He-shell material of 5\%}.
\label{fig:newfba2}}
\end{figure*}

\begin{figure*}[t!!]
\centering{
{\includegraphics[width=0.3\textwidth]{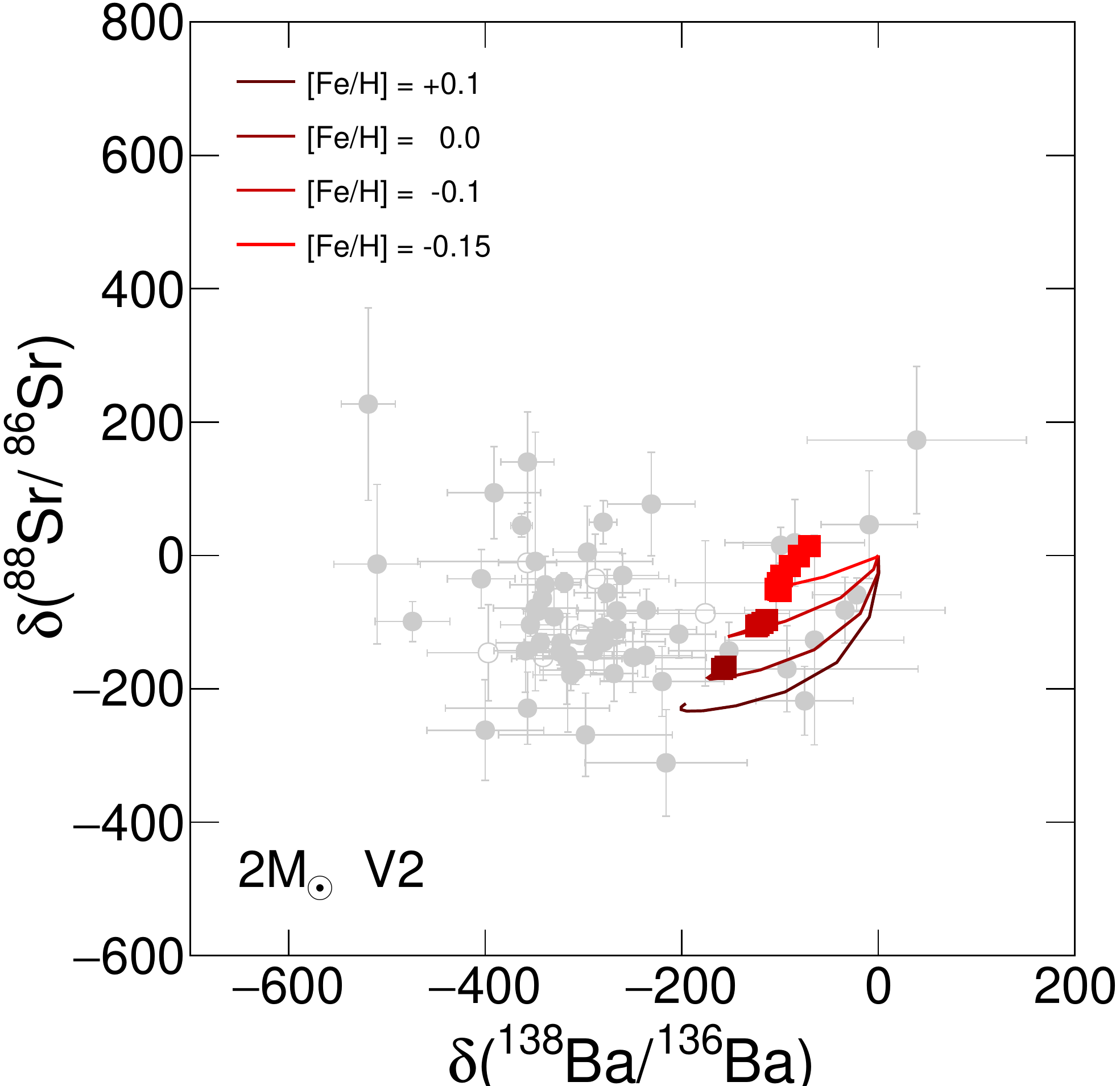}}
{\includegraphics[width=0.3\textwidth]{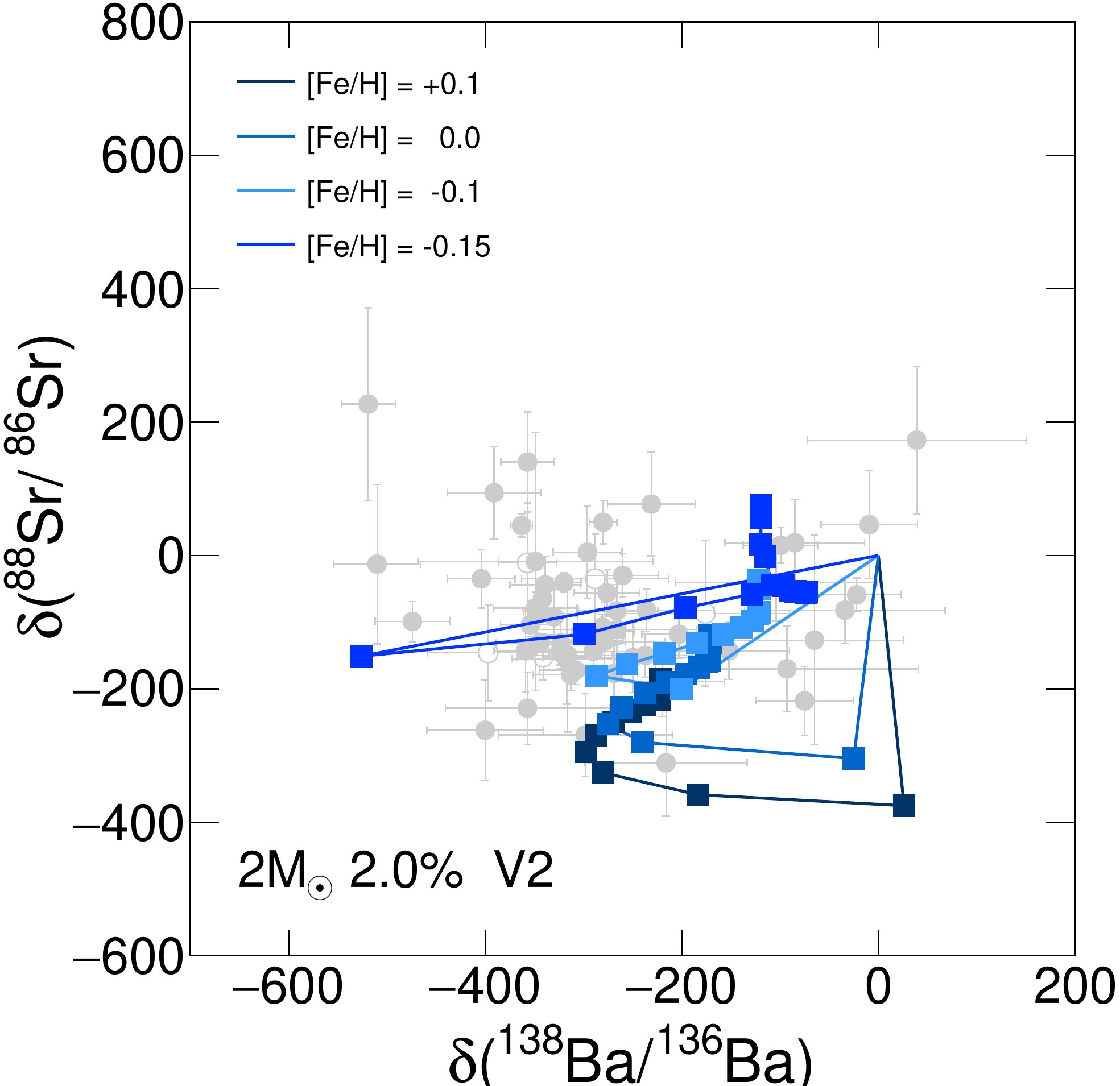}}
{\includegraphics[width=0.3\textwidth]{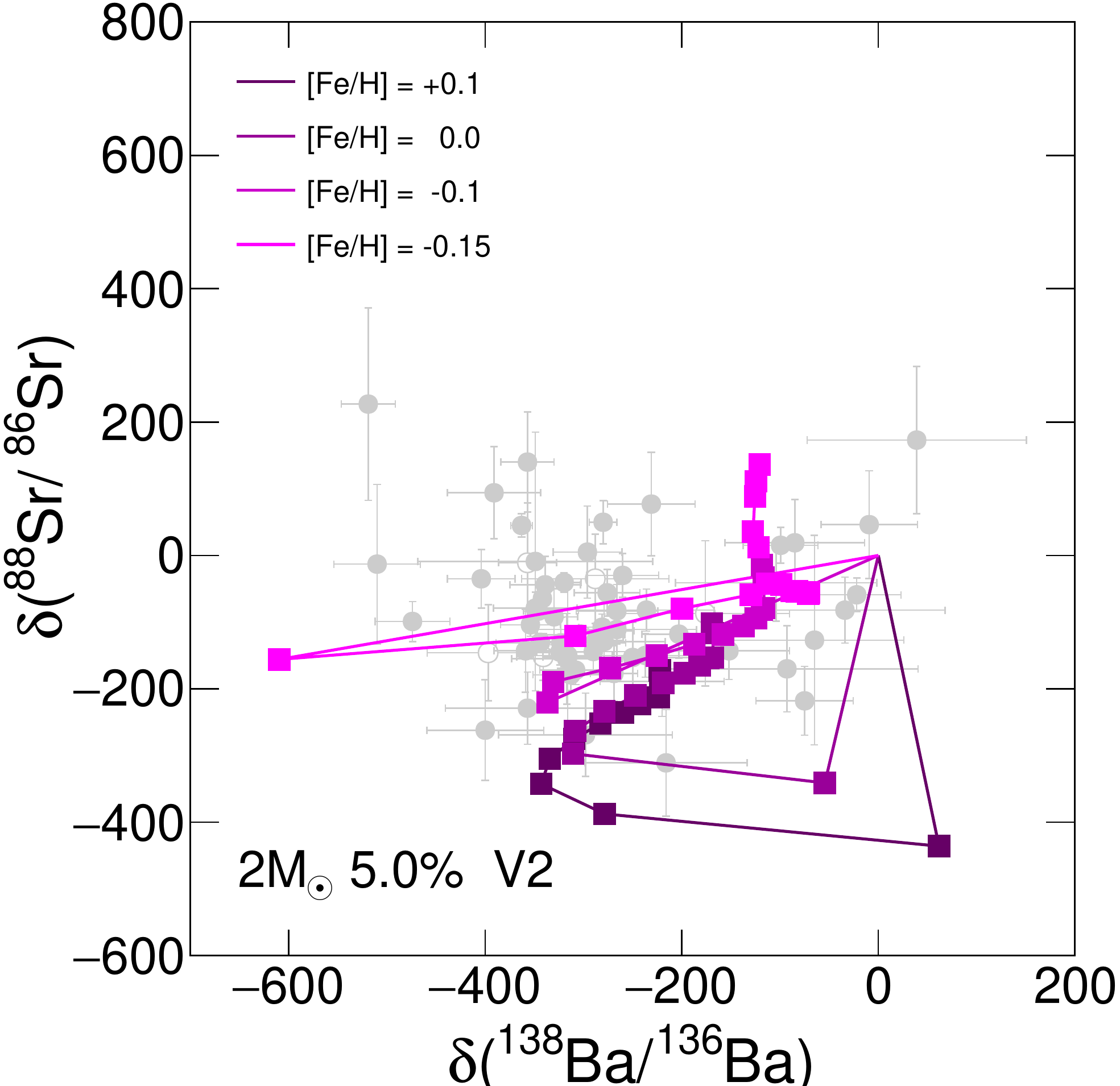}}
\caption{A plot similar to Figure \ref{fig:newfba2}, but for test models 
V2, with tentatively modified nuclear input parameters (see section \ref{sec:nuclear}).}
\label{fig:newfba3}}
\end{figure*}

\begin{figure*}[t!!]
\centering{
{\includegraphics[width=0.3\textwidth]{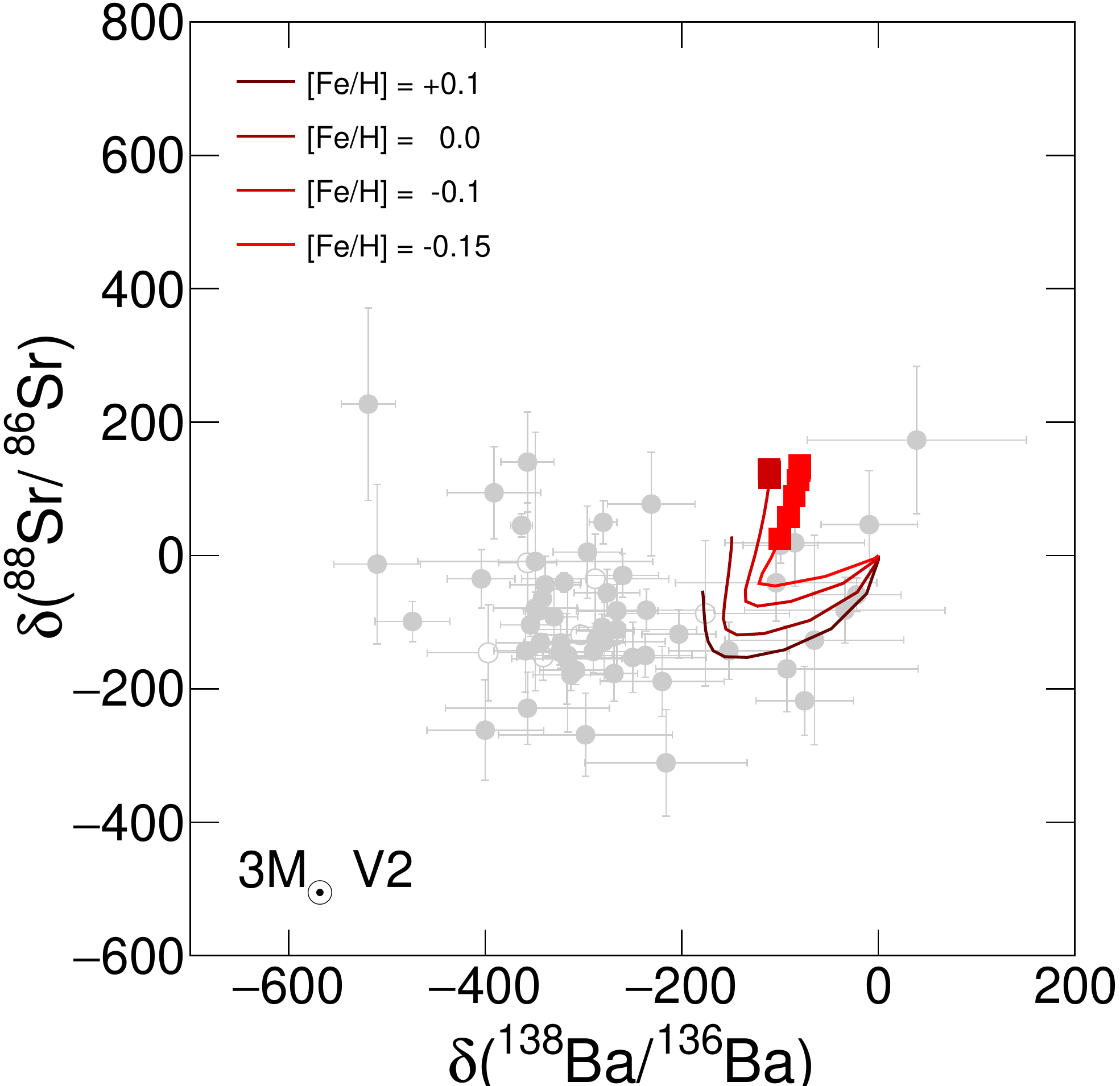}}
{\includegraphics[width=0.3\textwidth]{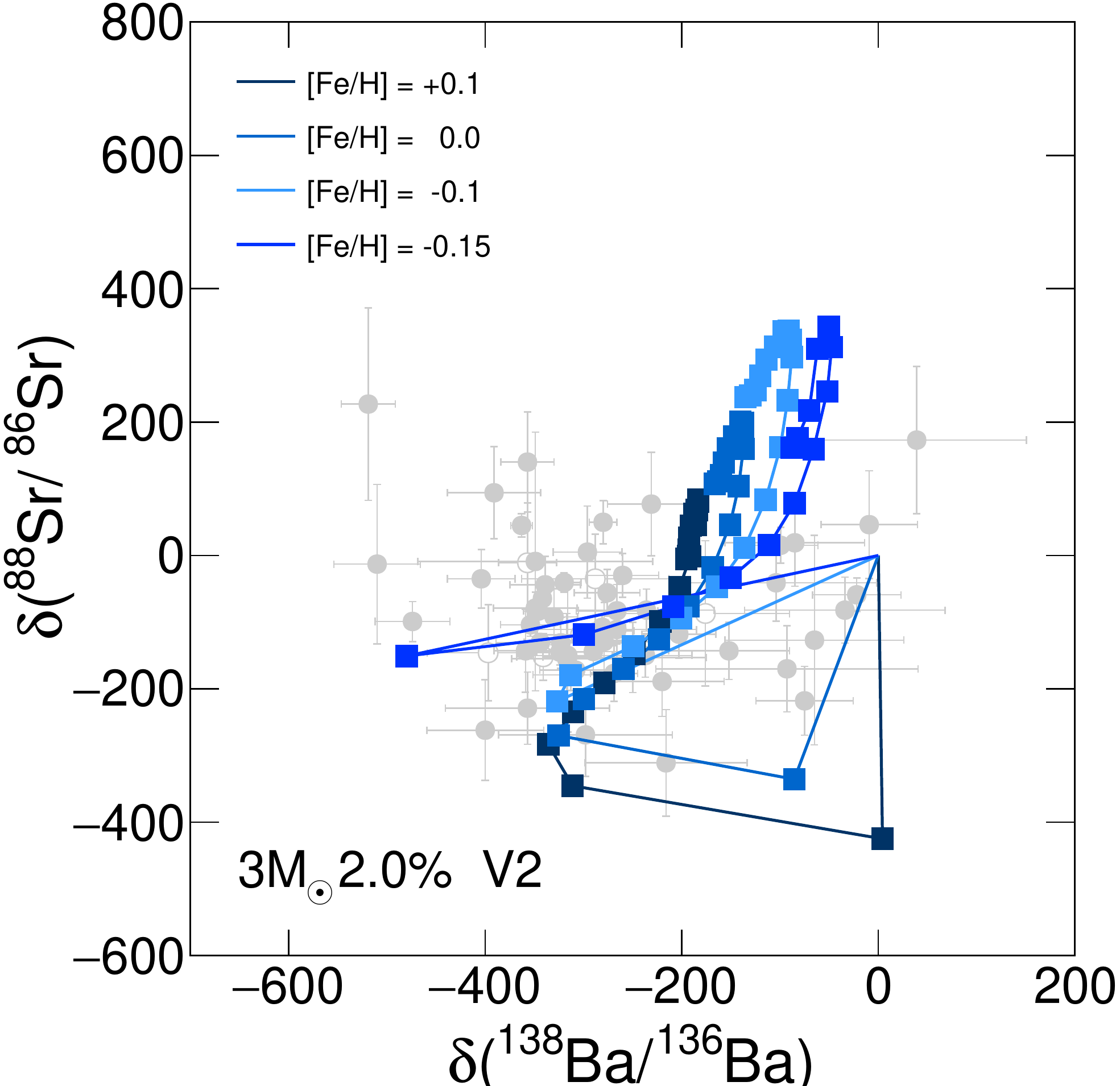}}
{\includegraphics[width=0.3\textwidth]{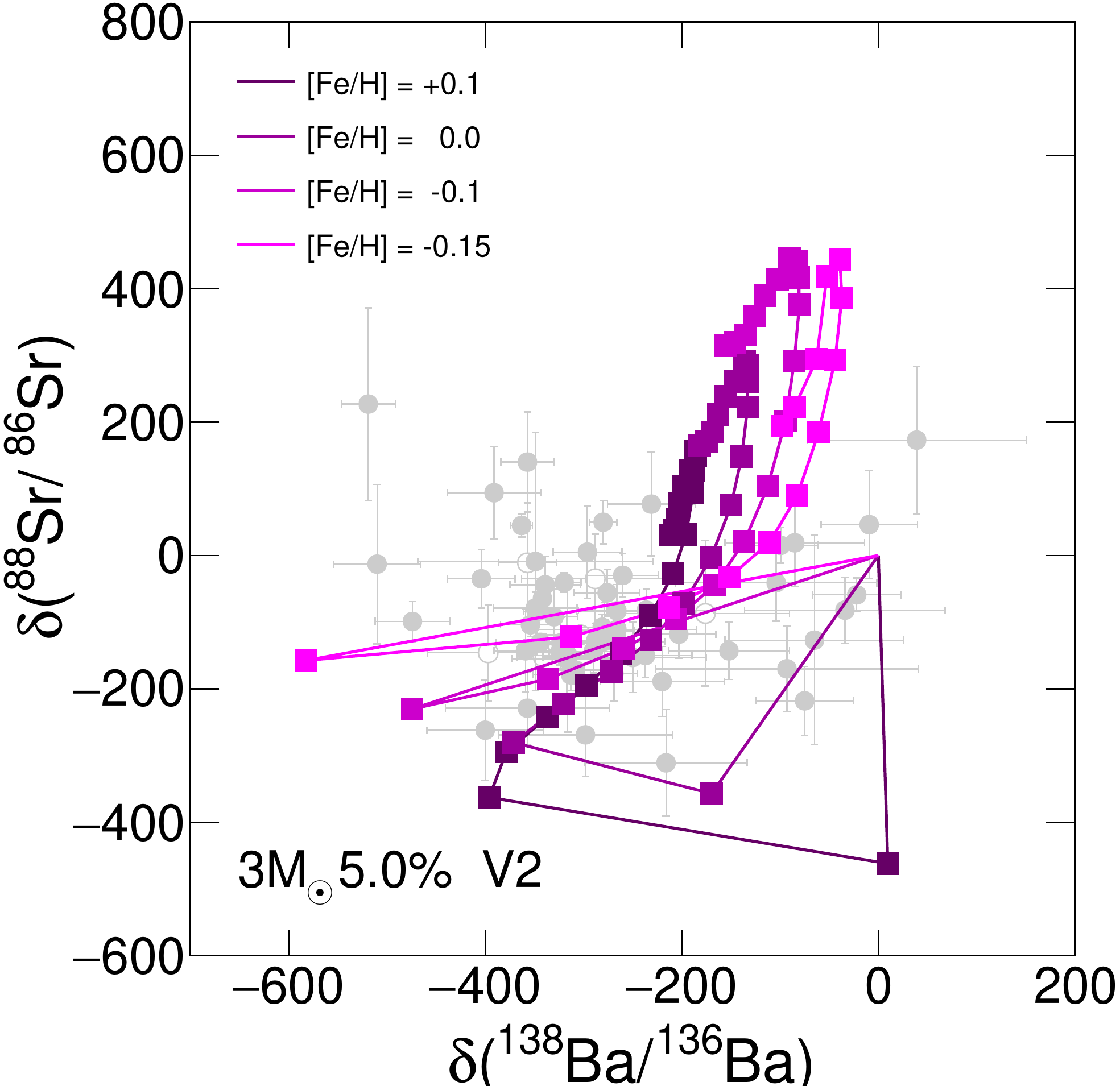}}
\caption{A plot similar to Figure \ref{fig:newfba3}, but for 3 \msb models.}
\label{fig:heavybasr}}
\end{figure*}

Figure \ref{fig:stroba1} illustrates
the case of the measured isotopic ratios $^{88}$Sr/$^{86}$Sr and $^{135}$Ba/$^{136}$Ba, as compared to predictions from our 2 \msb models and  reveals some interesting details. In general, the area covered by the measurements and the one covered by the models overlap largely (see especially the panel at the right side). This indicates that
the quality of the nuclear parameters, as adopted after the discussion of section \ref{sec:nuclear}, is already rather satisfactory. However, important improvements are possible. Let us analyze this situation in some detail. As shown by the left panel of Figure \ref{fig:stroba1}, models for the C-rich phases in the envelopes tend to produce intermediate delta values for the ratio $^{135}$Ba/$^{136}$Ba, not suited to cover the wide area of the measurements. This problem was common in previous approaches in the literature \citep[see e.g.][]{lugaro8}. When instead we consider the magnetized winds discussed in the text (see section \ref{sec:magwinds}), their compositions sample a wider distribution of $^{135}$Ba/$^{136}$Ba ratios, and cover increasing portions of the measurement area for increasing amounts of the He-shell material added in them. They can actually account for these SiC data almost completely, as shown in the right panel of Figure \ref{fig:stroba1}. This fact shows the relevance of this wind component, not considered so far in the literature. Anyhow, further improvements in the model predictions might be welcome if one wants to refer also to 3 \msb models. Their behavior with respect to the experimental constraints is illustrated in Figure \ref{fig:stroba2} for the same cases chosen in the previous plot. As is clear, with our standard choice of the nuclear parameters these models should be excluded. Indeed, when they cover the whole range of 
$^{135}$Ba/$^{136}$Ba data (see full lines with heavy dots in Figure \ref{fig:stroba2}), their Sr isotopic ratio in the ordinate  extends at too high $\delta$-values, where there are no measurements. This is due to the higher efficiency, in these more massive models (due to the higher pulse temperature) of the reaction branch that from $^{85}$Kr feeds 
$^{86}$Kr, $^{87}$Rb and $^{88}$Sr (see section \ref{sec:krbranch}). 
Can we exclude completely, on this basis, that SiC grains be formed, at least in part, by stars more massive than 2 \msb at the adopted metallicities? {The previous analysis by \citet{lugaro8}, on the basis of the same troubles with the Sr versus 
Ba plot, concluded in this way and suggested to move to higher-than-solar 
metallicities (where the reduced neutron exposure feeds $^{88}$Sr less efficiently).   We don't really know if this is a solution, but we think that such a conclusion might be too drastic, in the light of our previous discussion of Sr cross sections (see section \ref{sec:Sr}).
Indeed, with our test choice V2 of the parameters \citep[where, for the cross section of $^{88}$Sr, we adopted the value measured by][larger by 30\% with respect to the K1 recommendations]{katabuchi} the situation changes considerably, as shown in Figure \ref{fig:bastro}. 
As is illustrated there, in this case the range of predictions from 3 \msb models for the $^{88}$Sr/$^{86}$Sr ratio shrinks sharply. The right panel of Figure \ref{fig:bastro} would represent now a quite good reproduction of experimental data, without invoking excessively high metallicities, which we consider unlikely for presolar grains} (in this case the fits of Figure \ref{fig:stroba1} would not be modified largely). We leave for the moment this possible indication as a warning on the $^{88}$Sr neutron-capture cross section.

We can also notice that the $^{135}$Ba/$^{136}$Ba
ratios (and in general those for Ba isotopes of higher atomic mass)  are sensitive to further uncertain nuclear parameters, namely the (n,$\gamma$) cross sections for the radioactive Cs isotopes $^{134}$Cs and $^{135}$Cs, for which only theoretical values exist, and their $\beta^-$ decay rates. In addition to the already discussed case of $^{134}$Cs, indeed (see section \ref{sec:rates}), also the case of $^{135}$Cs is worth a dedicated study, as its dependence on temperature is large above 2$\cdot 10^8$ K and might well be different than assumed here from \citet{ty87}. A slightly longer half-life (so far estimated to be of the order of a few hundred years in {\it TP} conditions) would modify the ratio $^{135}$Ba/$^{136}$Ba in a complex way, with the effect of further stretching the area of the measurements covered by envelope and wind models with respect to what is obtained in Figures from \ref{fig:stroba1} to \ref{fig:bastro}. Similar effects would be induced by variations in the $^{135}$Cs neutron-capture cross section.  

Another crucial test for nucleosynthesis models and their nuclear parameters emerges from the 
$\delta$-values of $^{88}$Sr/$^{86}$Sr, when plotted as a function of the ones for 
$^{138}$Ba/$^{136}$Ba. Figure \ref{fig:newfba2} shows the situation for a 2 \msb star: here we illustrate the composition of the envelope and that of magnetized winds (with 2\% to 5\% of He-shell material admixed). Even in this 2 \msb case, in our estimates a considerable fraction of the
{\it TPs} (full dots) would produce in the winds 
values of $^{88}$Sr/$^{86}$Sr that are higher than measured (see in particular  the right panel). This fact may, again, play in favor of a larger $^{88}$Sr neutron capture cross section than recommended in the K1 database, as illustrated in Figure \ref{fig:newfba3}, where the shrinking in the dispersion of the ordinate is obtained by adopting the already mentioned choice V2 for the nuclear parameters. 

Even more evident in Figures \ref{fig:newfba2} and \ref{fig:newfba3} is that there is some excess of $^{138}$Ba 
in the models. This nucleus is at the $N = 82$ magic number and is fed effectively when the neutrons available are abundant. However, this is not the case in the computations shown here, which refer to solar-like metallicities (hence to stars with high contents of Fe), with rather small neutron exposures. A solution, especially for the 2  \msb case, would be to accept
that magnetized winds be generated with even larger fractions of He-shell matter than displayed in the figures, as the model curves perform a U-turn that reaches down to lower and lower values of the $^{138}$Ba/$^{136}$Ba ratio for increasing mixing fractions. This solution, however, would be restricted to the 2 \msb models, as those for the 3 \msb cases with high mixing ratios tend to produce values excessively positive in the Sr isotopic ratio in the ordinate, even in the test case V2. This is indeed illustrated in Figure \ref{fig:heavybasr}. The
best compromises for the 3 \msb cases seem to be limited to 1-2\% of 
mixing of He-shell matter in the winds. Alternatives exist, but must invoke further changes in the nuclear parameters. As mentioned, these might affect the decay rates of $^{135}$Cs and the cross sections of radioactive Cs isotopes.
%, while the cross sections for Ba isotopes are considered, in this moment, too accurate to be questioned

On the other hand, the nuclear data on barium neutron capture cross sections are satisfactorily precise
at large enough temperatures (T$\sim$ 300 MK), but are rather loosely constrained at lower energies (around 8 keV, i.e. the energy characterizing neutron release in the $^{13}$C pocket). Also in this case, refinements in the experimental data are urgently needed.

\subsection{Sr and Ba isotopes separately}\label{sec:srba}
\begin{figure*}[t!!]
\centering{
{\includegraphics[width=0.3\textwidth]{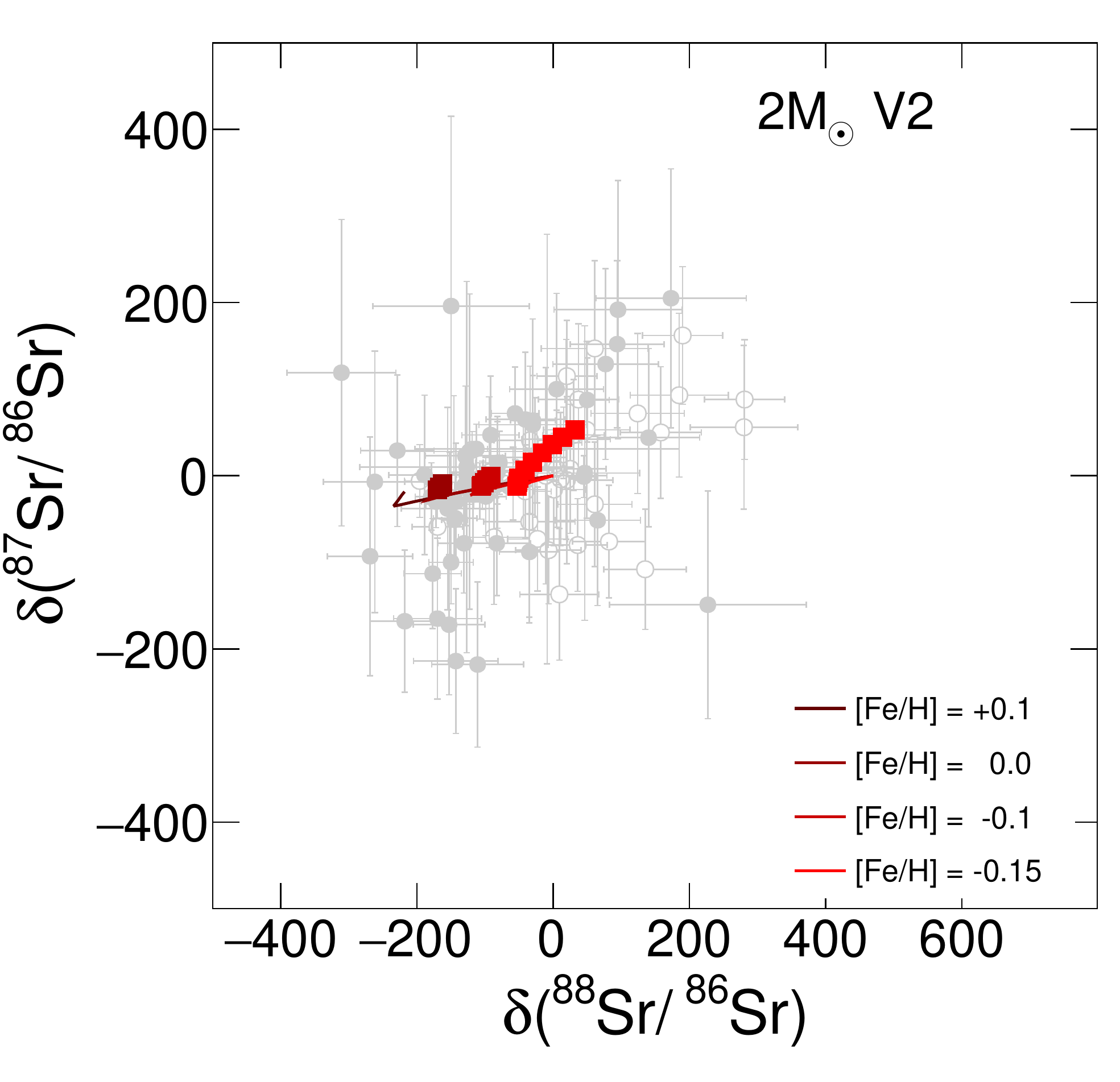}}
{\includegraphics[width=0.3\textwidth]{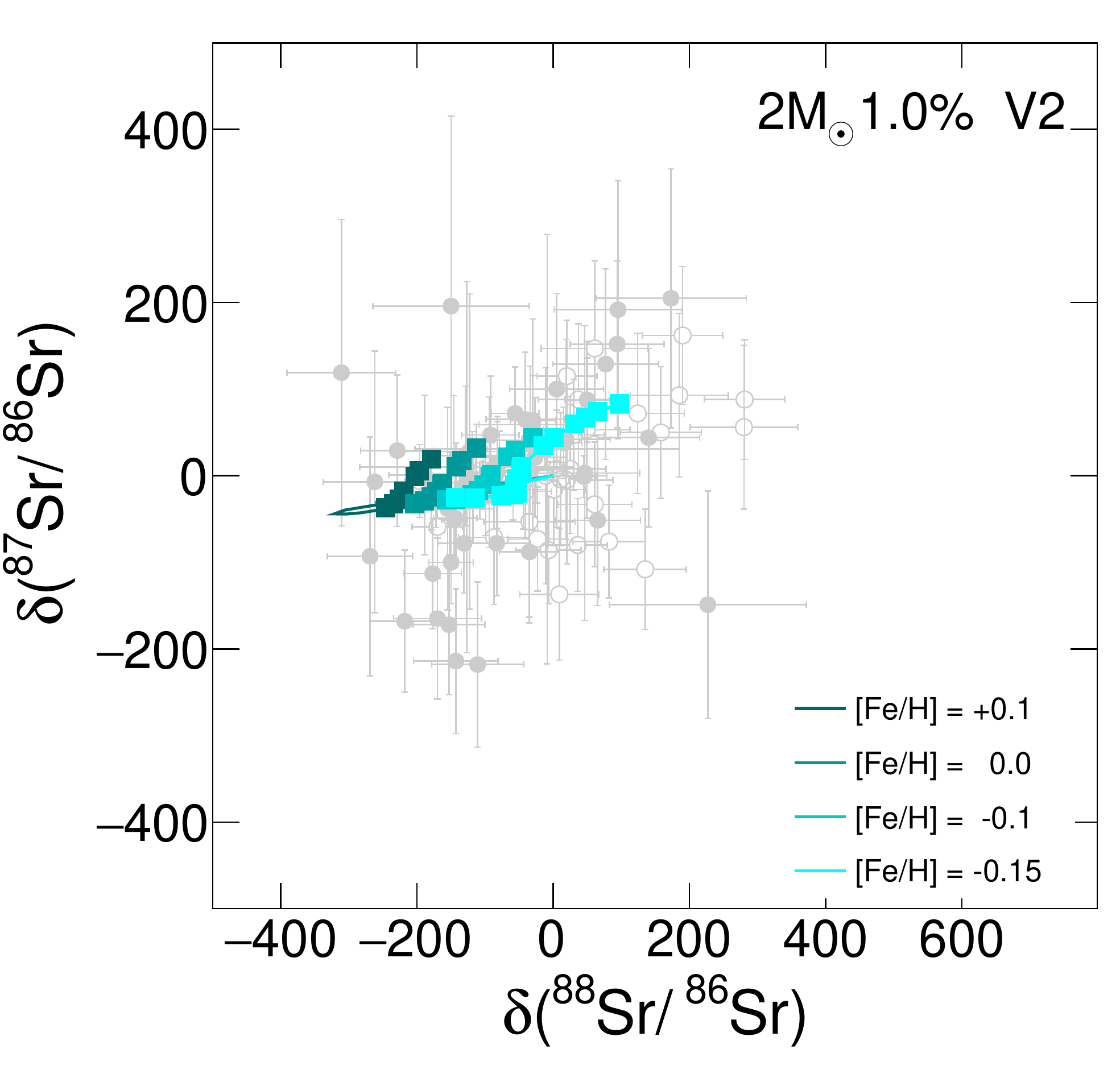}}
{\includegraphics[width=0.3\textwidth]{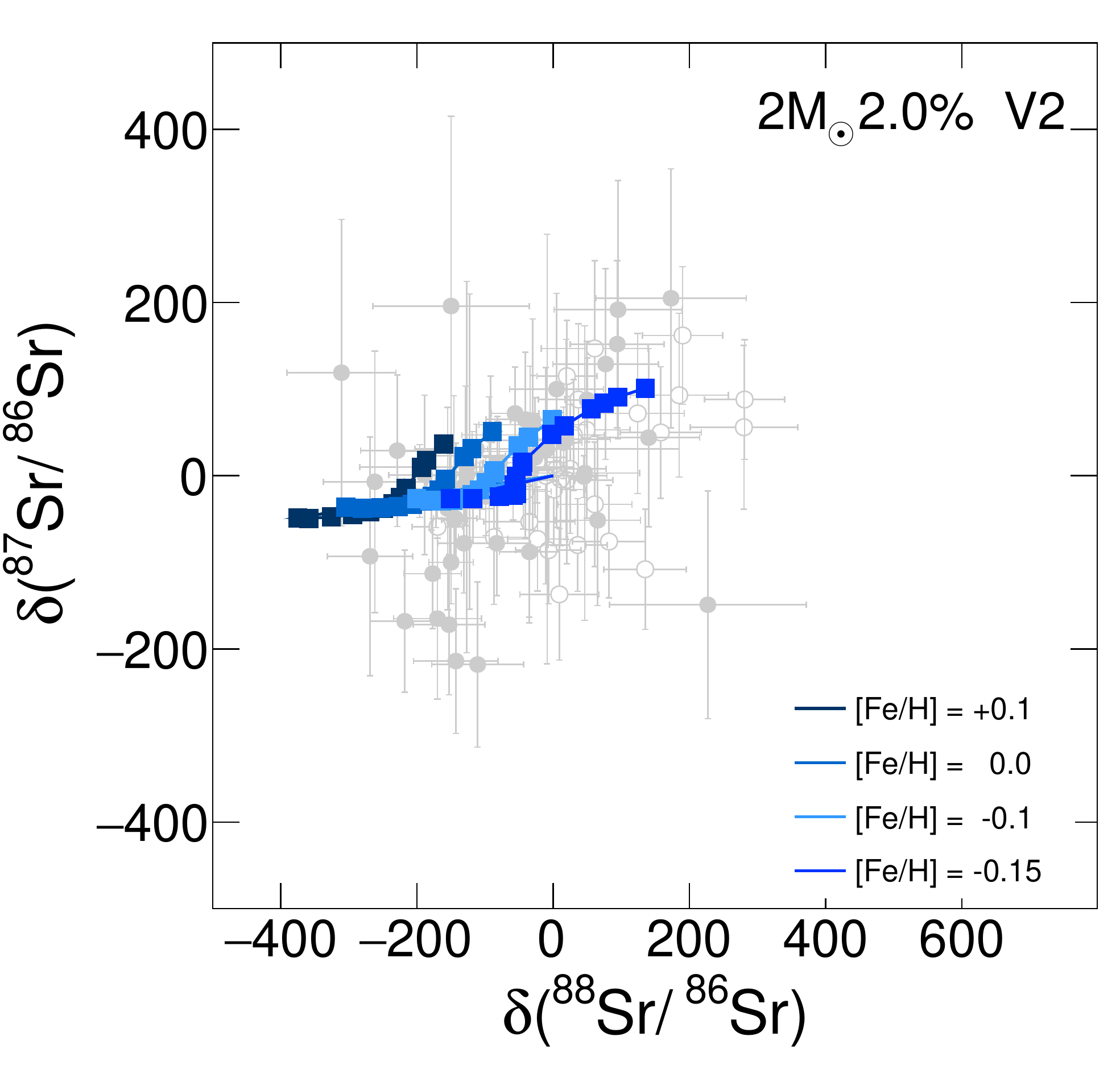}}
\caption{A comparison of model predictions from 2 \msb models of various metalliciities (full lines with heavy dots) with SiC data 
for Sr isotopes. The meaning of the symbols is the same as in previous figures, and the three panels
represent again the envelope (left) and magnetized winds with 2\% (central) and with 5\% (right) of He-shell material added.
\label{fig:lastsr2}}}
\end{figure*}

\begin{figure*}[t!!]
\centering{
{\includegraphics[width=0.3\textwidth]{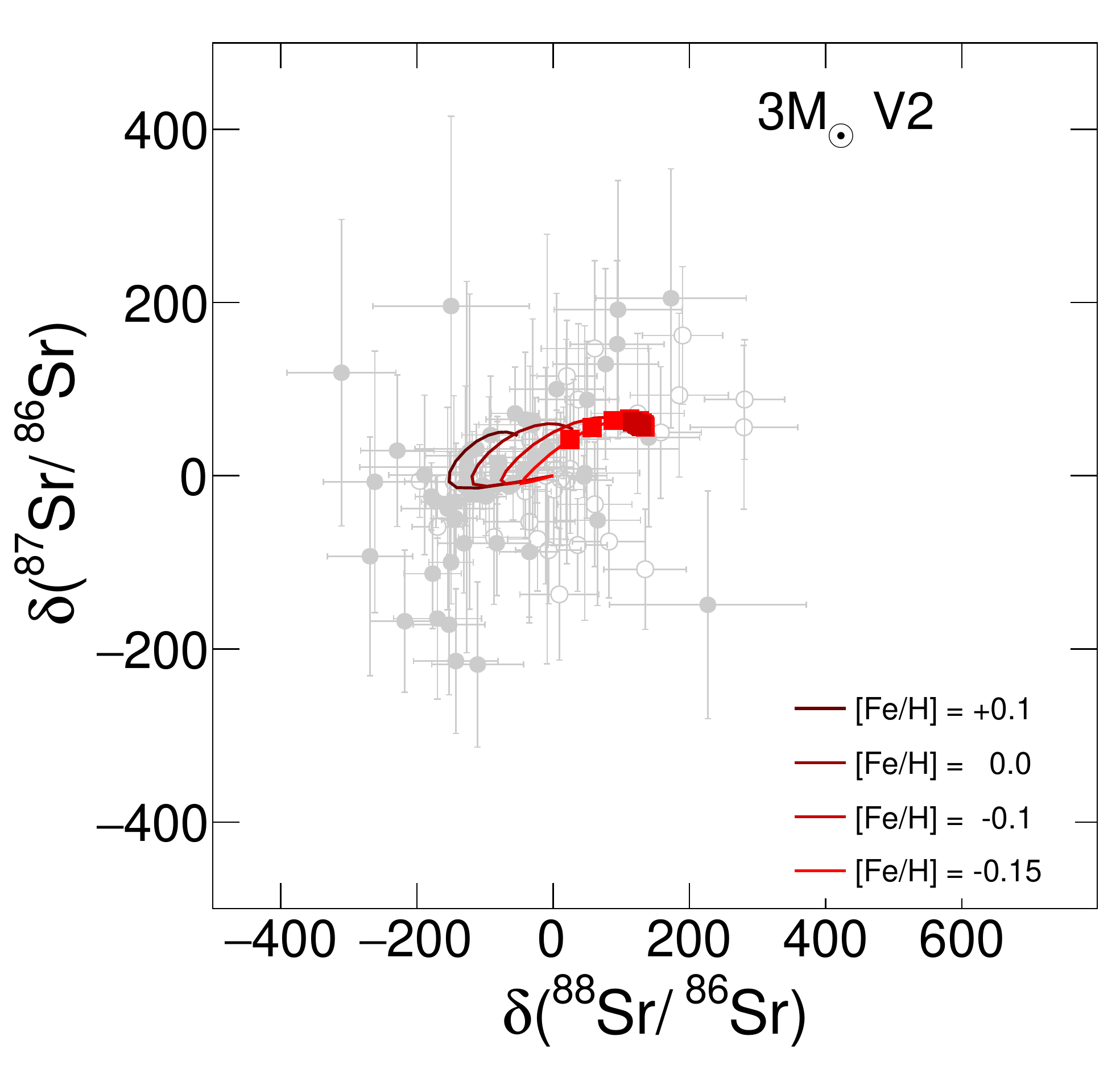}}
{\includegraphics[width=0.3\textwidth]{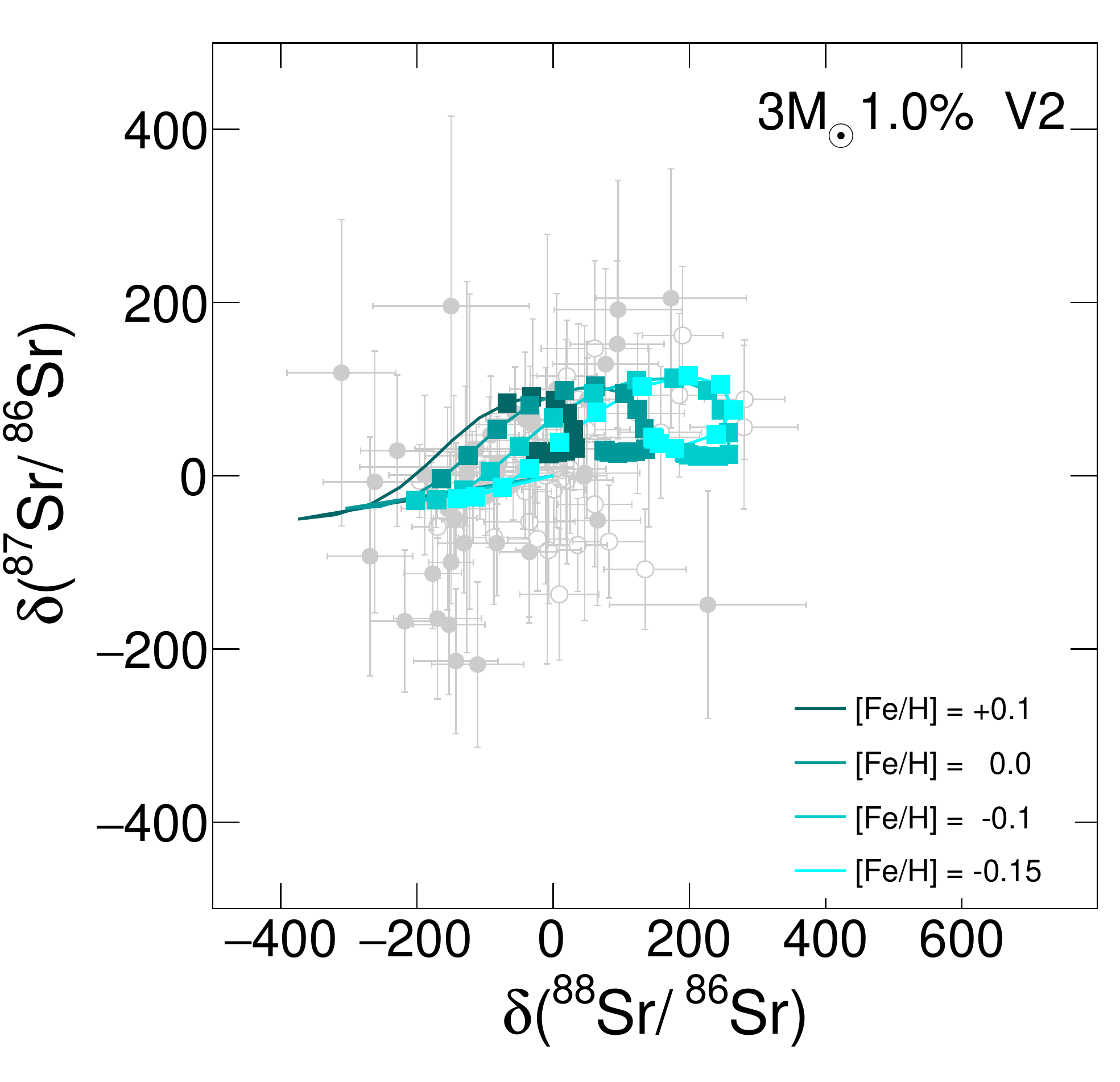}}
{\includegraphics[width=0.3\textwidth]{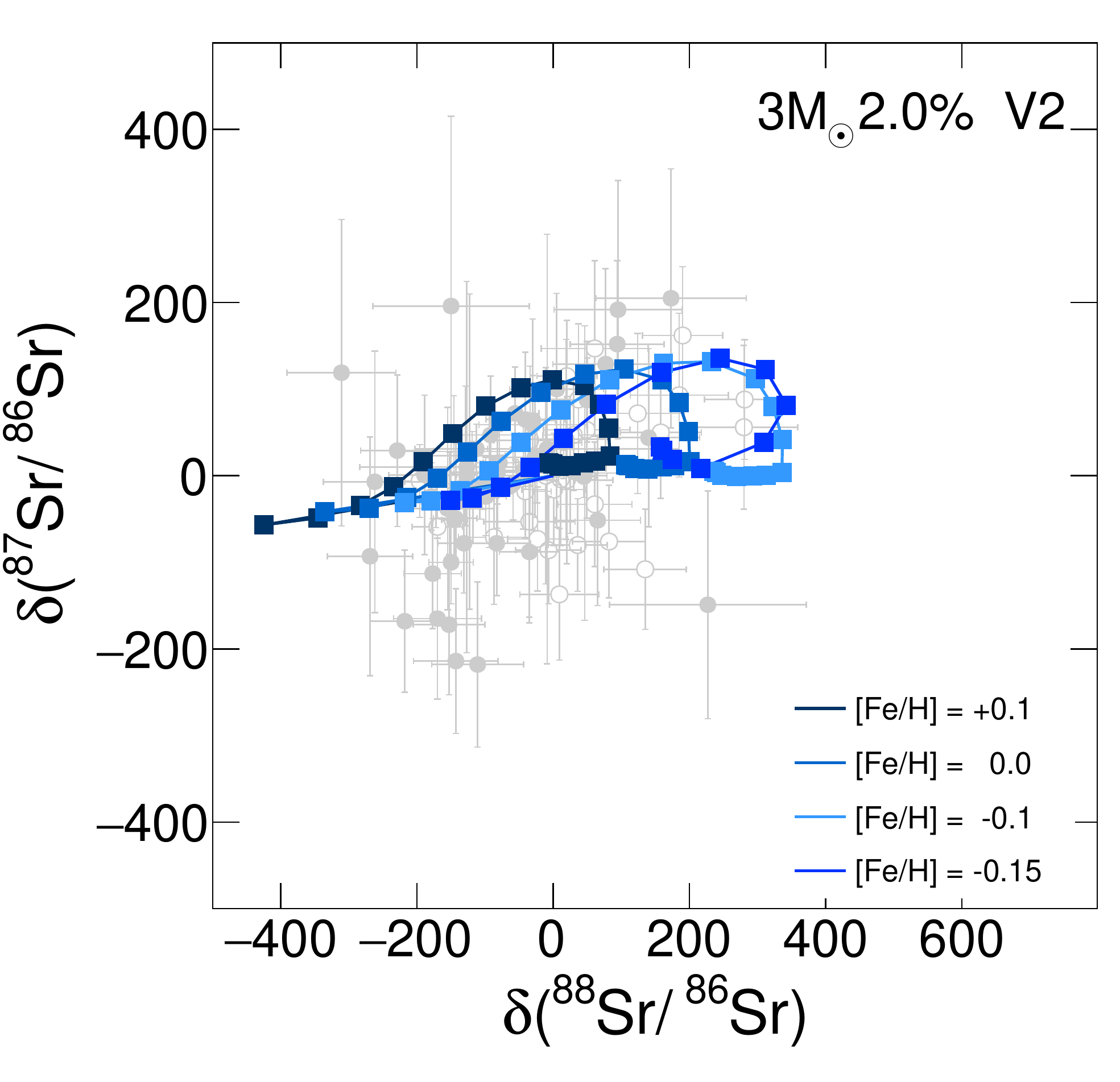}}
\caption{Same plots of Figure \ref{fig:lastsr2}, but for 3 \msb models. \label{fig:lastsr}}}
\end{figure*}

A comparison of model predictions with the measured $\delta$-values for Sr isotopes is shown 
in Figures \ref{fig:lastsr2} and \ref{fig:lastsr}, adopting for illustration a 2 and a 3 \msb model, { respectively}. We refer directly to the cases V2, discussed in the previous section, because the excess production of $^{88}$Sr there identified can be avoided.  As is made clear in the 
Figure \ref{fig:lastsr}, in a 3 \ms star the envelope alone (left panel) does not offer a good reproduction of the measurements, while referring to the magnetized winds allows for a much better agreement, which improves when
the percentage of He-shell material in the outflows increases (central and right panels). 

The cases for 2 \msb stellar models (Figure \ref{fig:lastsr2}) look similar, but with a lower
proportion of the area containing experimental data covered by models.
It can be noticed that the large spread in the measured data for $^{87}$Sr is not well matched by model curves. Actually, such a 
spread can best be interpreted as a consequence of a range of temperature values (i.e. of stellar masses), in which the complex path leading to $^{87}$Sr varies. In this respect, the unique value for the branching ratio at $^{84}$Kr to the isomer of $^{85}$Kr, which we are forced to assume
from the K1 repository, in the absence of different data, is certainly an oversimplified assumption (see section \ref{sec:krbranch}). Here we clearly need new nuclear physics measurements on the chains departing from Kr isotopes, in this case the one proceeding through $^{85}$Kr$^m$, $^{85}$Rb, $^{86}$Rb and $^{86,87}$Sr. In general, therefore, the 
evidence provided so far by SiC grains points to the needs of
revisions in the nuclear parameters discussed in sections \ref{sec:krbranch} and \ref{sec:Sr}.

\begin{figure*}[t!!]
\centering{
{\includegraphics[width=0.3\textwidth]{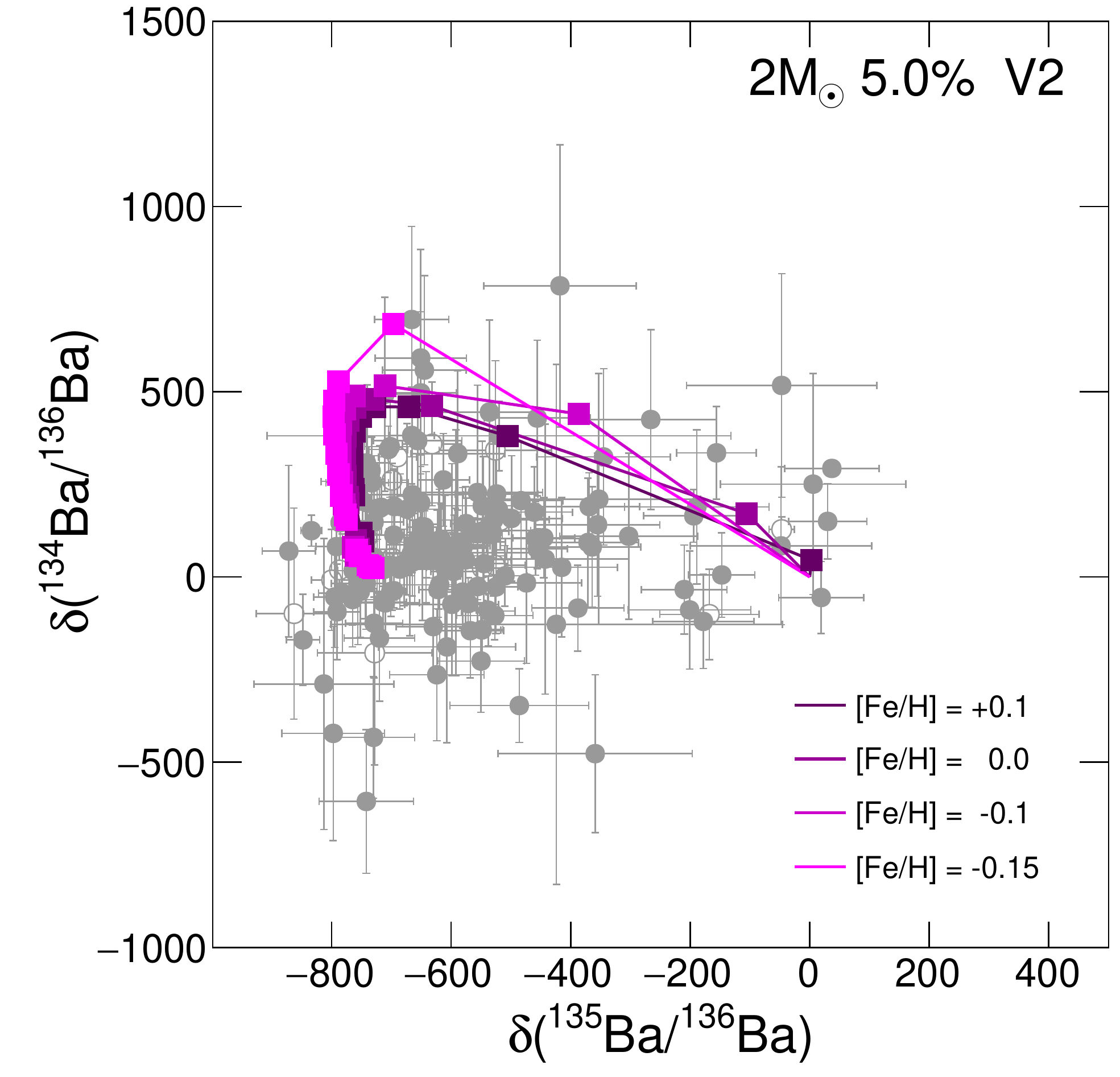}}
{\includegraphics[width=0.3\textwidth]{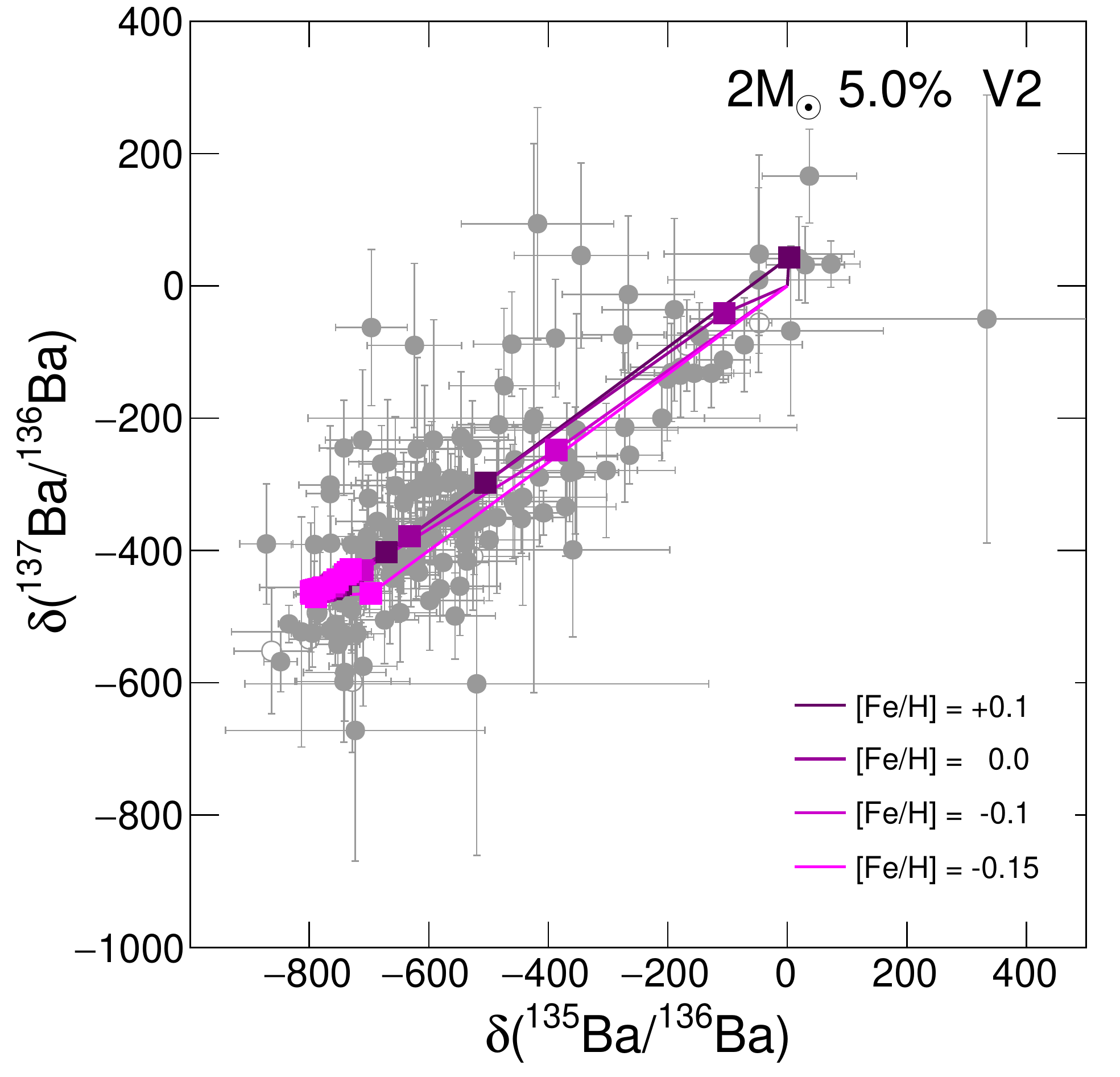}}
{\includegraphics[width=0.3\textwidth]{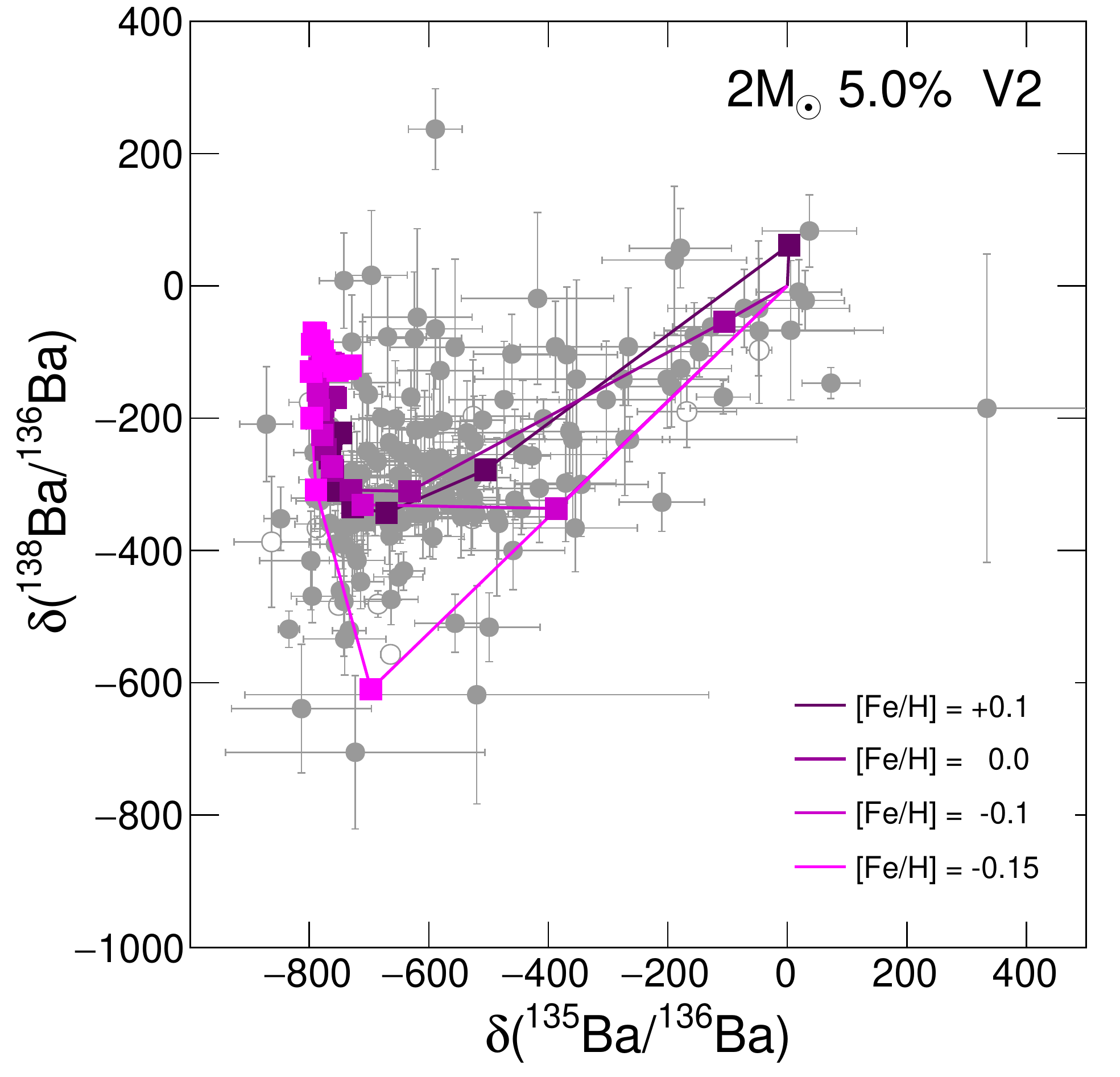}}
\caption{A comparison of model predictions from a representative 
case of 2 \msb models (full lines with heavy dots) with SiC data 
for various Ba isotopes and with the the choice V2 for nuclear parameters, including revisions for the $^{134}$Cs decay, as illustrated in section \ref{sec:rates}. The meaning of the symbols is the same as in previous figures. The three panels represent the composition of winds with 5\% of He-shell material added.
\label{fig:variousba2}}}
\end{figure*}

\begin{figure*}[t!!]
\centering{
{\includegraphics[width=0.3\textwidth]{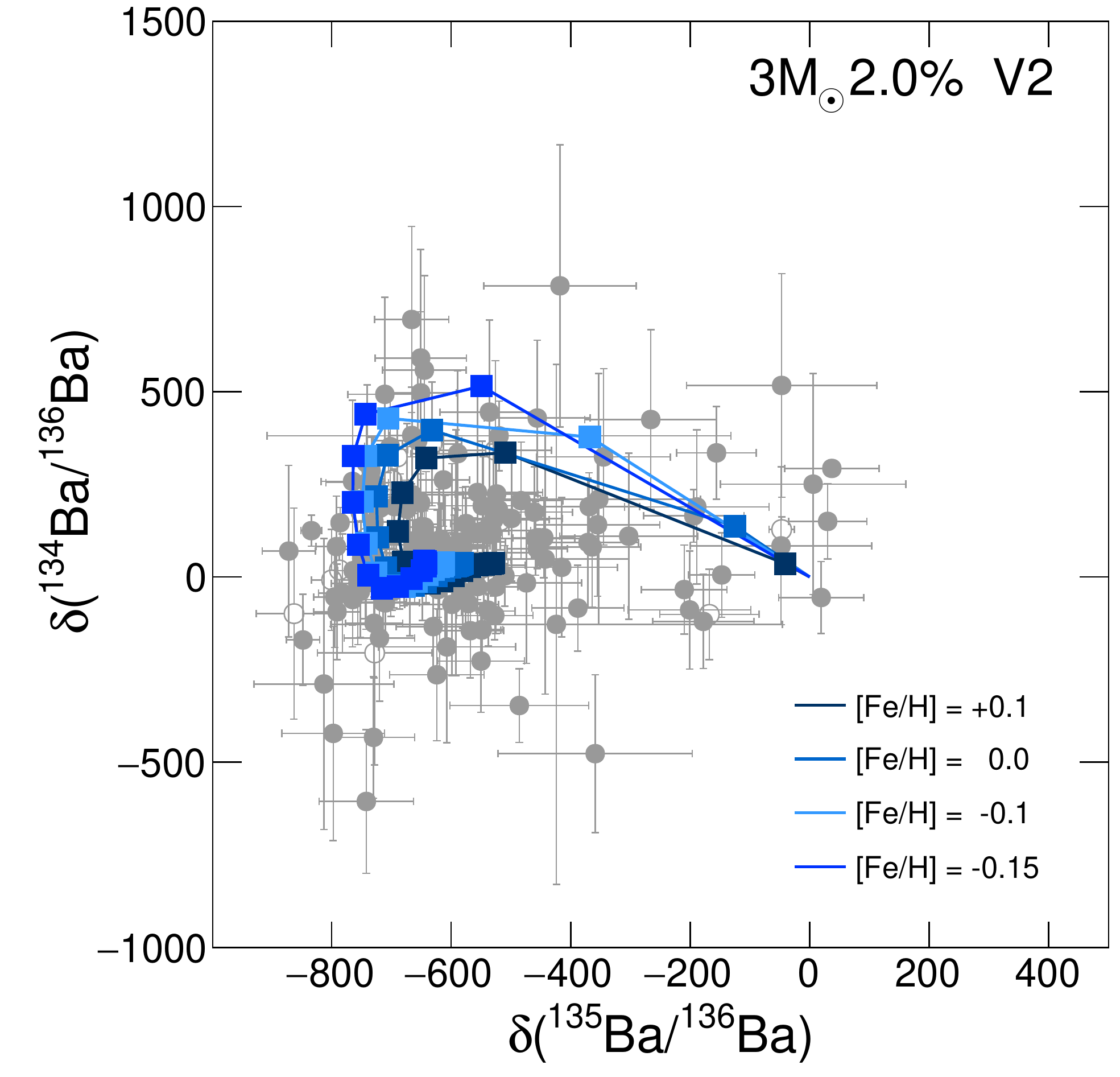}}
{\includegraphics[width=0.3\textwidth]{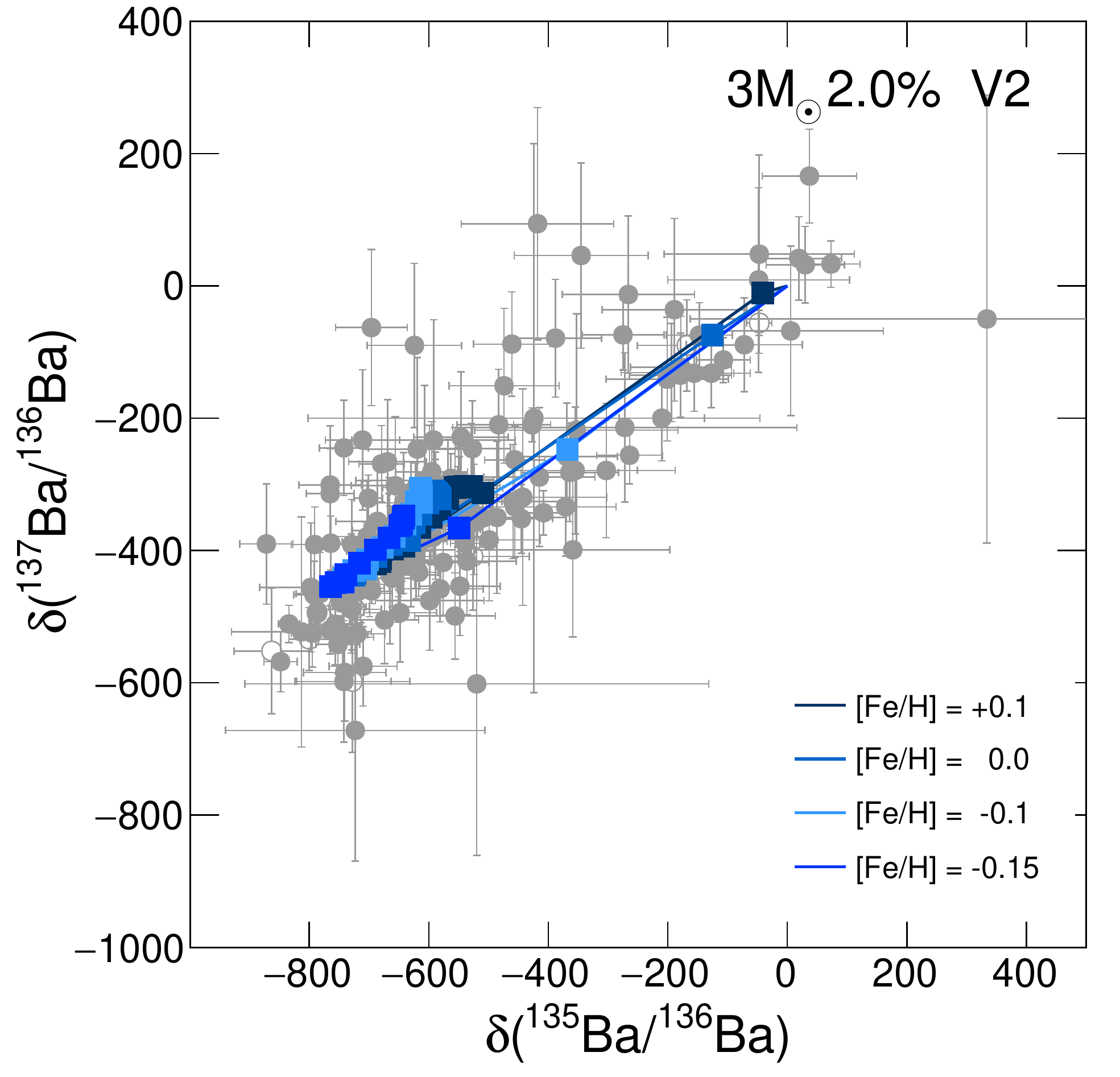}}
{\includegraphics[width=0.3\textwidth]{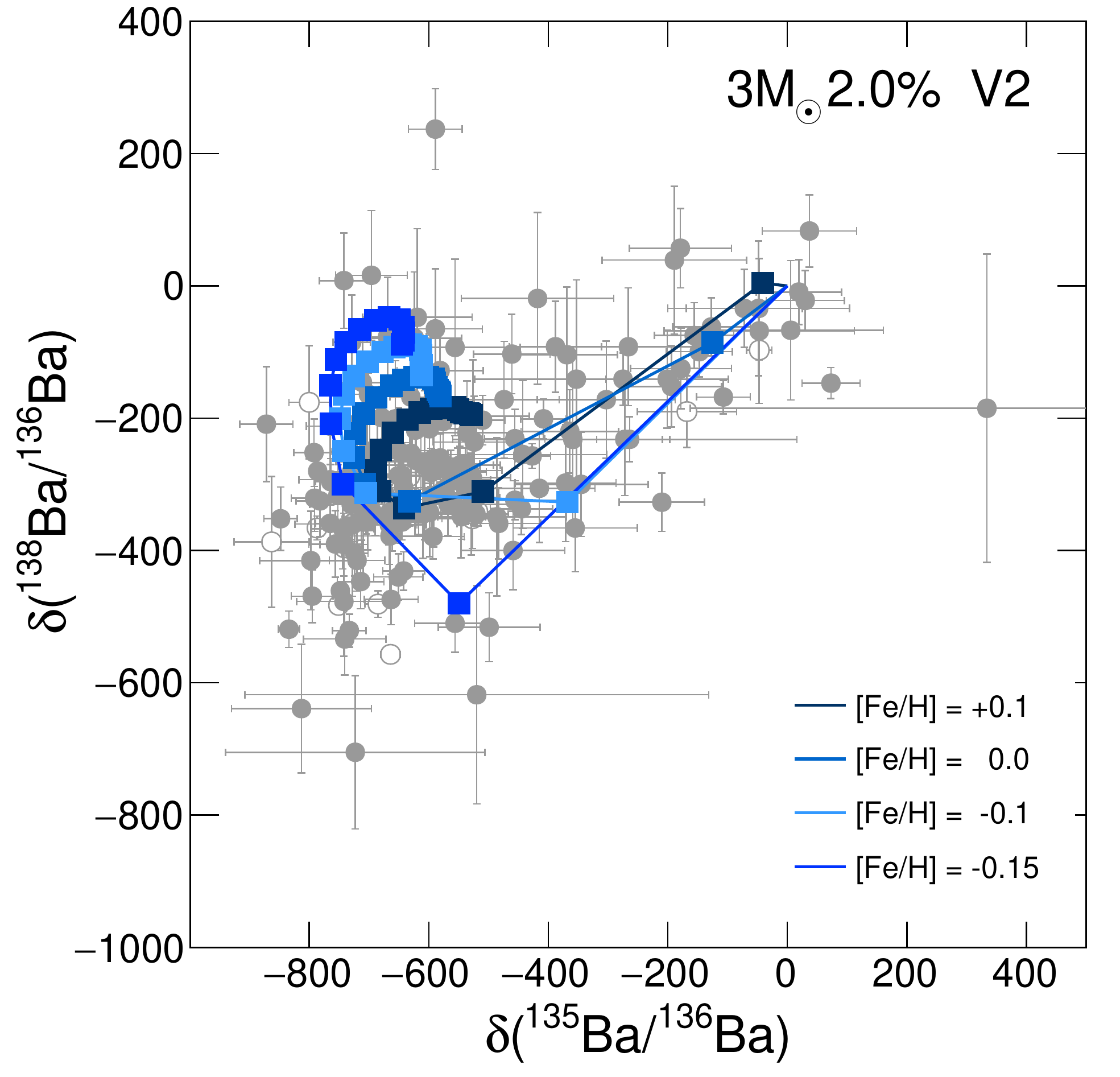}}
\caption{A plot similar to Figure \ref{fig:variousba2}, but for 3 \msb models and a dilution of He-shell material in the wind of 2\%. \label{fig:variousba}}}
\end{figure*}

Concerning the various Ba isotopes, Figures \ref{fig:variousba2} and \ref{fig:variousba} illustrate a synthesis of what can be obtained for them in  representative cases among
those discussed so far. In particular, we plot there situations pertaining to magnetized winds, 
with 5\% admixture of material from the He-shell for 2 \msb models and 2\% for 3 \msb ones. This latter case seems to
represent a rough average condition permitting to reproduce sufficiently well the three-isotope plots of barium, including the constraints 
from $^{134}$Ba, $^{137}$Ba and $^{138}$Ba. Despite the good accord, these plots confirm that small revisions in the decay rates or in the cross sections for Cs isotopes, leading to wider spreads in 
$^{137}$Ba and $^{138}$Ba, would be welcome.

\subsection{Isotopic ratios for Zr and Mo}\label{sec:zrmo}
\begin{figure*}[t!!]
\centering{
{\includegraphics[width=0.3\textwidth]{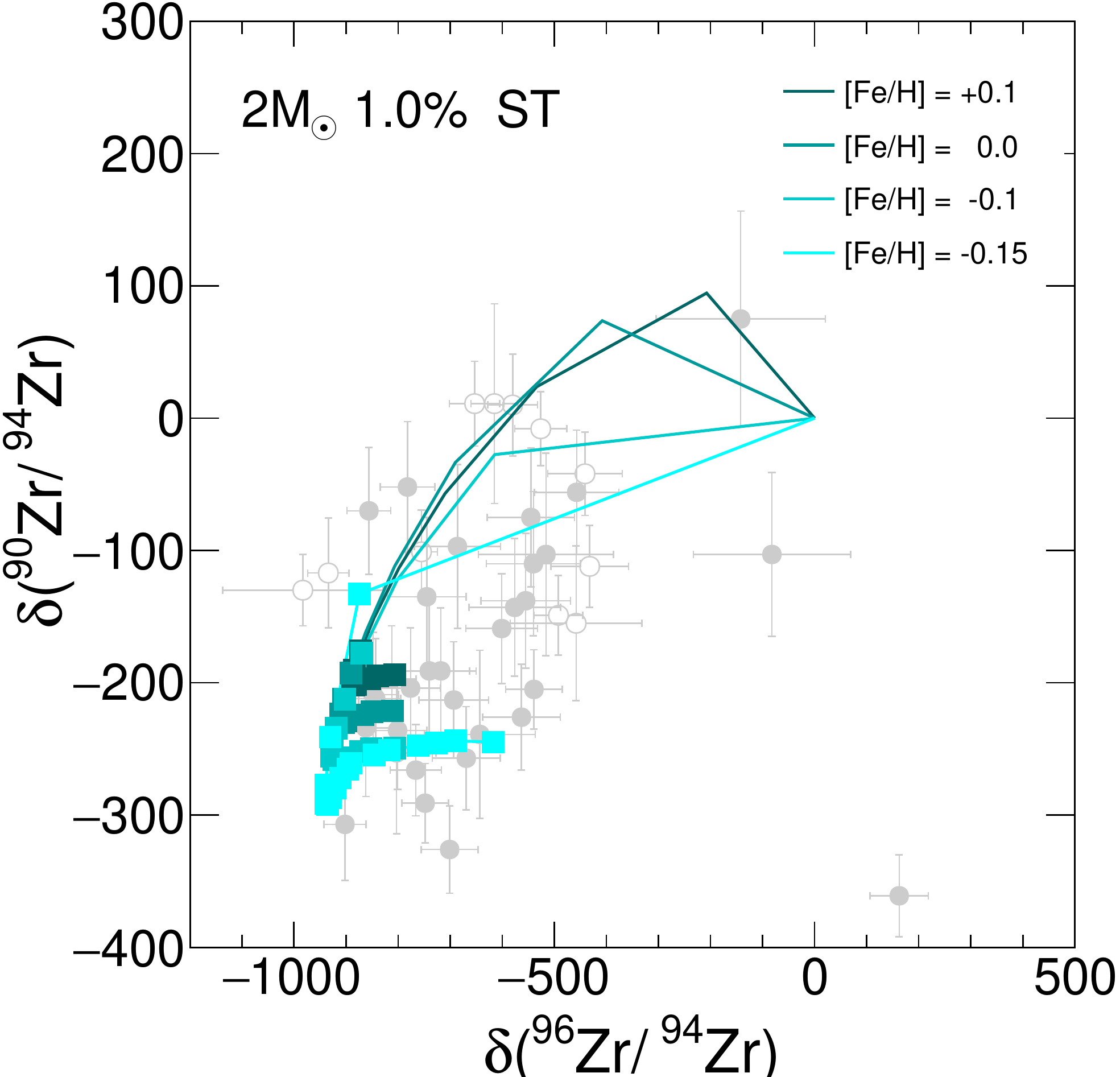}}
{\includegraphics[width=0.3\textwidth]{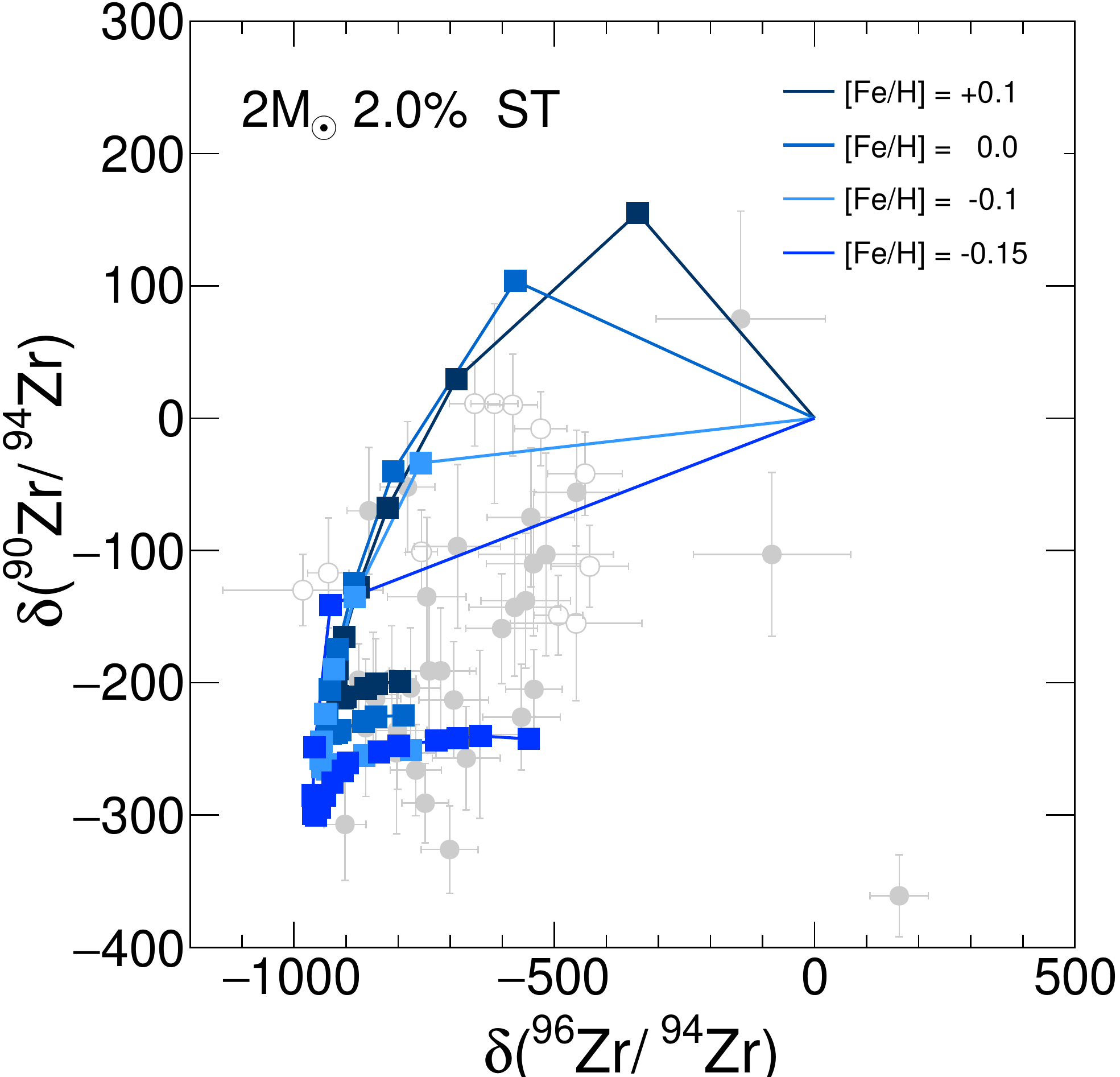}}
{\includegraphics[width=0.3\textwidth]{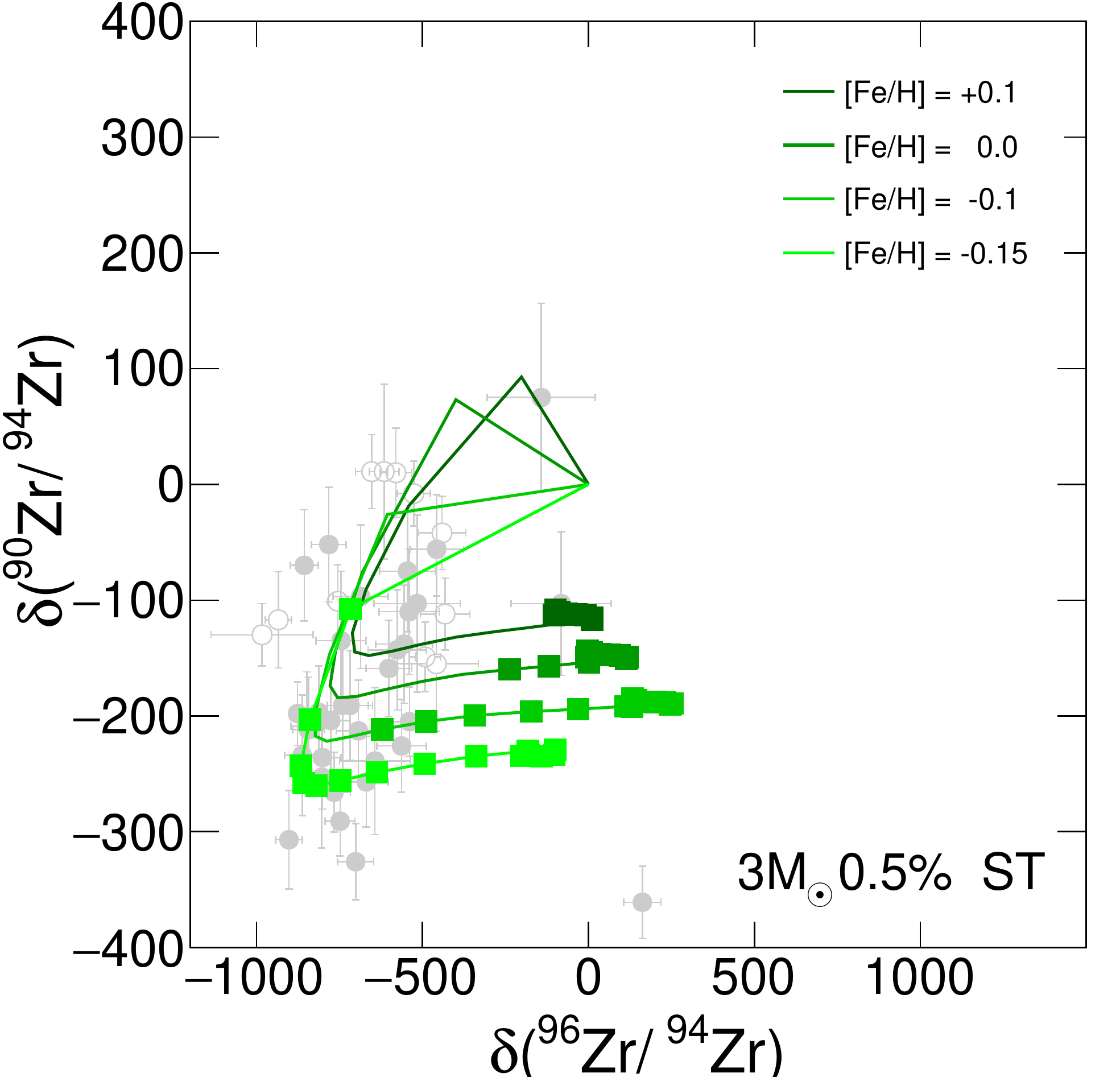}}
\caption{The comparison of measured SiC data involving $^{90}$Zr with 
model sequences for the winds of 2 \msb stars at 1\% and 2\% dilution of He-shell matter in the winds (left and center panels ) and for those of 3 \msb stars with a low percentage (0.5\%) of He-shell matter added (right panel).\label{fig:90zr}}
}
\end{figure*}

\begin{figure*}[t!!]
\centering{
{\includegraphics[width=0.3\textwidth]{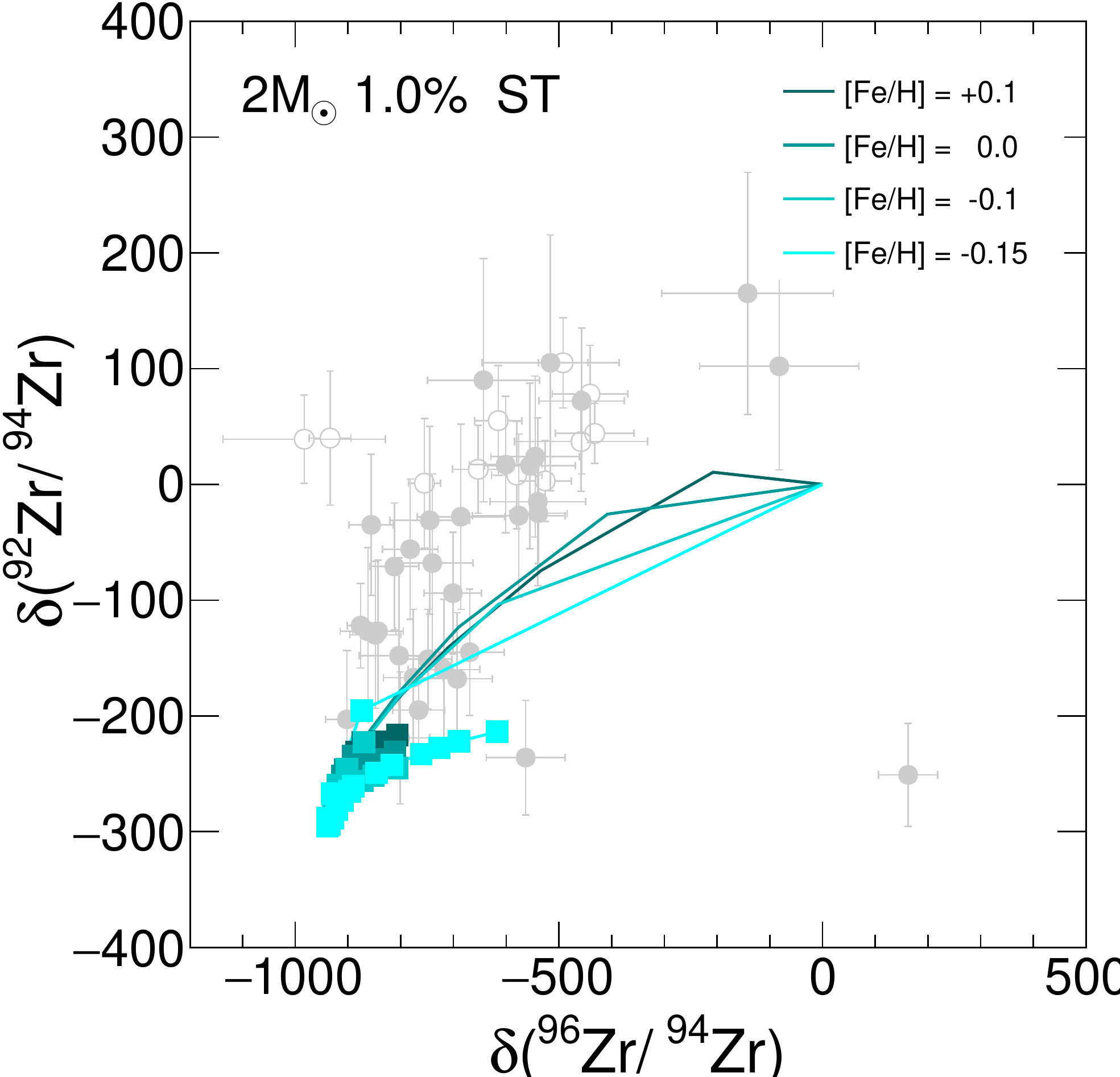}}
{\includegraphics[width=0.3\textwidth]{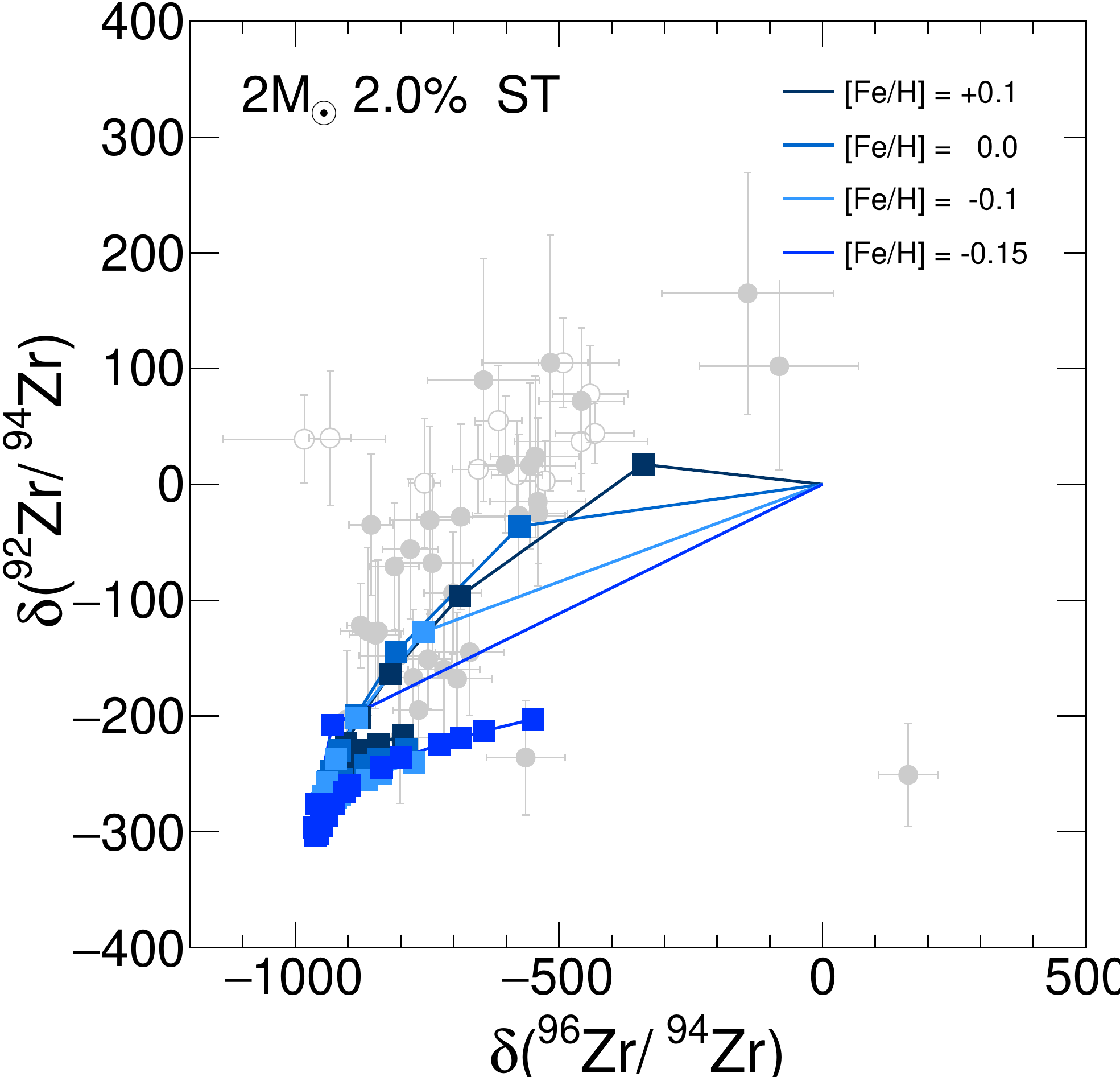}}
{\includegraphics[width=0.3\textwidth]{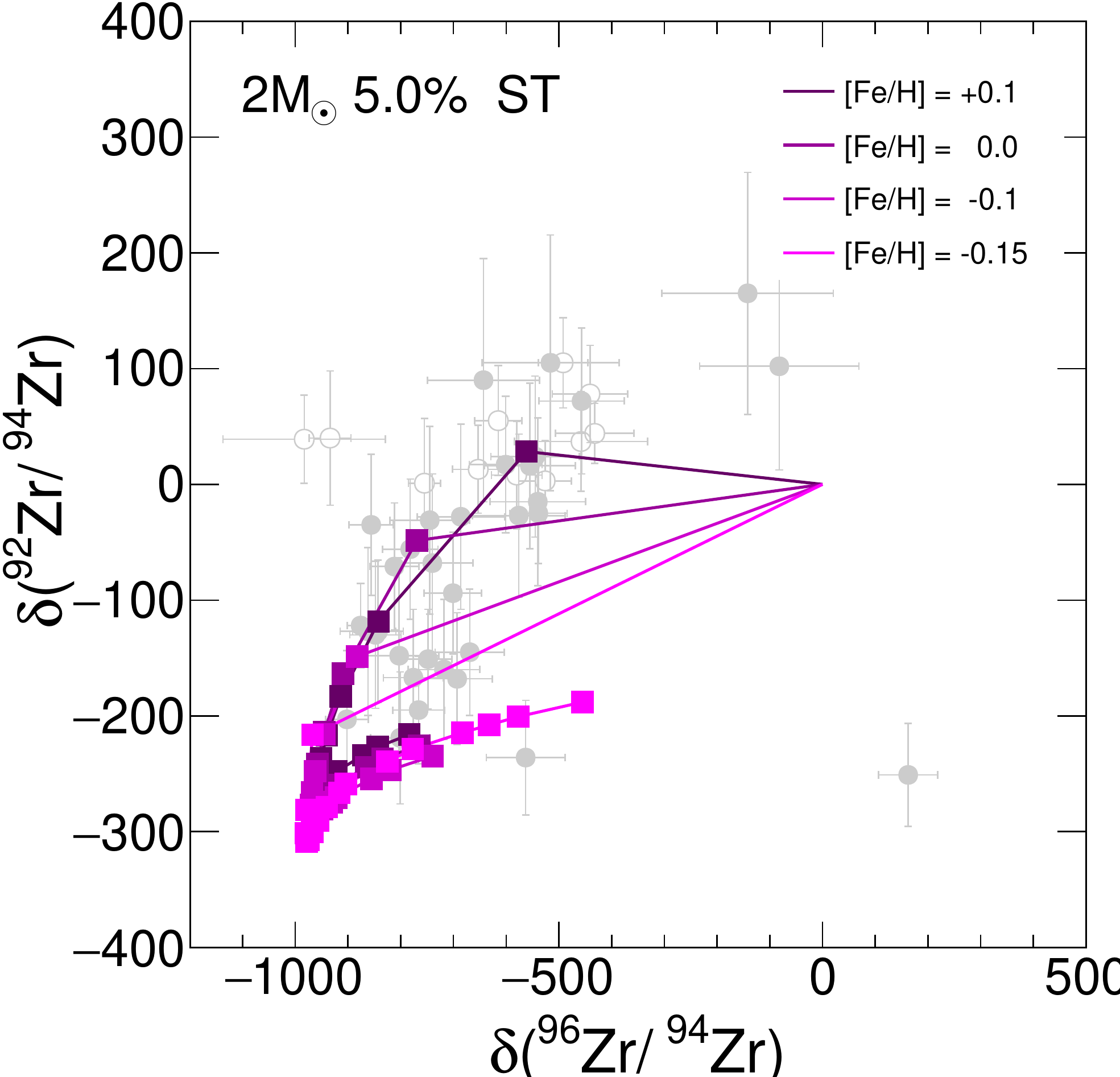}}
\caption{The comparison of measured SiC data involving $^{92}$Zr with 
model sequences for the winds of 2 \msb models, for progressively higher (from left to right) amounts of He-shell matter added.\label{fig:92zr}}
}
\end{figure*}

The isotopic ratios involving Zr, as measured in SiC grains, appear to pose much smaller problems to AGB models and the nuclear parameters seem to be in this case sufficiently good. This is so also for isotopes that were so far quoted by other groups as being problematic, like $^{92}$Zr \citep{lugaro4} and which are instead compatible with our magnetized winds. Also for the Zr ratios, therefore, including this wind component largely increases the mutual compatibility of data and models,
improving considerably over previously published analyses. We show here, in Figure \ref{fig:90zr} and in Figure \ref{fig:92zr} a couple of synthetic, representative cases, involving $^{90}$Zr/$^{94}$Zr, $^{92}$Zr/$^{94}$Zr, and $^{96}$Zr/$^{94}$Zr. 

Figure \ref{fig:90zr} presents the data for the $^{90}$Zr/$^{94}$Zr ratio as a function of $^{96}$Zr/$^{94}$Zr, compared to the outputs from a choice of our models for the winds. Panels at the left and at the center refer to 2 \msb models with 1\% and 2\% admixtures of He-shell materials, respectively, while the right panel represents the case of 3 \msb models at 0.5\% dilution. The three plots together show how the measurements
can be well accounted for by the ensemble of our models, with 2 \msb cases explaining better the vertical spread and 3 \msb ones complementing them for the horizontal distribution. No special change in the nuclear parameters seems necessary.  
 
Figure \ref{fig:92zr} then illustrates the situation with $^{92}$Zr. Previous attempts in the literature \citep{lugaro4} concluded that  
$\delta$-values of $^{92}$Zr in excess of $-50$ could not be explained by AGB models. Here, instead, we see that 2 \msb
models are sufficient to account for the data, provided we considered various He-shell admixtures. Grains with  higher $\delta$ values for $^{92}$Zr/$^{94}$Zr are explained by progressively higher percentages of mixing. { In the plots} provided we consider the composition of the winds for increasing He-shell admixtures, where grains with progressively higher $\delta$ values for $^{92}$Zr/$^{94}$Zr are found. 
{In this respect we recall the present complex picture of AGB winds illustrated by \citet{ho18}, as enhanced mass loss rates and dust production in phases preceding the final superwinds are probably needed to support our model. Flaring activity might actually lead to such enhanced mass loss rates and dust formation \citep{sk03}.}

\begin{figure*}[t!!]
\centering{
{\includegraphics[width=0.3\textwidth]{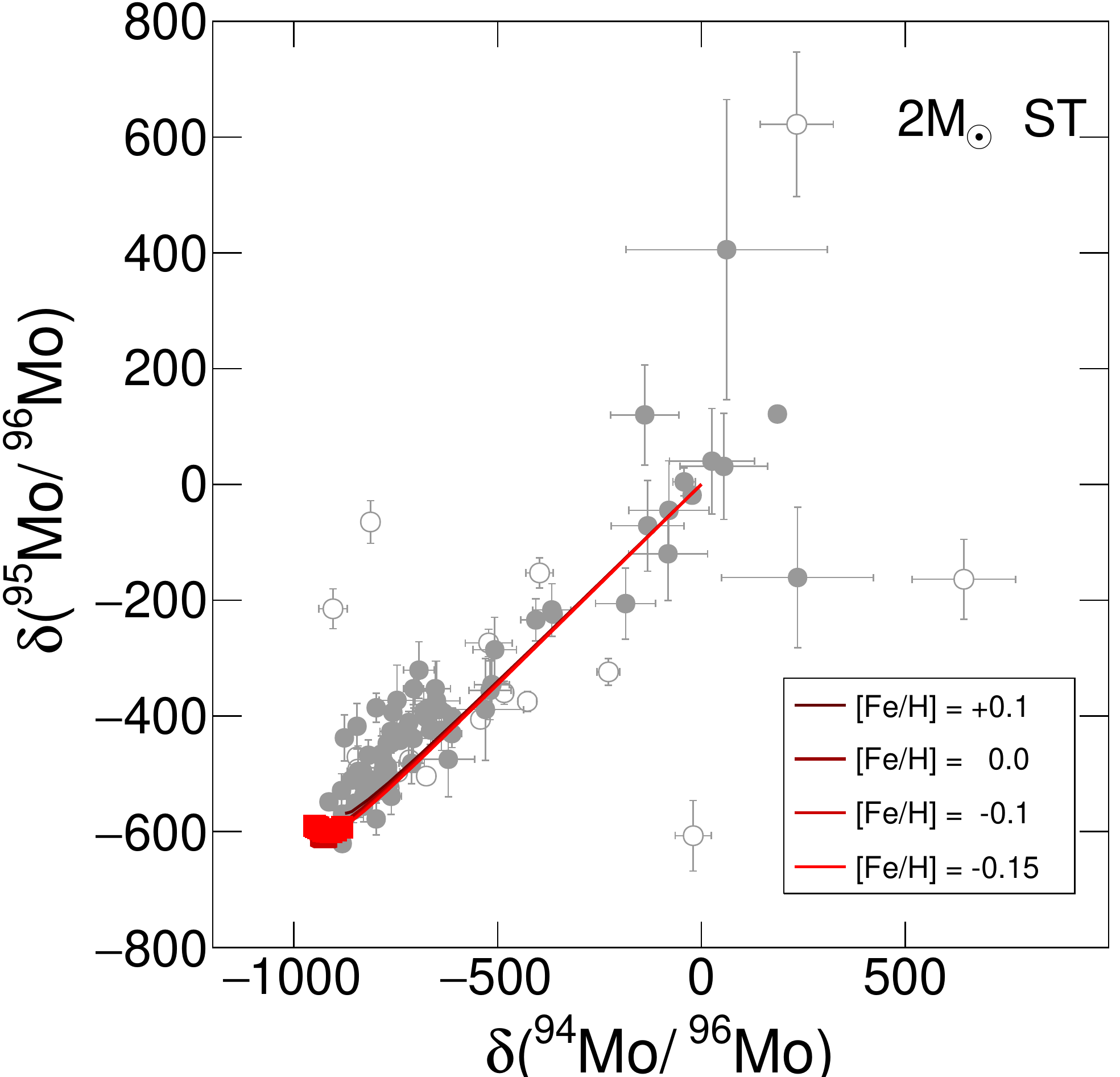}}
{\includegraphics[width=0.3\textwidth]{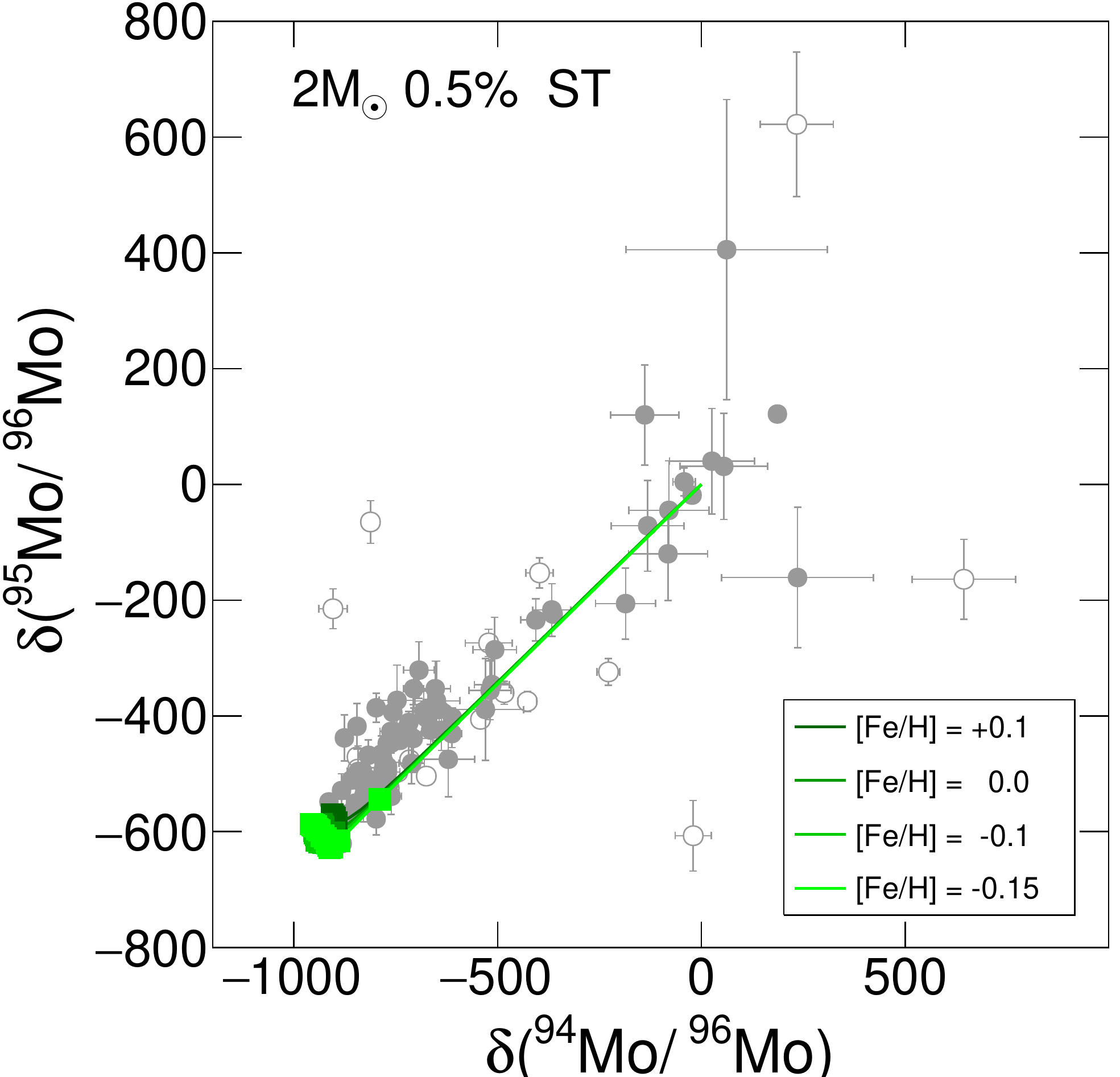}}
{\includegraphics[width=0.3\textwidth]{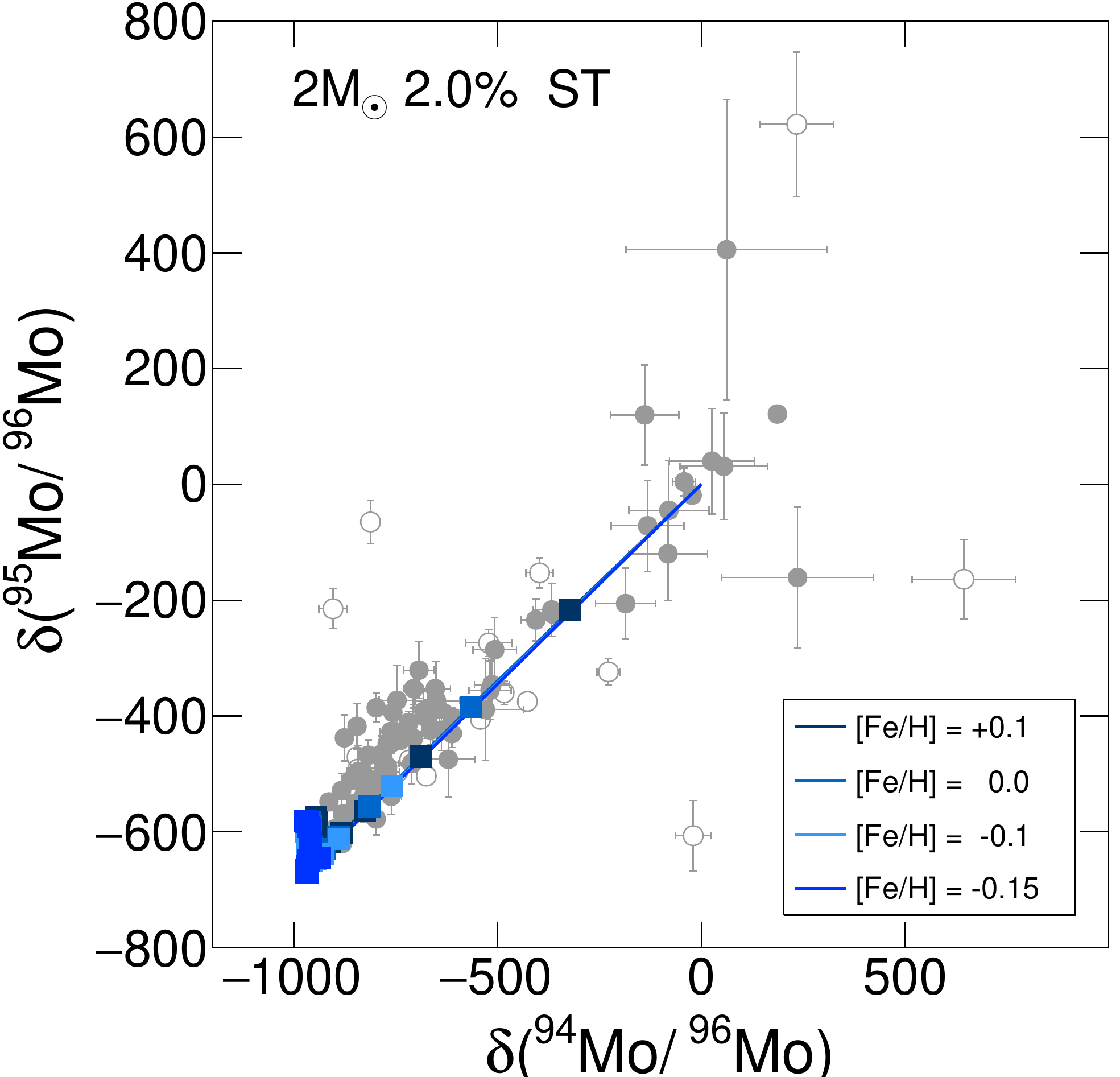}}
\caption{The measured SiC $\delta$ values for $^{95}$Mo/$^{96}$Mo 
and $^{94}$Mo/$^{96}$Mo as compared to model sequences for the envelope (left) and for two mixing cases (center and right) in the magnetized winds of 2 \msb stars.\label{fig:Mo01}}
}
\end{figure*}

\begin{figure*}[t!!]
\centering{
{\includegraphics[width=0.3\textwidth]{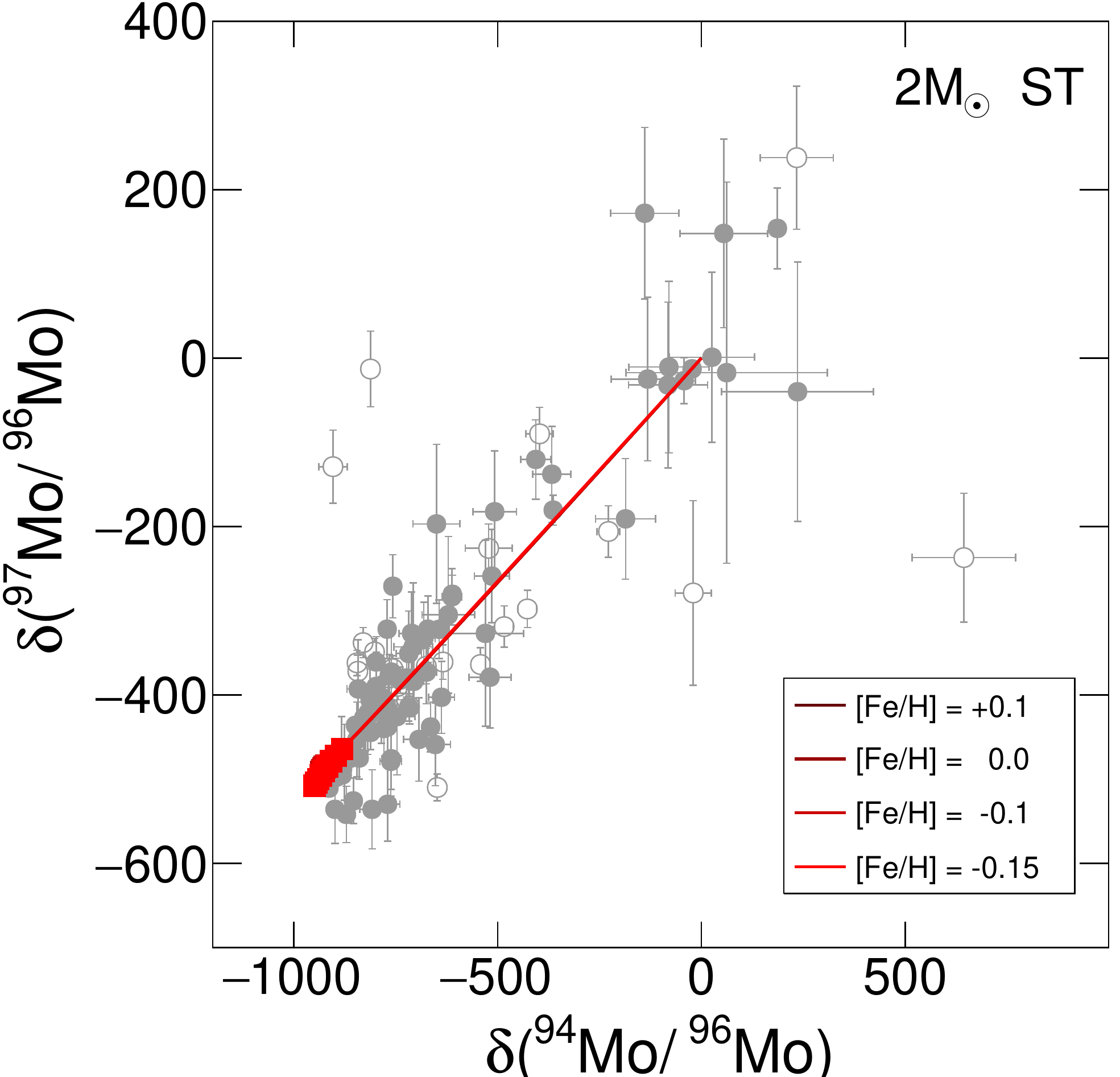}}
{\includegraphics[width=0.3\textwidth]{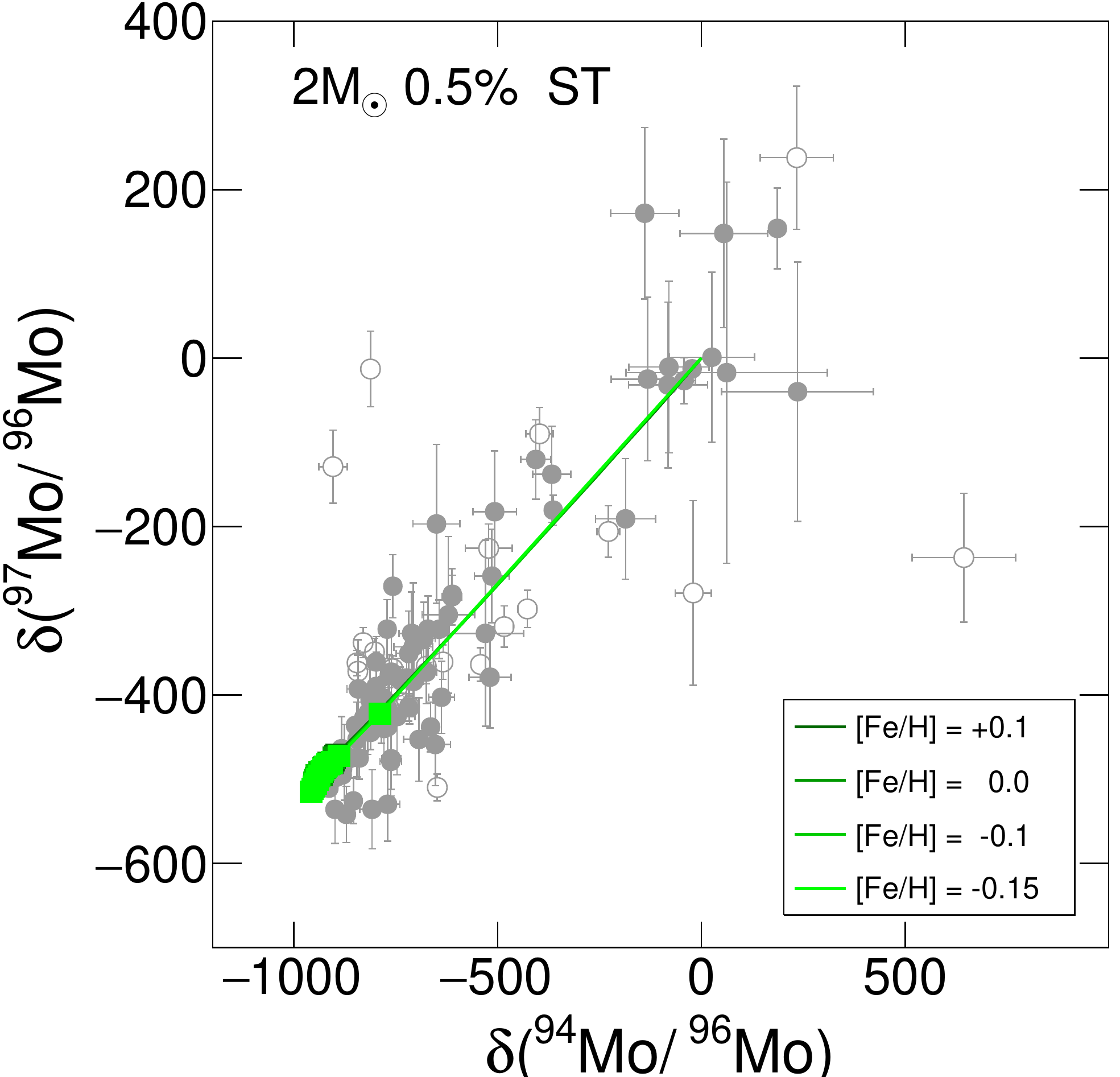}}
{\includegraphics[width=0.3\textwidth]{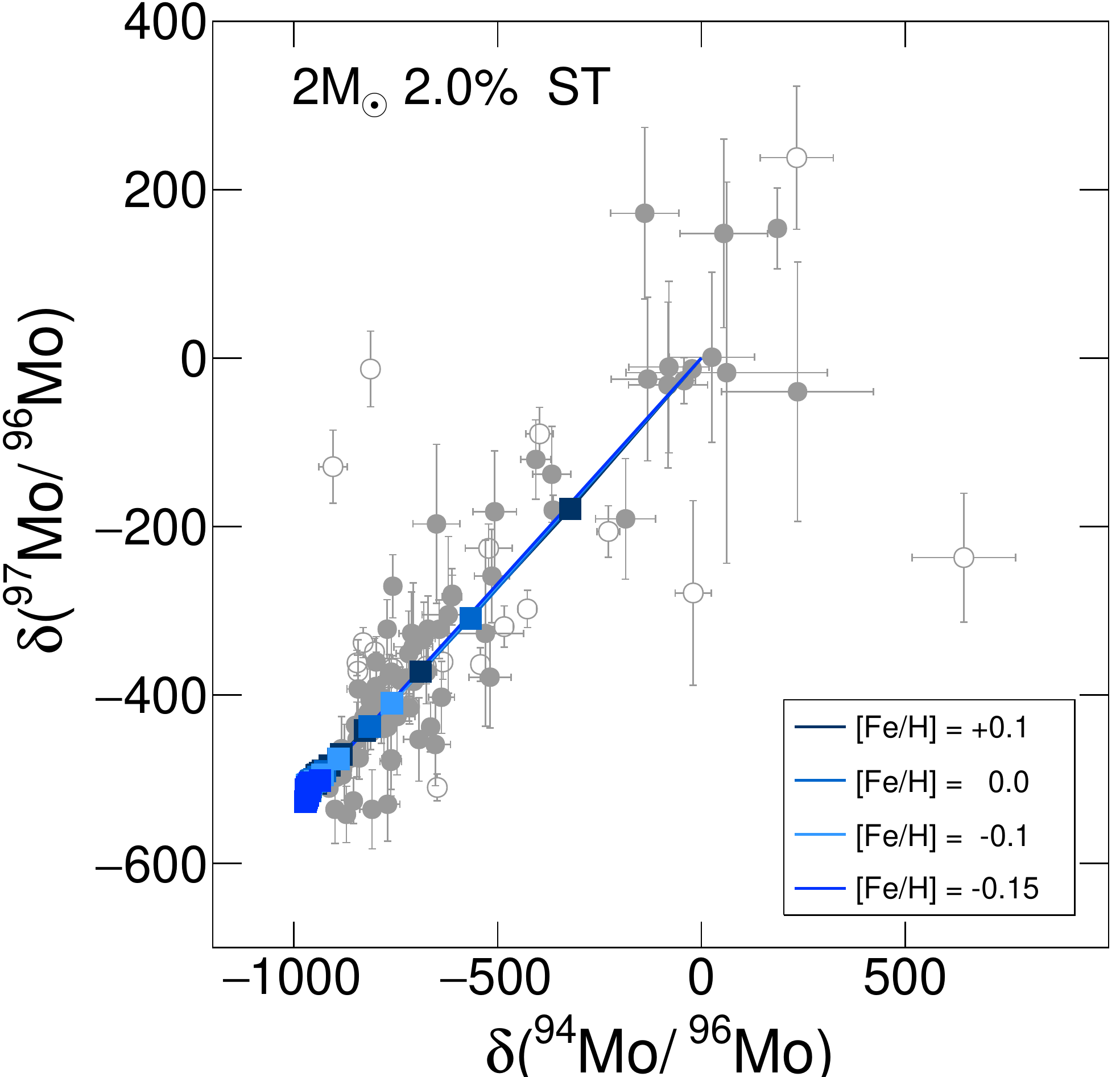}}
\caption{Same as Figure \ref{fig:Mo01}, but with the ratio $^{97}$Mo/$^{96}$Mo in the ordinate.\label{fig:Mo02}}
}
\end{figure*}

\begin{figure*}[t!!]
\centering{
{\includegraphics[width=0.3\textwidth]{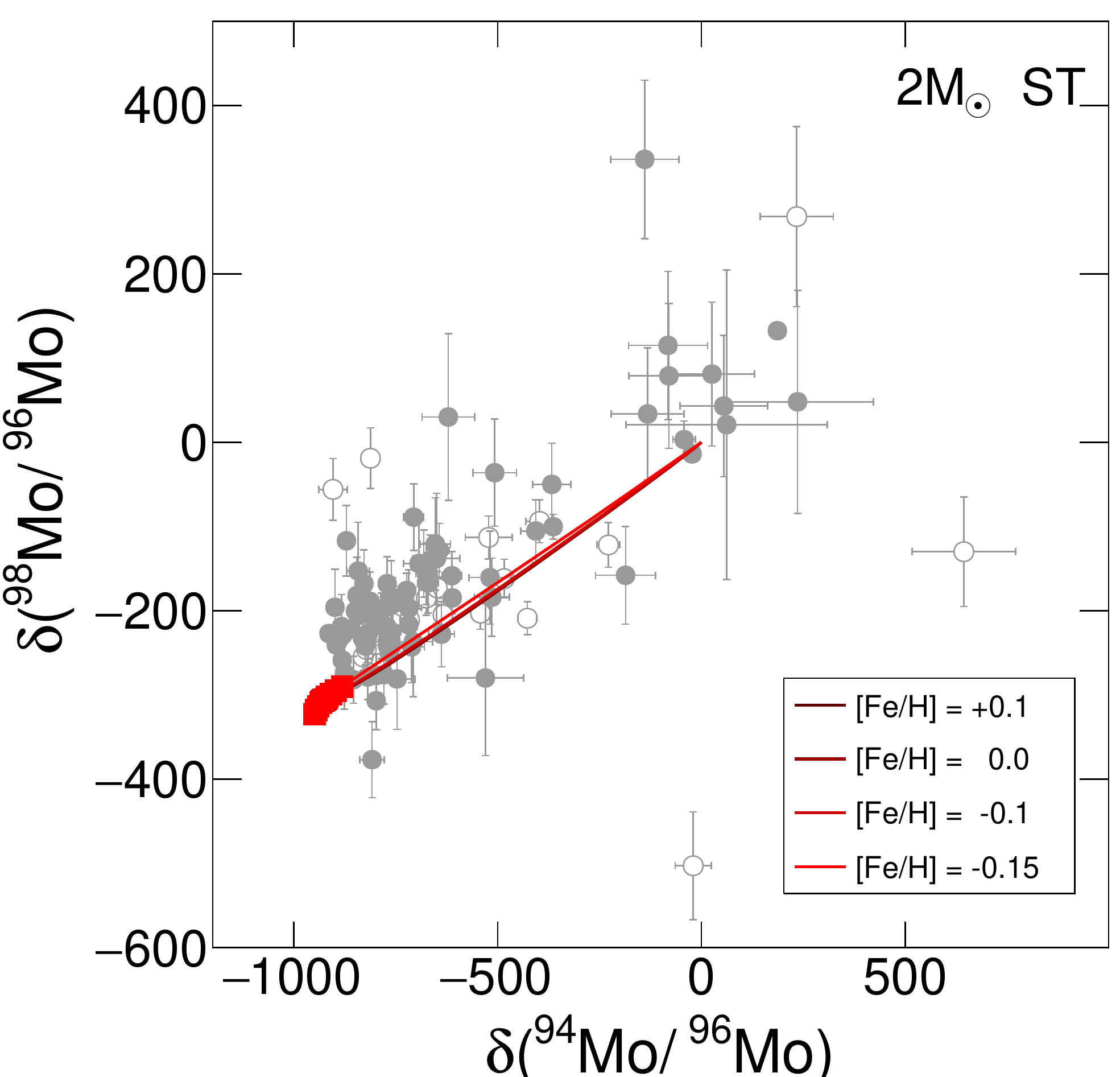}}
{\includegraphics[width=0.3\textwidth]{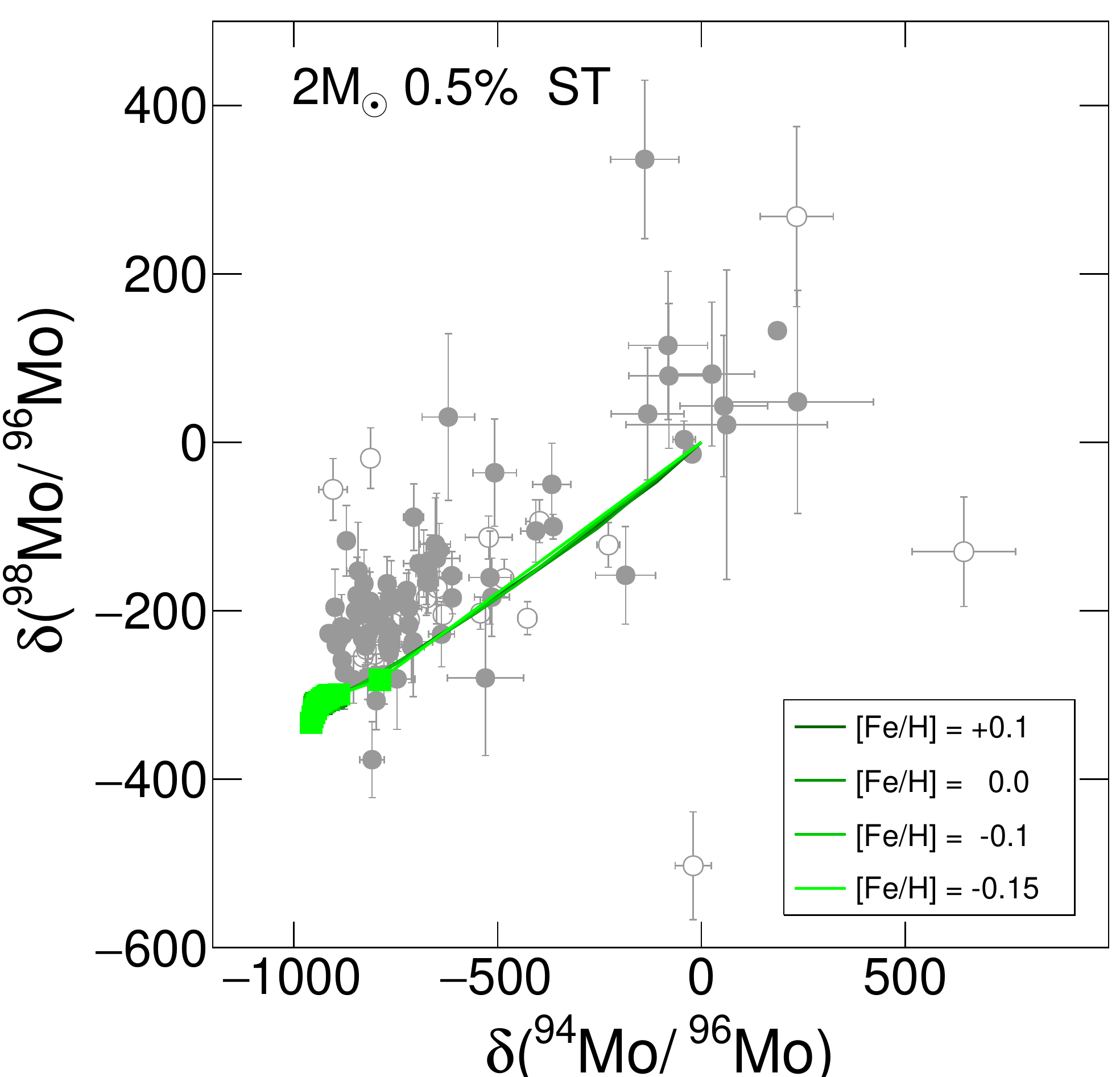}}
{\includegraphics[width=0.3\textwidth]{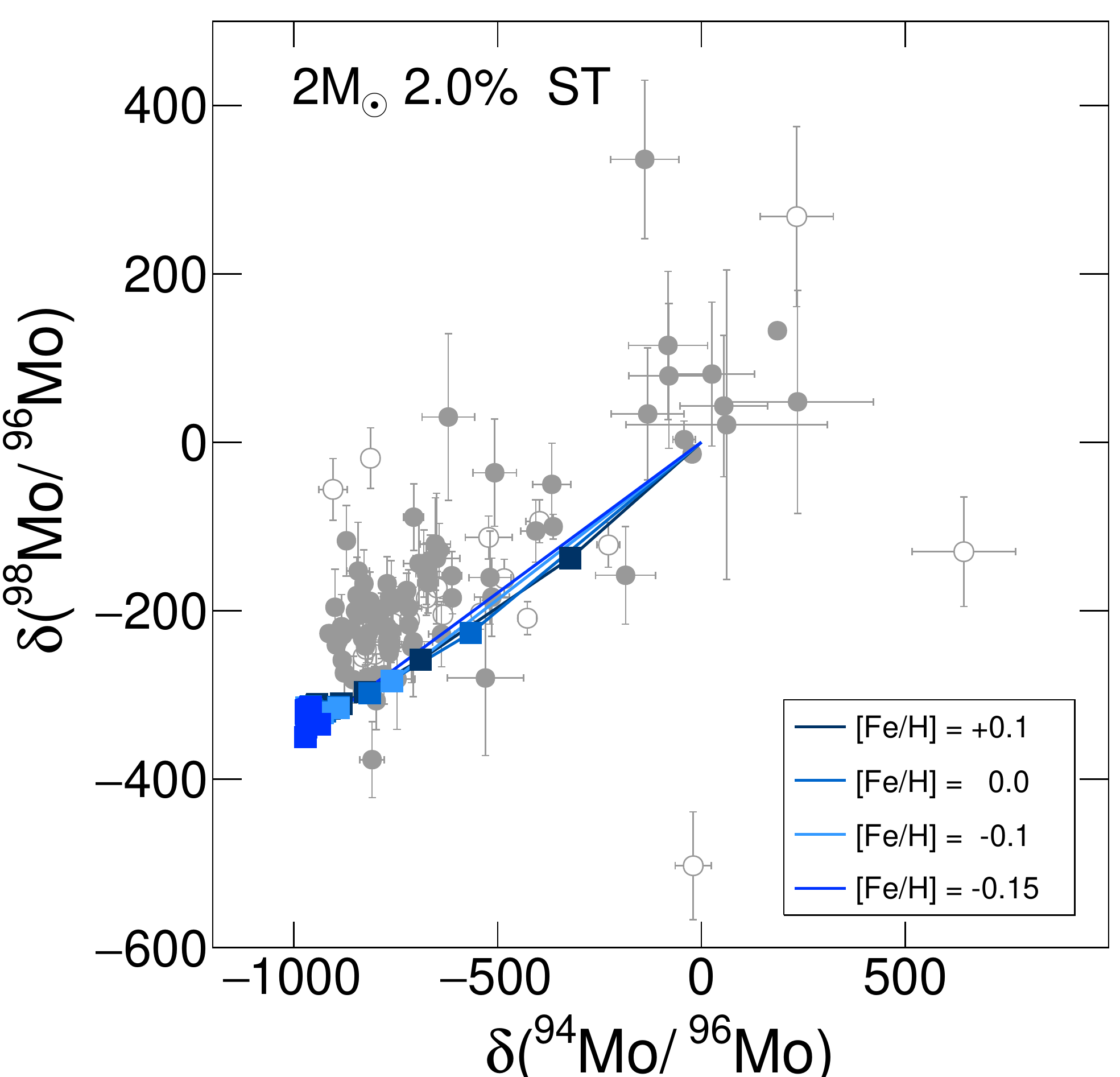}}
\caption{Same as Figure \ref{fig:Mo01}, but with the ratio $^{98}$Mo/$^{96}$Mo in the ordinate.\label{fig:Mo03}}
}
\end{figure*}

Concerning Mo, \citet{ste19} showed how its isotopic mix seems to derive from a combination of essentially { only two} components, one purely s-process controlled, the second of isotopically solar composition. In this
condition, as is known, the data in three-isotope plots distribute
along almost straight diagonal lines. Beyond this simple trend, however, there are tiny details, from which new lessons can be learned.

\begin{figure*}[t!!]
\centering{
{\includegraphics[width=0.3\textwidth]{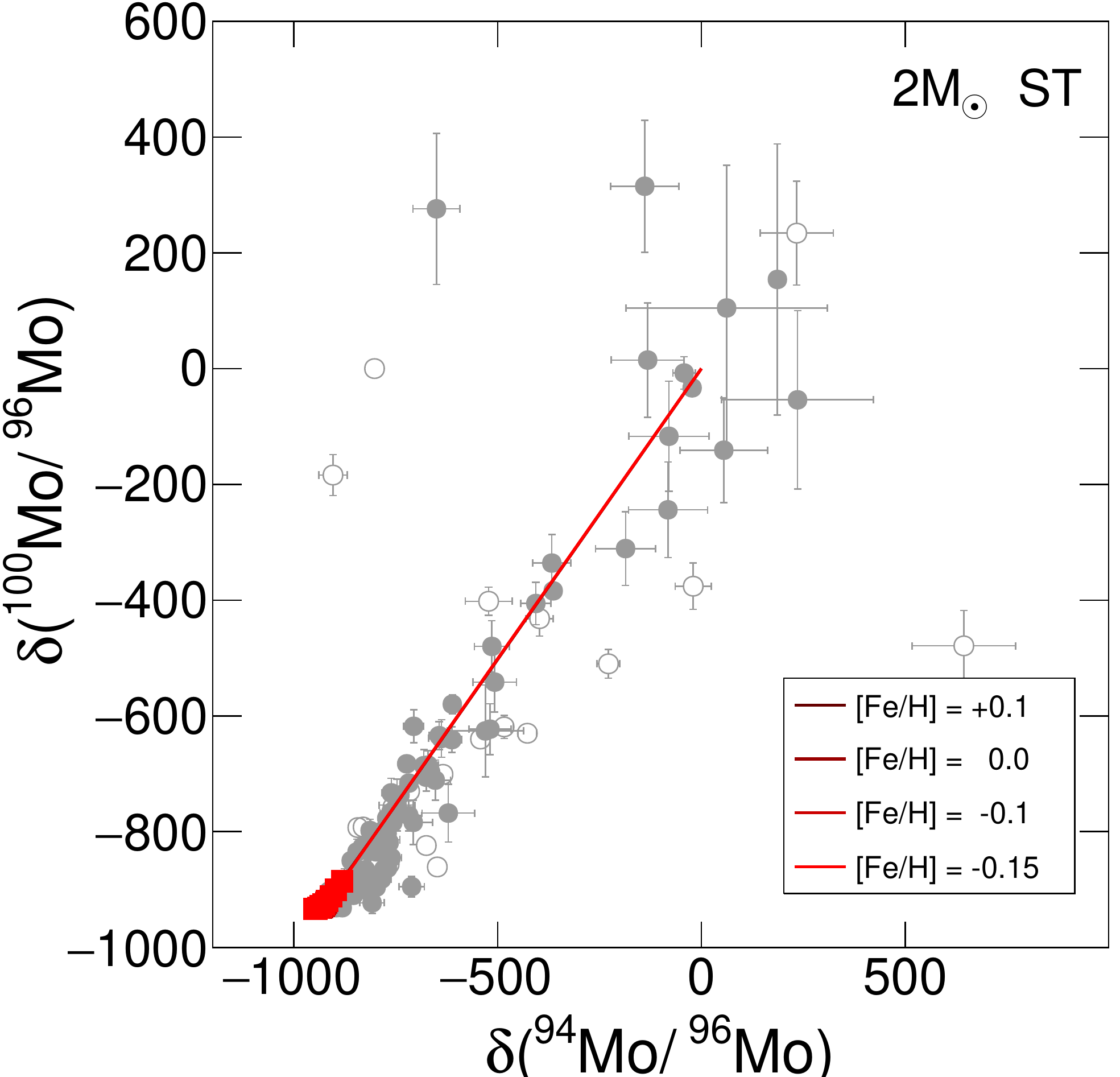}}
{\includegraphics[width=0.3\textwidth]{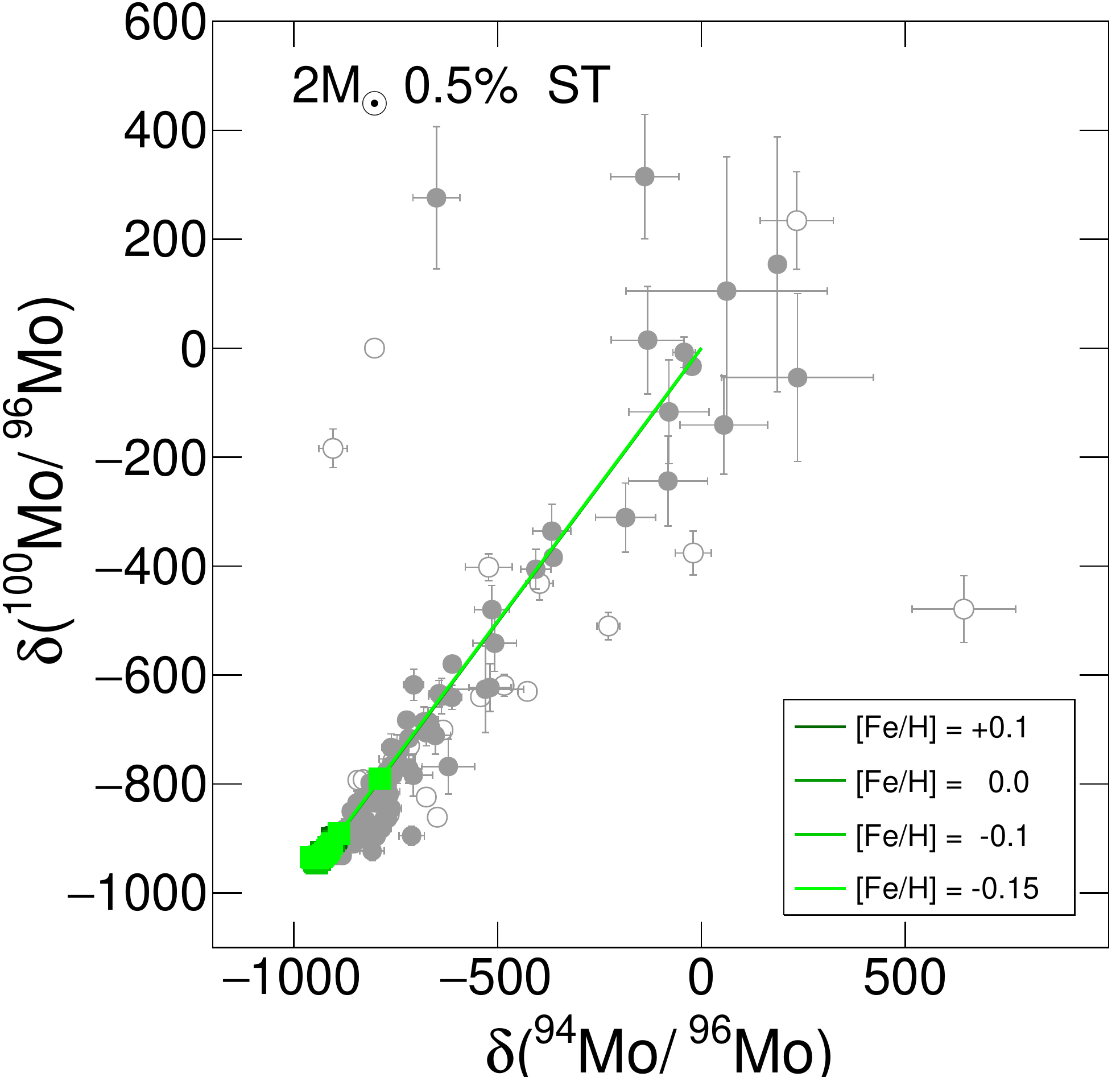}}
{\includegraphics[width=0.3\textwidth]{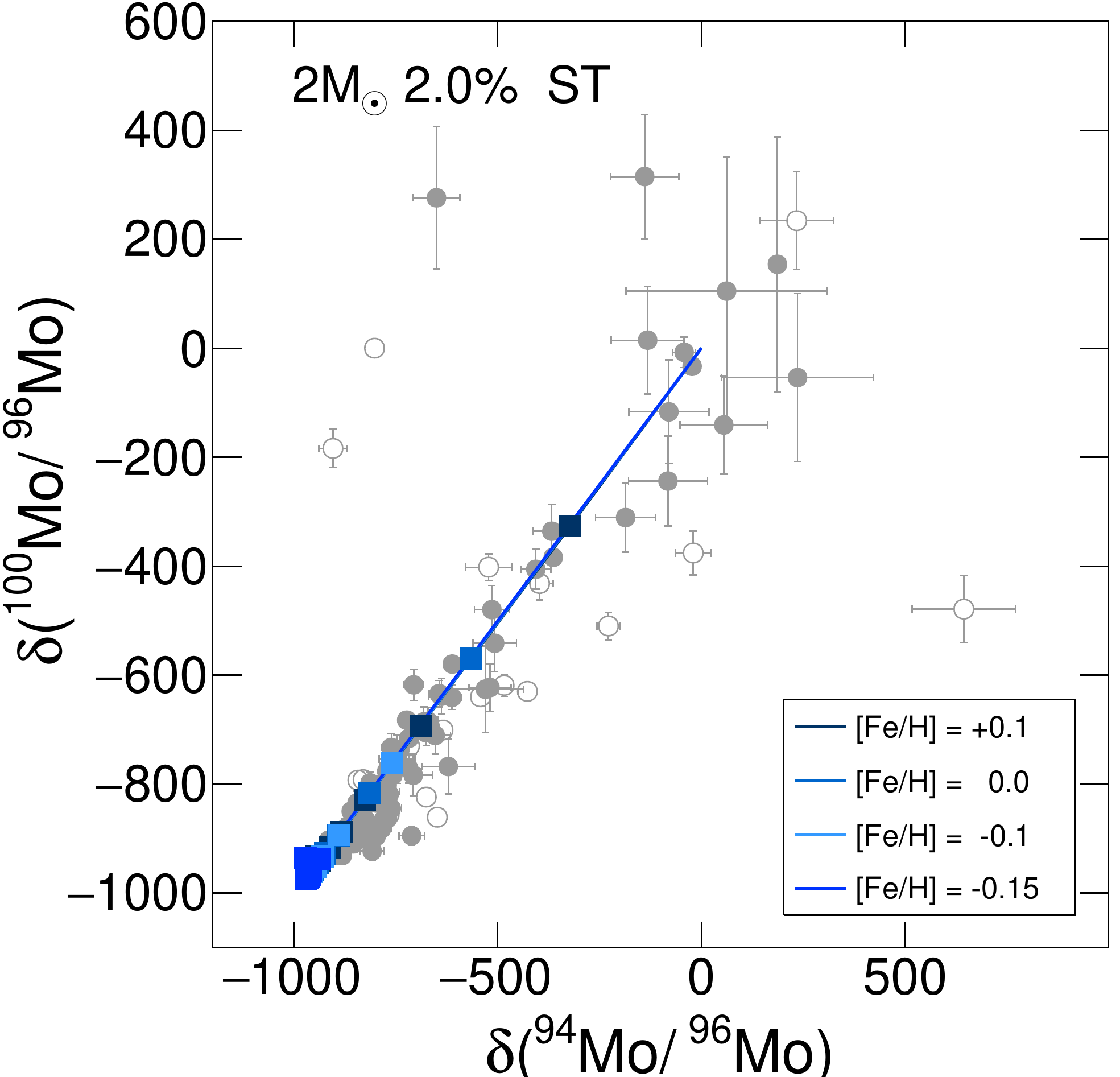}}
\caption{Same as Figure \ref{fig:Mo01}, but with the ratio $^{100}$Mo/$^{96}$Mo in the ordinate.\label{fig:Mo04}}
}
\end{figure*}

\begin{figure*}[t!!]
\centering{
{\includegraphics[width=0.3\textwidth]{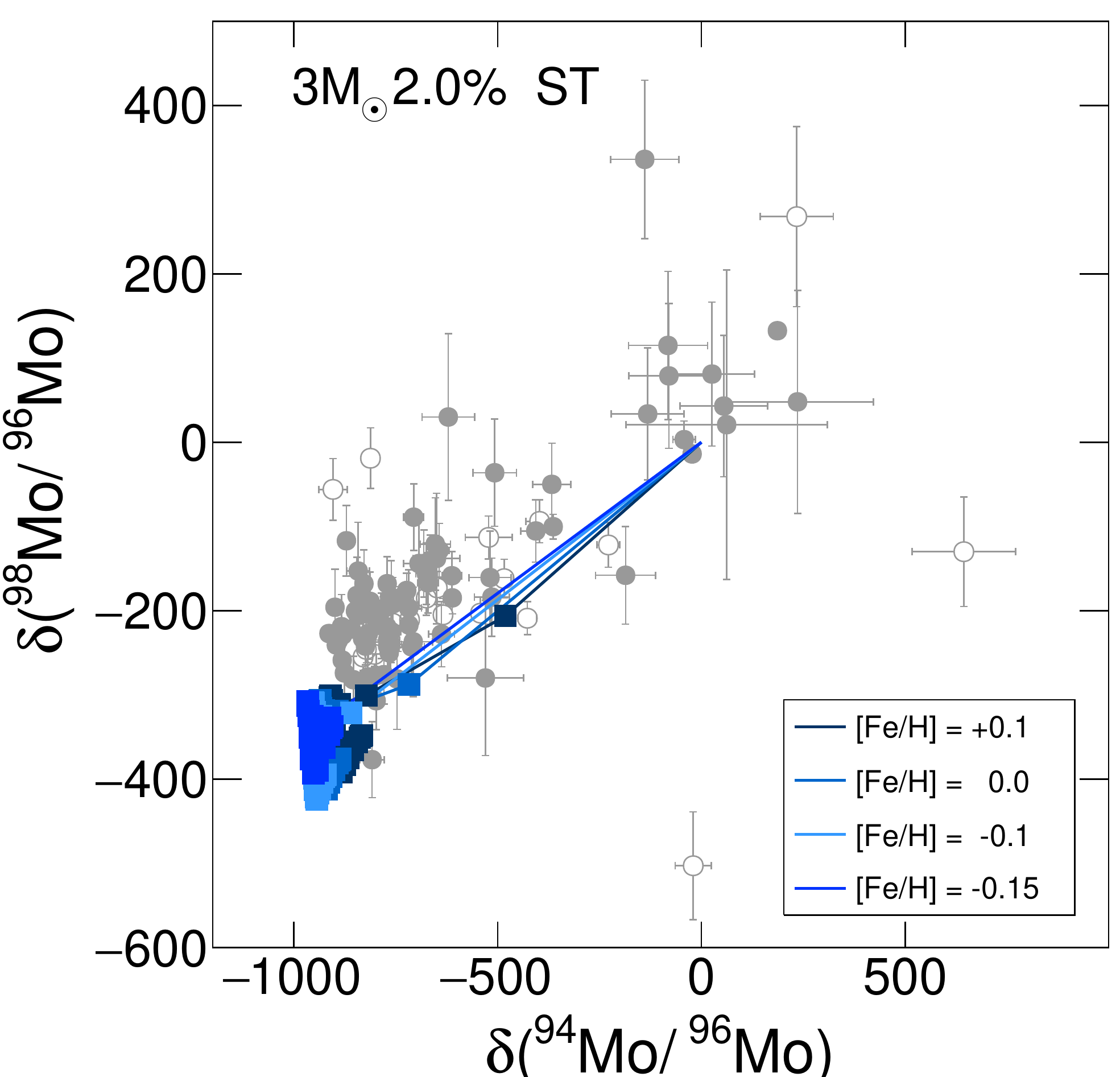}}
{\includegraphics[width=0.3\textwidth]{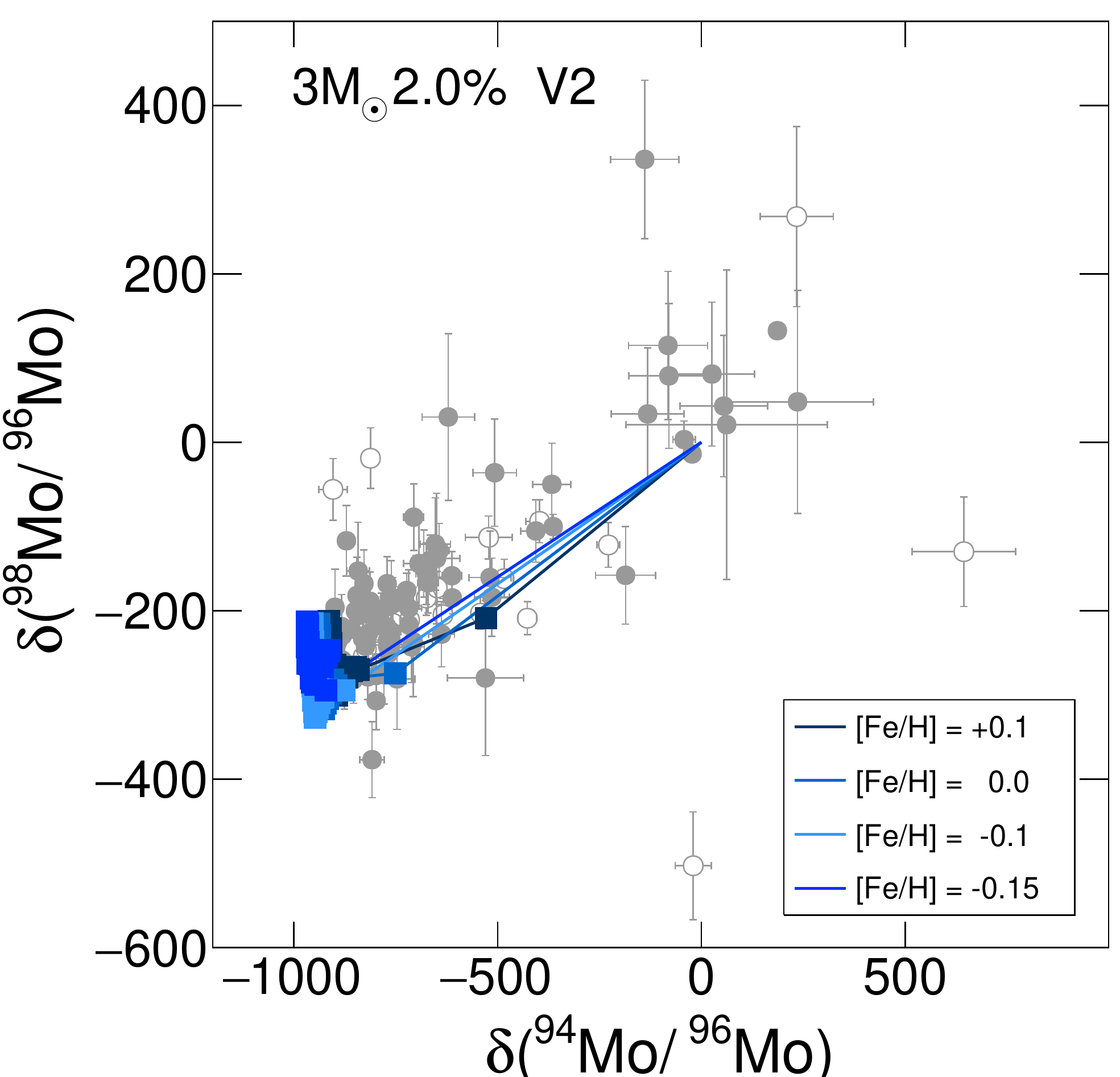}}
{\includegraphics[width=0.3\textwidth]{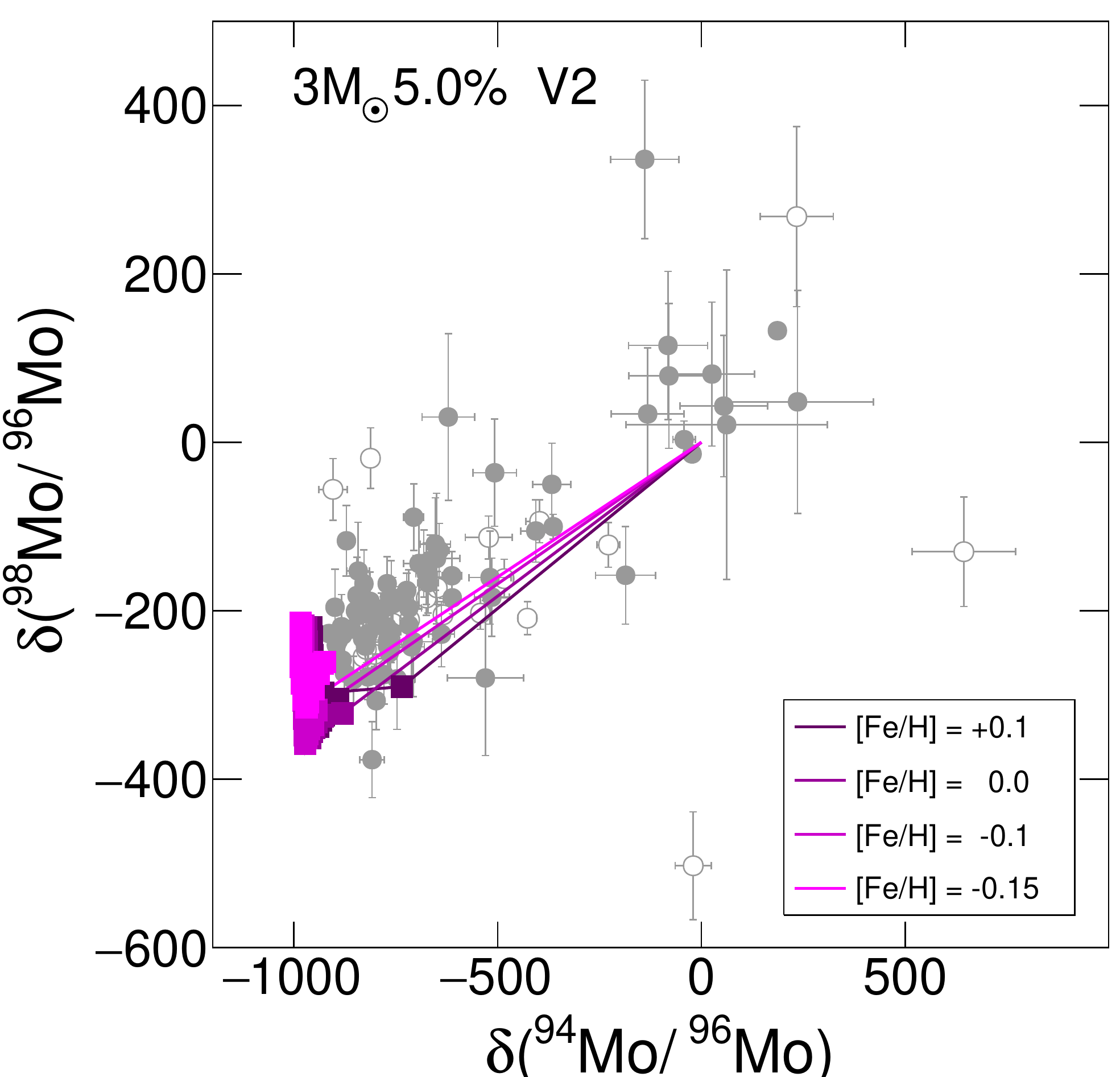}}
\caption{Same as Figure \ref{fig:Mo03}, but for a 3 \msb star, in which the left panel shows the standard wind case, for a dilution of 2\%, while the center and right panels show the V2 case, for dilutions of 2\% and 5\%, respectively.\label{fig:Mo06}}
}
\end{figure*}

\begin{figure*}[t!!]
\centering{
{\includegraphics[width=0.3\textwidth]{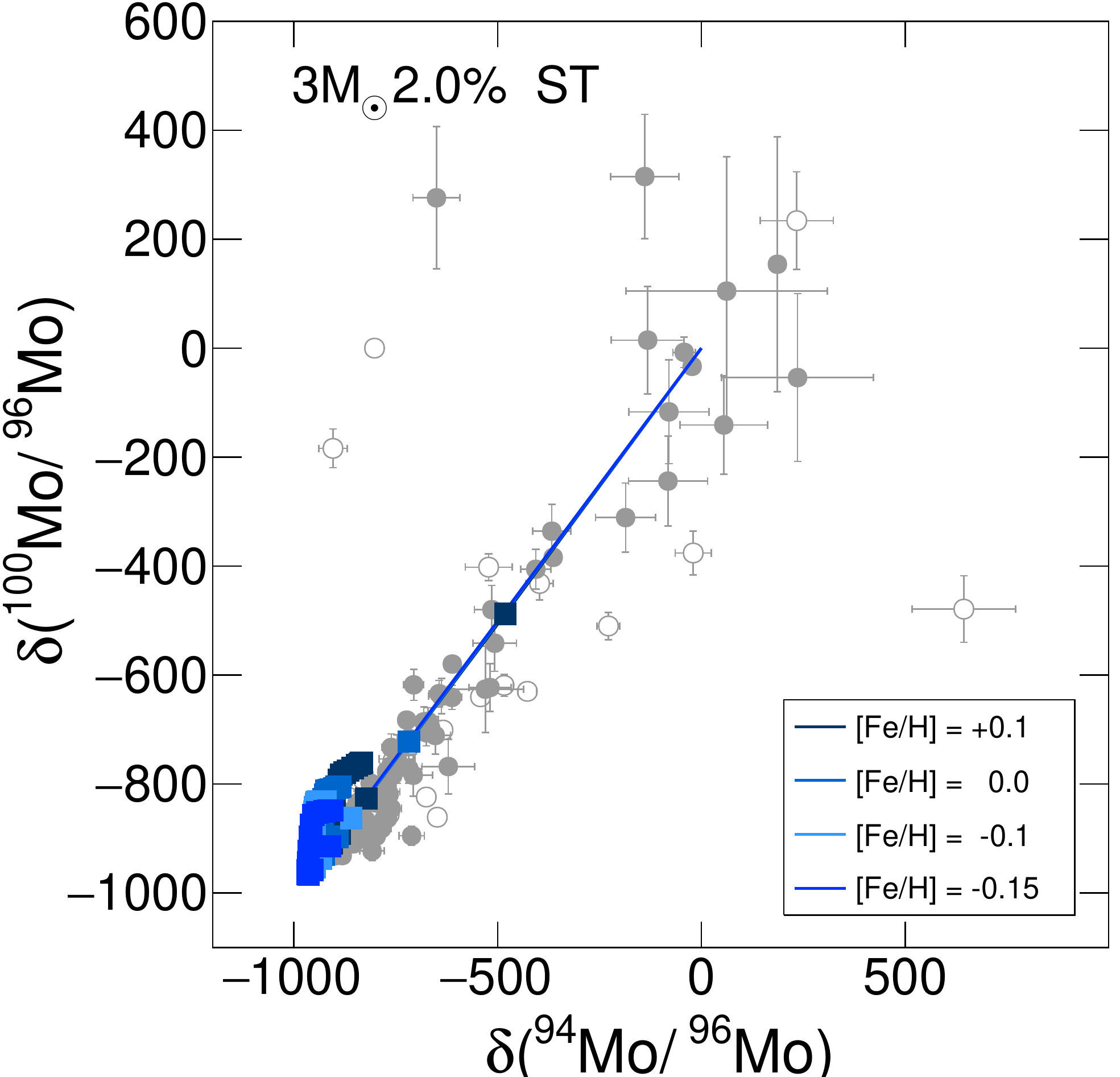}}
{\includegraphics[width=0.3\textwidth]{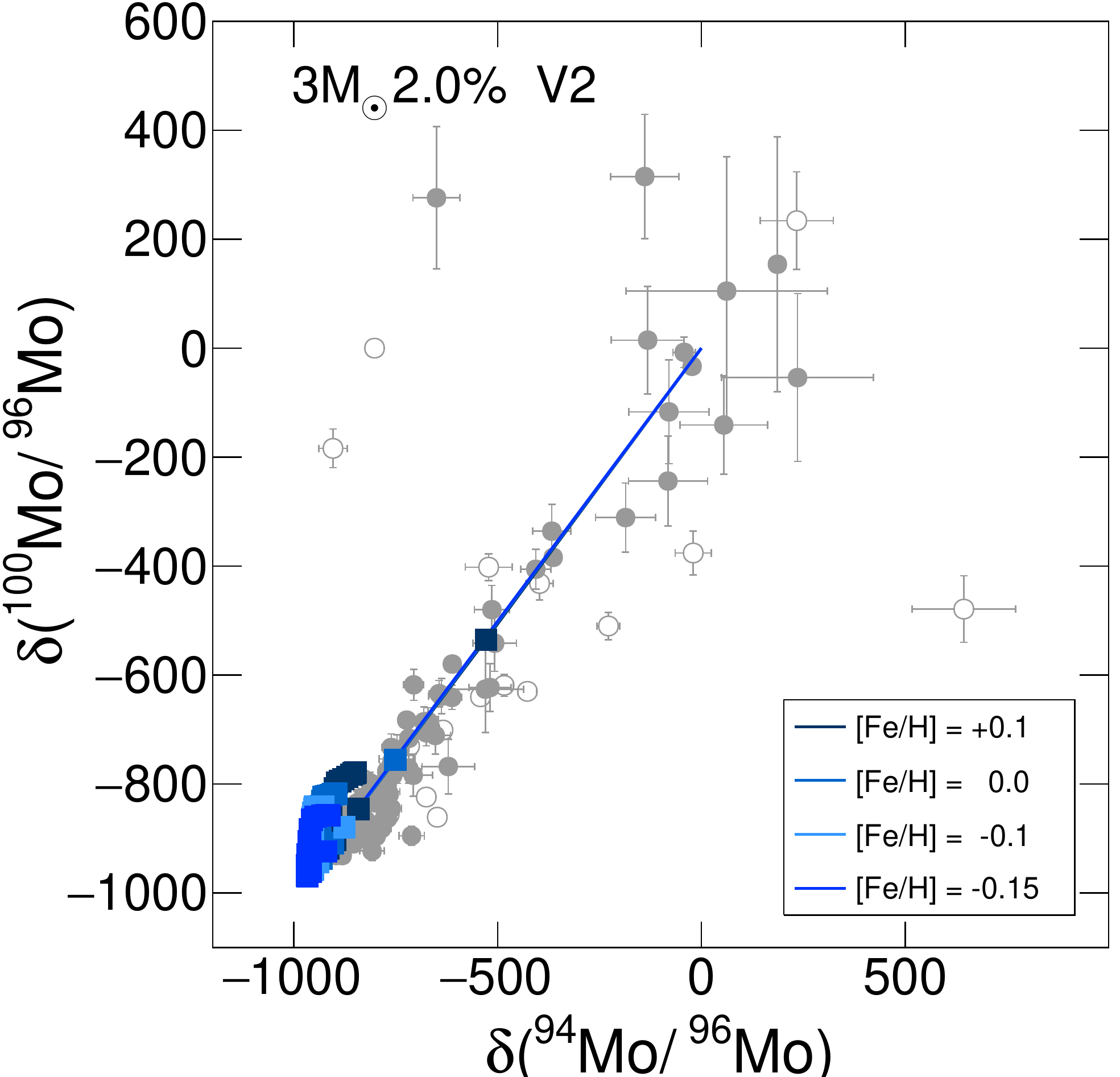}}
{\includegraphics[width=0.3\textwidth]{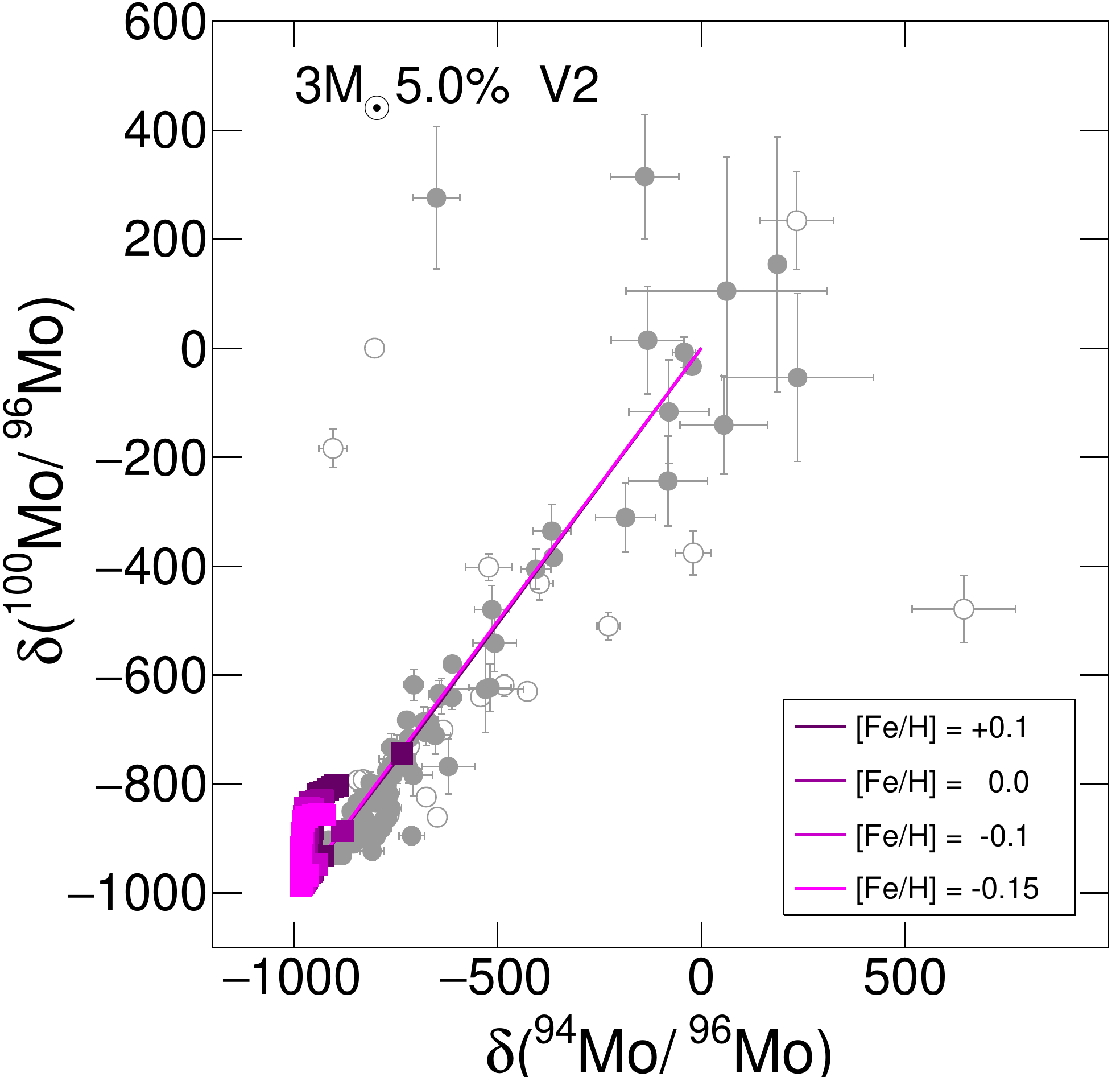}}
\caption{Same as Figure \ref{fig:Mo04}, but for a 3 \msb star, in which the left panel shows the standard wind case, for a dilution of 2\%, while the center and right panels show the V2 case, for dilutions of 2\% and 5\%, respectively.}\label{fig:Mo07}
}
\end{figure*}

Models for the C-rich envelopes tend to represent rather { exstreme}
admixtures, very close to the $s$-process end-member \citep[see the discussion in][]{liu+19}
The composition of the magnetized winds, instead, allows us to cover larger and larger portions of the area filled by the data, for increasing efficiencies of He-shell matter mixed. This is so because in the winds, through the mechanism sketched in Figure \ref{fig:cartoon}, one samples small portions of the C-rich 
component transported by magnetized structures and large portions 
of the envelope, even in phases preceding the final ones of the C-star formation.
As mentioned, small but important details reveal, even here, that
something needs to be improved. Let us review these details, starting with inspection of Figures from \ref{fig:Mo01} to \ref{fig:Mo04}, all
representing the envelop composition (left panels) and winds of 2 \msb models, enriched at the 0.5\% and the 2\% in C-rich material (central and right panels, respectivelly). We choose the ratio $^{94}$Mo/$^{96}$Mo as a common abscissa and the first evidence is that the
models at the TDU episodes (full dots) crowd at $\delta$ values for
$^{94}$Mo which are slightly, but clearly, too low with respect to
the measured  data. Some increase in $^{94}$Mo would solve the
problem and we believe that this is an indication either of a 
smaller value of its neutron-capture cross section than recommended in the K1 repository, { or of the need to invoke for $^{94}$Mo the subtle contributions from the chain passing through $^{93}$Nb and  
$^{94}$Nb.} %, which can be activated in a binary {\it bi-intrinsic} AGB star, as outlined in section \ref{sec:mo}}. 
Another evidence is shown by the heaviest isotopes $^{97}$Mo, $^{98}$Mo and $^{100}$Mo, see Figures from \ref{fig:Mo04} to \ref{fig:Mo07}; the last two comparing the data with 3 \msb models, the previous one with 2 \msb models. Here,
the straight diagonal line represents an average trend, over which
some significant vertical spread exist, at different levels for the different isotopes. This behavior is not accounted for with our
standard choice of the nuclear parameters, but can instead be interpreted by referring to our test case V2. Among the changes 
there introduced (see section \ref{sec:zrmo}), we enhanced the cross section of $^{95}$Zr to its upper limit allowed by the K1 recommendations, to take into account the suggestions for a larger cross section contained in the {\it BNL} repository. We also adopted, for $^{99}$Mo, the larger value provided by the same {\it BNL} site. Figures \ref{fig:Mo06} and \ref{fig:Mo07} show the effects of these changes on the plots involving $^{98}$Mo and $^{100}$Mo, in our 3 \msb models. In particular, Figure \ref{fig:Mo06} shows the effects, on $^{98}$Mo, of the first of
the mentioned changes. Increasing the (n,$\gamma$) cross section of $^{95}$Zr has the effect of feeding more efficiently the chain $^{96}$Zr, $^{97}$Zr, $^{97}$Nb, $^{97}$Mo, at the
expense of $^{96}$Mo. As a consequence, this last is slightly reduced. Since it is at the denominator of the isotopic ratio in the ordinate, its decrease induces higher values of the latter. Indeed, as the figure shows, in 3 \msb models one has originally a downward spread (left panel) in the model trend, worsening the fit with respect to the
2 \msb cases, while with the V2 choice the direction of the spread is reverted. 
Should one correct slightly $^{94}$Mo in the way already discussed, this would result 
in fitting quite well even the small dispersion of the observed points. A very similar change is induced by the test case V2 on $^{100}$Mo, this time being the 
direct effect of the larger cross section adopted for $^{99}$Mo, as illustrated 
in Figure \ref{fig:Mo07}.  

{We also recall that, in the case in which the abundance of a nuclide in 
grains is extremely poor, thus making very small also a connected isotopic ratio, a minimum sample pollution  (for example if a grain is coated by a very thin residue of solar matter) could affect the measured value in such a way to mimic a dilution of the progenitor stellar winds with further unpolluted solar material \citep{l+17}. This fact may account for the composition of grains that show Mo isotopic admixtures
close to the solar values.}

\section{Conclusions}\label{sec:the-end}
In this paper we analyzed the isotopic composition of heavy neutron capture 
elements measured in presolar MS-SiC grains of AGB origins, interpreting them 
as the outcome of $s$-process nucleosynthesis occurring in the AGB stages
of stars in the mass range 2$-$3 \ms, with close to solar metallicities (with 
[Fe/H]  values from $-$0.15 to +0.1). In the reference models we adopted the 
scheme of magnetically-induced mixing discussed in previous paper from our group and the composition of the stellar winds were simulated taking into
account the effects of flares and/or mass ejection from magnetic structures
ascending from the He-intershell and breaking in the outer layers of the atmosphere, depositing there C-rich matter and $s$-processed heavy elements.  

The comparisons made use of nuclear parameters available in the present literature, in particular adopting most neutron capture cross sections from the Kadonis 1.0
compilation (K1) and weak interaction rates from \citet{ty87}.

We found that our specific mixing scheme, discussed here in detail for the first 
time, greatly enhances the general agreement between model predictions and observed 
isotopic ratios (expressed in the $\delta$ notation) and on this basis we suggest that magnetic fields in AGB stars are important not only for driving neutron capture processes, but also for mixing their products into the circumstellar envelopes.

When the details of the comparisons (involving isotopic ratios of $s$-process elements near the $N = 50$ and $N = 82$ neutron magic numbers) are concerned, an
analysis of the input nuclear parameters and of their uncertainties  leads us to 
identify a series of crucial points on which new measurements would be welcome,
both for the neutron capture cross sections and for the decay rates of radionuclei 
in stellar plasmas. On this last point, we presented an anticipation from a work 
to be submitted, in which we showed how more precise theoretical studies of 
crucial decays can lead to remarkable changes in the rates commonly used in stellar models, available through the \citet{ty87} work. We showed in particular
how, in the example case of $^{134}$Cs, the temperature enhancement of its rate
is less steep than so far assumed by important factors (from 2.5 to about 30). On the basis of simulations of the operation of the plasma trap PANDORA (now under construction), we also suggested that variations of the type found in our  theoretical approach should be easily verified experimentally in the next few 
years.

Using test computations with ad-hoc modified input nuclear parameters, we also 
indicated where future nuclear physics efforts should concentrate. This includes re-evaluations of the  branching ratios on $^{84, 85}$Kr and of neutron captures for $^{88}$Sr, $^{95}$Zr, the Mo and Ba isotopes and the unstable nuclei $^{134, 135}$Cs. We also suggested that
measurements of the weak interaction rates of $^{94}$Nb and $^{135}$Cs in ionized plasmas should immediately follow those of $^{134}$Cs.

On the basis of the results possible with slightly modified values of the crucial input parameters, which are presently in the range permitted by uncertainties,
we find that there is no need to invoke stars considerably more metal-rich than the
Sun as sources for presolar grains, as sometimes suggested in the literature.

\acknowledgments{The PANDORA collaboration acknowledges the support of the Third Committee of the Italian Institute for Nuclear Physics (INFN-CSN3). 
DV acknowledges the financial support from the German-Israeli Foundation (GIF No. $I-1500-303.7/2019$)}

\bibliography{sicbiblio1}{}
\bibliographystyle{aasjournal}

\end{document}